\documentclass{article}

\usepackage[english]{babel}
\usepackage{physics}
\usepackage{subcaption} 

\usepackage[letterpaper,top=2cm,bottom=3cm,left=3cm,right=3cm,marginparwidth=1.75cm]{geometry}
\usepackage{amsmath, amsfonts, amssymb, amsthm}
\usepackage{enumerate, color}
\usepackage{multido}
\usepackage{url}
\usepackage{mathrsfs}
\usepackage{mathtools}
\usepackage{verbatim}
\usepackage{setspace}
\usepackage{float}

\usepackage{authblk}

\usepackage{tcolorbox}

\usepackage{tikz}
\usetikzlibrary{arrows.meta, decorations.markings, positioning, shapes.misc, calc}

\usepackage{graphicx}
\usepackage[colorlinks=true, allcolors=blue]{hyperref}
\usepackage{xcolor}
\usepackage{xspace}
\ExplSyntaxOn
\NewDocumentCommand{\dotsrectarray}{m m m O{solid}}{

  \fill[blue] (\clist_item:nn {#2}{1}, #1) circle (\dotsize);
  \fill[blue] (\clist_item:nn {#2}{2}, #1) circle (\dotsize);
  \fill[blue] (\clist_item:nn {#2}{3}, #1) circle (\dotsize);
  \fill[blue] (\clist_item:nn {#2}{4}, #1) circle (\dotsize);

  \fill[red] (\clist_item:nn {#2}{5}, #1) circle (\dotsize);
  \fill[red] (\clist_item:nn {#2}{6}, #1) circle (\dotsize);
  \fill[red] (\clist_item:nn {#2}{7}, #1) circle (\dotsize);
  \fill[red] (\clist_item:nn {#2}{8}, #1) circle (\dotsize);

  \draw[
    blue,
    line~width=\rectth,
    #4
  ]
  (\clist_item:nn {#3}{1}, #1-0.22)
  rectangle
  (\clist_item:nn {#3}{2}, #1+0.22);

  \draw[
    red,
    line~width=\rectth,
    #4
  ]
  (\clist_item:nn {#3}{3}, #1-0.22)
  rectangle
  (\clist_item:nn {#3}{4}, #1+0.22);
}
\ExplSyntaxOff
\newtheorem{theorem}{Theorem}
\newtheorem{lemma}[theorem]{Lemma}
\newtheorem{remark}[theorem]{Remark}

\newtheorem{proposition}[theorem]{Proposition}

\newtheorem{assumption}{Assumption}
\newtheorem{corollary}[theorem]{Corollary}

\newcommand{\caE}{\mathcal E}
\newcommand{\caC}{\mathcal C}
\newcommand{\caR}{\mathcal R}
\newcommand{\caI}{\mathcal I}

\newcommand{\caA}{\mathcal A}
\newcommand{\caF}{\mathcal F}
\newcommand{\caP}{\mathcal P}

\newcommand{\caG}{\mathcal G}
\newcommand{\caZ}{\mathcal Z}
\newcommand{\caN}{\mathcal N}

\newcommand{\sss}{w}

\title{Many-body Localization and Poisson Statistics in the Quantum Sun model}

\author[1]{Wojciech De Roeck}
\author[2]{Amirali Hannani}
\affil[1]{Institute for Theoretical Physics, KU Leuven, 3000 Leuven, Belgium \thanks{wojciech.deroeck@kuleuven.be}}
\affil[2]{Section of Mathematics, University of Geneva, 1205 Genève, Switzerland \thanks{amirali.hannani@unige.ch}}

\begin{document}
\maketitle

\begin{abstract}
The Quantum Sun model is a many-body Hamiltonian model of interacting spins arranged on the half-line (or the line).   
Spins at distance $n$ from the origin are coupled to the rest of the system via a term of strength $\alpha^n$, with $\alpha \in  (0,1)$.
From theoretical and numerical considerations, it is believed that this model undergoes a localization-delocalization transition at the critical value $\alpha=\frac{1}{\sqrt{2}}$.  We prove that, for $\alpha \ll \frac{1}{\sqrt{2}}$, the model is localized and that its spectral statistics are Poissonian. 
The main novelty of this result is that the model considered here is a genuine many-body model, in which the interaction has a profound effect. 

\end{abstract}

\section{Model and results}

\subsection{Introduction}\

The localization/delocalization phenomenon has been a central theme in mathematical physics ever since the pioneering work of P.\ Anderson \cite{Anderson}. The question is whether certain random linear operators have eigenvectors that are largely supported on a few basis vectors, or equally distributed over essentially all basis vectors, with respect to some natural basis.   Initially, this question was mainly studied in the context of Schr{\"o}dinger operators with random potentials, see \cite{Sinai1987,kunz1980spectre,Frohlich1983,Aizenman1993}, but in the last decades it has also been considered in many other models of random operators or matrices. 
Besides the eigenvector properties, the localization/delocalization dichotomy is often discussed on the level of spectral statistics, with localization corresponding to Poissonian statistics, whereas delocalization corresponds to GOE or GUE statistics, depending on the symmetry class. 

Over the past 20 years, the focus in the physics community has been on localization/delocalization in \emph{many-body systems} \cite{Basko2006,Gornyi2005interacting,oganesyan2007localization}. In the language of quantum mechanics, the original work on the Schr{\"o}dinger operator was concerned with a single quantum particle (or a wave), i.e.\ a single degree of freedom.  In many-body models, the number of degrees of freedom is proportional to spatial volume, and hence it tends to infinity.    Discussion of the localization/delocalization phenomenon in this framework requires more setup and subtlety. However, the biggest challenge is that the mathematics seems considerably harder, and rigorous results are scarce. Indeed, the results fall short of clearly establishing a localized or delocalized phase, though some preliminary steps were taken in \cite{ChulaevskySuhov2009MultiParticle,AizenmanWarzel2009LocalizationBounds} and in \cite{Imbrie2016a,deroeck2025absencenormalheatconduction}.   Another approach consists of considering a many-body system close to its ground state, or, more generally, close to a spectral edge. This has been done in classical mechanics, where the most emblematic model is the discrete nonlinear Schr\"odinger equation with disordered potential, see \cite{frohlich1986localization, poschel1990small, johansson2010kam,DeRoeck_2026};  for quantum systems, see \cite{Mastropietro2017, elgart2022localization, beaud2017low,yin2024eigenstate}.   Yet another approach is to consider mean-field models \cite{laumann_pal_2014, baldwin_laumann_2017,manai_warzel_2023}. These results are conceptually very close to the aim of our paper, and yet, on a technical level, they seem to be only weakly related. 
For example, consider many-body models that are analyzed near spectral edges, more specifically at energies that are not away from a spectral edge by an extensive amount. In such a situation, the dimension of the effective Hilbert space mostly does not grow exponentially, but rather polynomially, in the number of degrees of freedom. This is not the case in our model: the effective Hilbert space dimension is exponential.  It is precisely this circumstance that brings the possibility of an avalanche-driven (see below) phase transition, and this was also the rationale for introducing the Quantum Sun model. 
 In the present work, we prove for the first time localization and Poisson statistics in a genuine many-body model in the above sense, away from spectral edges.  
Finally, we mention that there are many (often rigorous) related works \cite{benettin1988nekhoroshev,wang2009long,fishman2012nonlinear,fishman2009perturbation,cong_long-time_2021,cong_2024,PhysRevB.80.115104,basko2011weak,Huveneers_2013,de2015asymptotic,DeRoeck2013,DEROECK2023129245,mcdonough2025nonperturbativelyslowspreadquantum,DeRoeck2023LongPersistence} that exhibit many-body localization effects for large but finite times that do not scale with the number of degrees of freedom.  Such results have been labeled by some authors as ``asymptotic localization" or ``quasi-localization" or ``prethermalization".   These results are of utmost relevance from the point of view of experimental physics, but, again, they pose rather different technical challenges compared with the question of true localization which in principle is about the behavior at infinite time.

  \subsection{The  Model}\label{sec: the model}
The ``Quantum Sun model" treated here was conceived as a class of toy models for many-body localization and delocalization  \cite{Roeck2016,luitz2017small,Goihl2019, Gopalakrishnan2019, Potirniche2019, Crowley2020, Suntajs2022, Crowley2022}. The model should be thought of as consisting of a small Hamiltonian bath coupled to a chain of quantum spins that, in the absence of the bath, would be many-body localized in a trivial manner. \vspace{2mm}
  
Given $n \in \mathbb{N}$, we consider the Hilbert space $\mathcal H= \otimes_{i=1}^n \mathbb{C}^2$. The standard inner product is denoted by $\langle\cdot,\cdot\rangle$.  We write $Z_i,X_i$ for operators acting 
as the Pauli matrices 
$ Z = 
\begin{bmatrix}
1 & 0 \\
0 & -1
\end{bmatrix}
$ and 
$
X= \begin{bmatrix}
0 & 1 \\
1 & 0
\end{bmatrix} $ 
on the $i$-th leg of the tensor product (the $i$-th copy of $\mathbb{C}^2$), and as the identity on the other legs.   
With these conventions,  the Hamiltonian is the following self-adjoint matrix on $\mathcal{H}$:
\begin{equation} \label{eq: main model1}
    H_n = H_B+ \sum_{i=n_B+1}^{n} \left( h_i Z_i + \alpha^{i-n_B} g_i X_1 X_i  \right), 
\end{equation}
where
\begin{enumerate}
    \item $n_B \in \mathbb{N}$ and $n_B<n$. The ``bath Hamiltonian" $H_B$ is a self-adjoint matrix that acts non-trivially on the spins $1,\ldots, n_B$, and as the identity on the spins $i$ with $i>n_B$. For convenience we assume that $\|H_B\| \leq C_B $ for an absolute constant $C_B>1$, where $\|\cdot \|$ denotes the operator norm.   The parameters $ n_B \in \mathbb{N}$ and $ C_B>1$ are arbitrary but fixed. In particular, they do not grow with $n$.
    \item The parameter $\alpha$ is a number in $(0,1)$. Our results concern the regime $\alpha \ll 1$.
\item  The collection $\{h_i,g_i,\dots\}_{i=1}^{\infty}$ is a sequence of  i.i.d.\ random variables on a probability space $(\Omega,\caF,\mathbb{P})$. The expectation w.r.t. $\mathbb{P}$ is denoted by $\mathbb{E}$. For simplicity, the variables $h_i$ and $g_i$ are chosen to be uniform on $[-\frac12,\frac12]$. 
\end{enumerate}
Notice that $H_B$  is a deterministic matrix, which is independent of the random variables above. For our results, the bath Hamiltonian $H_B$ will play almost no role, save for making the model non-trivial. However, in the literature on the model, and in particular in the numerical studies, the bath is included, and therefore we include it as well. We comment further on the bath Hamiltonian $H_B$ and the interaction term $X_1X_i$ in Section \ref{rmkL intro}.\\
The model is believed to have a phase transition at a finite value of the coupling constant $\alpha$. For $\alpha > 1/\sqrt{2}$, the model is believed to be delocalized, and in this regime, one believes that, in the relevant energy range,  the spectral statistics are best described by those of a GOE matrix of dimension $2^n$. We will not further describe the delocalized phase, for which no rigorous results exist. 
For $\alpha < 1/\sqrt{2} $, the model is believed to be localized (in the basis of $\alpha=0$ eigenstates), with Poissonian spectral statistics, and we prove this for $\alpha \ll 1$.  
Even though the model is many-body, we find it interesting to compare it with a certain random matrix model, namely the \emph{ultrametric ensemble}.  This model was introduced in \cite{fyodorov2009anderson} and its localization/delocalization phases were rigorously exhibited in \cite{vonSoosten2018,vonSoosten2019}.  The two models have a rather similar hierarchical structure; see \cite{vsuntajs2024similarity} for a discussion.

\subsubsection{Motivation}
Given the recent conception of the Quantum Sun model, the motivation behind our work may not be obvious, especially to a mathematical audience.  Let us briefly comment on this issue. First, we begin by recalling the influence of this model on the many-body localization (MBL) community; see e.g.\ \cite{Thiery2018, Dumitrescu2019, Morningstar_2020}:

\begin{enumerate}
 \item The delocalization witnessed in this model is the main reason why one believes that systems in spatial dimensions greater than one cannot be many-body localized.  
\item This delocalization is tied to the so-called \emph{avalanche instability}. This instability has had a significant impact on theories of the many-body localization transition in $d=1$ spin chains.
\end{enumerate}

We now explain, at a heuristic level, how the Quantum Sun model is related to the stability problem for many-body localization. Consider a disordered one-dimensional spin chain, for instance with random-field terms $h_i Z_i$, where the $h_i$ are independent random variables uniformly distributed in $[-W,W]$. With small but non-zero probability, the disorder configuration contains a region where the field is anomalously flat. Such a region is not expected to behave as a localized system. Rather, it is naturally viewed as a small ergodic inclusion. In accordance with the eigenstate thermalization hypothesis, its Hamiltonian may then be modeled by a random matrix. This is the role of the finite-dimensional bath in the Quantum Sun model.

Suppose that this inclusion is adjacent to a long stretch of chain with a typical disorder realization. The avalanche question is whether the inclusion remains a finite thermal spot, or whether it thermalizes the neighboring degrees of freedom successively. To isolate this mechanism, one assumes that the typical stretch is many-body localized. We perform a quasi-local change of basis to a basis in which the long stretch is localized. The local coupling between the ergodic spot and the stretch, in this basis, becomes a quasi-local coupling, with matrix elements decaying exponentially with the distance from the inclusion. After this rotation, keeping only this leading structure gives precisely the architecture of the Quantum Sun model: an ergodic seed coupled to a chain of localized spins, with couplings decreasing geometrically away from the seed. In this correspondence, the parameter $\alpha$ measures the effective localization length of the surrounding MBL region.

This heuristic correspondence translates the stability problem for MBL into the study of a localization--delocalization transition in the Quantum Sun model. In particular, the expected absence of stable MBL in dimensions $d>1$ is reflected in the conjecture that, in the corresponding higher-dimensional version of the model, the critical value satisfies $\alpha_c=0$.

Our localization result, together with Poisson statistics, gives a rigorous realization of the stable side of this picture. It shows that, in one dimension and for sufficiently small $\alpha$, the thermal inclusion does not trigger an avalanche. In this effective sense, it supports the stability of strongly disordered one-dimensional MBL against the avalanche mechanism.

\begin{figure}\label{fig: quantum sun model}
\centering
\begin{tikzpicture}[>=Stealth, scale=1.2, every node/.style={scale=0.8}]

  \node[draw, fill=pink!60, rounded corners=2pt, minimum width=1.5cm, minimum height=0.8cm] (bath) at (0,0) {\textbf{Bath}};
  
  \node at (0, -0.5) {\(\underbrace{\hspace{1.5cm}}_{\text{\(n_B\) sites}}\)};

  \foreach \i [evaluate=\i as \x using {1.8 + (\i-2)*0.8}] in {2,...,13} {
    \node[circle, fill=cyan!70, minimum size=8pt, inner sep=2pt] (site\i) at (\x, 0) {};

    \ifodd\i
      \draw[blue, ->] ($(site\i.center)+(0,-0.25)$) -- ($(site\i.center)+(0,0.25)$);
    \else
      \draw[blue, ->] ($(site\i.center)+(0,0.25)$) -- ($(site\i.center)+(0,-0.25)$);
    \fi
  }

  \draw[->, line width=0.9mm,shorten >=1mm] (bath.north east) to [bend left=45] (site5.north);
  \draw[->, line width=0.6mm, shorten >=1mm] (bath.north east) to [bend left=45] (site8.north);
  \draw[->, shorten >=1mm] (bath.north east) to [bend left=45]  (site11.north);
  \draw[->, thin, shorten >=1mm] (bath.north east) to [bend left=45](site13.north);

  \node[rectangle, fill=white] at ($(site11.north)+(0,0.5)$) {\(g_{\ell+n_B} \alpha^{\ell}\)};

  \node at ($(site2.south)+(0,-0.4)$) {\scriptsize \(n_B + 1\)};
  \node at ($(site11.south)+(0,-0.4)$) {\scriptsize \(n_B + \ell\)};

\end{tikzpicture}
  \caption{Quantum Sun model: The farther to the right, the weaker the coupling of the spins to the bath. In the absence of the coupling to the bath, the spins on the right do not interact.}
\end{figure}





\subsection{Fock-space Localization} \label{sec: fock space localization}

By ``localization", we refer to the situation in which most eigenstates are predominantly supported on one or a few eigenstates of the uncoupled system, i.e.\ with $\alpha=0$.  Our first theorem states that we have localization in this sense.  The uncoupled model is denoted by $H_n^{(\mathrm{free}')}$ and it is displayed in \eqref{eq: h free prime}.
To describe its eigenvectors, we introduce some notation. 
Let $\Sigma_n=\{\pm 1\}^n$ be the set of \emph{configurations}. To each configuration $\sigma=(\sigma(i))_{i=1,\ldots,n} \in \Sigma_n$, we can associate a one-dimensional projector $P_\sigma$ given by
$$
P_{\sigma}= \prod_{i=1}^n\delta_{Z_i-\sigma(i)},
$$
where $\delta_{x}$ is the Kronecker delta, i.e.\   $\delta_0=1$ and $\delta_x=0$  if $x \neq 0$, and the above expression should be understood in the sense of functional calculus.   
Similarly, we let
$$P_{\sigma_{\geq\ell}}= \prod_{i=\ell}^n \delta_{Z_i-\sigma(i)}.  $$
Obviously the spectral projectors of $H_n^{(\mathrm{free}')}$ are of the form $P^{(B)}_{j}\otimes P_{\sigma_{\geq n_B+1}}$ for some eigenprojector $P^{(B)}_j$  of $H_B$ (after choosing an eigenbasis of $H_B$) with $1\leq j \leq 2^{n_B}$, and some $\sigma \in \Sigma_n$ of which the first $n_B$ components are irrelevant. \vspace{2mm}

For $1 \leq j \leq 2^n$, let $\psi_j$ be unit eigenvectors of $H_n$, ordered according to increasing eigenvalue.   

\begin{theorem}\label{thm: fock space localization}
 For any $0 \leq \alpha< 1/8$, and any $\theta \in (2/3,1)$ such that $\alpha^{\theta}<1/4$, and $\alpha^{1-\theta}<1/2$, the following holds:  There is a random variable $\ell_*$ with values in $\mathbb{N}_0=\{0,1,\dots\}$ satisfying 

$$\mathbb{P}(\ell_* \geq \ell) \leq C_{\alpha,\theta,B} (4\alpha^{\theta})^{\ell} $$ 
and such that the following holds: there are bijections $ \Sigma_n \to  \{1,\ldots, 2^n\} : \sigma \to j(\sigma)$ for $n \in \mathbb{N}$, such that, for all $n>\ell>\ell_* $ and for all $ \sigma \in \Sigma_n$,   
$$  \langle \psi_{j(\sigma)},P_{\sigma_{\geq\ell}} \psi_{j(\sigma)} \rangle \geq 1-C_{\alpha,\theta,B}(2\alpha^{(1-\theta)})^{\ell},   $$
where, in the above expressions, $C_{\alpha,\theta,B}>0$ depends on $\alpha,\theta,n_B,\|H_B\|$,  and is independent of $\ell$ and $n$. 
\end{theorem}
The auxiliary parameter $\theta$ could actually be taken in $(0,1)$, but we restrict it to $(2/3,1)$ for the sake of simplicity in later sections.  
This theorem thus states that we can associate to any configuration $\sigma \in \Sigma_n$  an eigenvector $\psi_{j(\sigma)}$  and that this assignment is, in particular, meaningful from spin $\ell_*$ onwards.  To make the connection with more traditional measures of localization explicit, we introduce the inverse participation ratio  $\mathrm{IPR}(\psi_j)$ w.r.t.\ the basis defined by $\sigma\in \Sigma_n$
\begin{equation}\label{eq: participation ratio}
\mathrm{IPR}(\psi_j)= \left(\sum_{\sigma \in \Sigma_n}  |\langle \psi_j, \sigma \rangle|^4 \right)^{-1}.
\end{equation}
Here we slightly abuse notation by writing $\sigma$ for the unit vector in the range of the 1d projector $P_\sigma$.
Our result above implies that 
\begin{equation}\label{eq: participation ratio result}
\mathrm{IPR}(\psi_j) \leq   C\,{2^{\ell_*}}  
\end{equation}
uniformly in the choice of eigenvector $\psi_j$, where $\ell_*$ is the random variable mentioned in Theorem \ref{thm: fock space localization}. 
In contrast, in the delocalized phase, one expects that  $\mathrm{IPR}(\psi_j)$ is essentially of the order of the dimension of the full Hilbert space, i.e.\  $2^n$, for most eigenvectors $\psi_j$ and with high probability.

The proof of Theorem \ref{thm: fock space localization} is short and intuitive; see Sections \ref{sec: term} and \ref{sec: proof of thm loc};  the bound \eqref{eq: participation ratio result} follows in a straightforward way.  In fact, to prove Theorem \ref{thm: fock space localization} and   the bound \eqref{eq: participation ratio result}, we do not need the variables $g_i$ to be random. In particular, if $g_i=g$ for some fixed $g$ with $|g|\leq 1$, then these statements hold just as well, as one can easily verify by inspection of the short proof.


Phrased in this way, the localization-delocalization dichotomy is reminiscent of the picture that is expected for random Schr\"odinger operators or that is proven for some types of random matrices.   This is the reason that the localization described here would rather be called ``Fock-space localization" in the context of many-body models. 
In many-body models where the interaction does not decay along the chain (say, in the disordered XXZ model), one expects that $\mathrm{IPR}(\psi_j)$ is exponentially large in $n$ on both sides of the transition, but with different exponential rates, see the 
references given in Section \ref{sec: the model} for a more detailed discussion.

\subsection{Level statistics} \label{sec: level statistics}
First, we state the assumption on the bath Hamiltonian $H_B$. 
\begin{assumption}[Non-degeneracy assumptions on $H_B$] \label{ass: nondegeneracy}
$H_B$ is non-degenerate and non-anti-degenerate:  Let us denote the eigenvalues of $H_{B}$ by ${E^{(B)}_i}$ with $i=1,\ldots, 2^{n_B}$. Then our assumption reads
\begin{equation} \label{eq: assumption 1}
\min_{i \neq j} |E^{(B)}_i-E^{(B)}_j|>0, \qquad \min_{i,j} |E^{(B)}_i+E^{(B)}_j|>0.
\end{equation}
\end{assumption}
We now turn to the eigenvalue process of $H_n$, locally around the origin. 
Let us denote the eigenvalues of $H_n$ by  $\{E_i\}_{i=1}^{2^n}$, ordered such that $E_i \leq E_{i+1}$.   We define the following point process: 
\begin{equation} \label{sec: stat def: point process}
    \xi_n := \sum_{j=1}^{2^n} \delta_{(E_j/\sss_n)}(dx), 
\end{equation}
where
\begin{equation} \label{sec: stat sn}
    \sss_n:= \mathsf{s} \cdot 2^{-n}\sqrt{2 \pi n}, 
\end{equation}
is the mean eigenvalue spacing around energy $0$, where $\mathsf{s}^2= \frac{1}{12}$ is the variance  of $h_i $. Moreover, $\delta_{a}(dx)$ denotes the Dirac delta measure at $a$.

Let $\bar \xi$ be the standard Poisson point process  on  $\mathbb{R}$ with intensity one and denote the corresponding expectation by $\bar{\mathbb{E}}$. Recall that the expectation with respect to $\mathbb{P}$ is denoted by $\mathbb{E}$. 
Our main result states that $\xi_n$ converges to $\bar\xi$ in an appropriate way.  To state this more precisely, let  $ C_{K}^+(\mathbb{R})$ be the space of compactly supported non-negative continuous functions on $\mathbb{R}$.  For any point process $\xi$ and $ \varphi \in C_K^+(\mathbb{R})$, we abbreviate
\begin{equation} \label{eq: sec level xiphi}
    \xi(\varphi):= \int \varphi(x) \xi(dx). 
\end{equation}

\begin{theorem}\label{thm: main theorem}  Let Assumption \ref{ass: nondegeneracy} hold.     There is $0<\alpha_0<1/2$ such that the following holds for all $0\leq\alpha\leq \alpha_0$: For all $\varphi \in C_K^+(\mathbb{R})$
  \begin{equation} \label{sec: stat Laplace convergence}
    \lim_{n \to \infty} \mathbb{E}\left[\exp(-\xi_n(\varphi)) \right] = \bar{\mathbb{E}}\left[ \exp(-
    \bar{\xi}(\varphi))\right].
\end{equation}
This implies that  $\xi_n$ converges to $\bar\xi$ in distribution with respect to the vague topology on Radon measures.
\end{theorem}
The convergence of point processes in the vague topology is equivalent to \eqref{sec: stat Laplace convergence}; see e.g.\ \cite{minami1996local} Section 1, and we will not discuss this further.  

Let us explain the role of Assumption \ref{ass: nondegeneracy}. The non-degeneracy of the bath $H_B$ is necessary because we do not have any assumptions that ensure that the bath is sufficiently well-coupled to the rest of the chain. In particular, let us take $H_B =Z_1$ with $n_B>1$, such that spins indexed by $i=2,\ldots,n_B$ are not coupled to the rest of the chain.  Then, every eigenvalue of $H_n$ with $n\geq n_B$, has a degeneracy that is at least $2^{n_B-1}$-fold and hence the spectral statistics cannot be Poissonian.

The  anti-degeneracy assumption, i.e.\ the second condition in \eqref{eq: assumption 1}, is more subtle. To understand it, let us take $n_B=0$ and $\alpha=0$, that is, we have the trivial Hamiltonian
$$
H^{(\mathrm{free})}_n = \sum_{i=1}^n h_i Z_i .
$$
Somewhat surprisingly, for this Hamiltonian, Theorem \ref{thm: main theorem} is not valid. 
To see this,  note that the spectrum of $H^{(\mathrm{free})}_n$ is invariant under reflections through the origin; hence, the statistics cannot be Poissonian at the origin. It would be Poissonian at the origin if we shifted the eigenvalues, i.e.\ if we replaced $E_j$ in \eqref{sec: stat def: point process} by $E_j-E$, with $E$ any non-zero, and $n$-independent value. 

Alternatively, if we consider the Hamiltonian (still without coupling between the bath and the rest of the chain)
\begin{equation}\label{eq: h free prime}
H^{(\mathrm{free}')}_n = H_{B} +  \sum_{i=n_{B}+1}^n h_i Z_i,  
\end{equation}

with $H_B$ satisfying Assumption \ref{ass: nondegeneracy}; then the  statement of Theorem \ref{thm: main theorem} does hold for $H_n^{\mathrm{(free')}}$. The symmetry of the spectrum has been shifted from the origin to the values $E^{(B)}_i$ with $i=1,\ldots,2^{n_B}$, and it has been weakened to a symmetry between subsets of the spectrum, i.e.\ for each $i$, the corresponding subset of the spectrum is symmetric around $E_i^{(B)}$.

Furthermore, we note that Assumption \ref{ass: nondegeneracy} could be avoided by choosing $H_B$ randomly from some appropriate random matrix ensemble, but we find that this just obscures the situation, and therefore we keep the assumption explicit.

We already mentioned that rigorous results concerning many-body localization are scarce. The situation is even more drastic for Poisson spectral statistics in disordered many-body models, and we are not aware of any results in this direction. However, we find it insightful to compare our result to the literature on \textit{classical random spin systems} \cite{borgs2009proof, borgs2009proof2,concetti2026rem, bovier2006local, bovier2007poisson}. In \cite{borgs2009proof, borgs2009proof2},  the Poisson spectral statistics are proved for a model which, in our notation, is exactly $H_n^{(\mathrm{free})}$. \cite{bovier2006local} proves a similar result for a wide class of classical spin systems, such as short-range spin glasses and mean-field spin glasses of the SK type. For more recent results and other types of Poissonian behavior in such systems, we refer to \cite{concetti2026rem, arous2007new} and references therein.  
Let us emphasize that our model differs from the works mentioned above at both the physical and technical levels.
On the physical level, we treat quantum spin systems consisting of mutually non-commuting terms, in particular kinetic energy terms. 
On the technical level, the fact that the Hamiltonian consists of non-commuting terms blocks direct access to the spectrum. The bulk of our paper is devoted to overcoming this difficulty.

{}

\subsection{Dissolution of resonances} \label{sec: intro dissolution statement}
We state a third theorem, which needs more setup.  Let us fix a parameter $\theta \in (2/3,1)$. 
Given the eigenvalues $E_i$, indexed in increasing order, with possible repetitions,  we can partition the index set $\{1,\ldots, 2^n\}$ into so-called resonant patches:   
Two consecutive levels $E_{i},E_{i+1}$ belong to the same resonant patch if and only if $E_{i+1}-E_i \leq \alpha^{\theta n}$. 
This fixes  a partition $\caP_n$ of  $\{1,\dots,2^n\}$, see Figure 
\ref{fig: partition resonant patches}. 

{By a \textit{multiset} we mean a set where repetitions are allowed. Then $\{E_1,\dots,E_{2^n}\}$ is the multiset of eigenvalues of $H_n$ with $2^n$ elements. Here, degeneracy is allowed.  $\mathcal{P}_n$  imposes a unique partition of the multiset $\{E_1,\dots,E_{2^n}\}$  into multisets: $E_i$ and $E_{i+1}$ belong to the same multiset if $E_{i+1}-E_i \leq \alpha^{\theta n}$. By abusing the notation, we use the same symbol $\caP_n$ for this partition. Let us mention that given Assumption \ref{ass: nondegeneracy}, we prove in Lemma \ref{lemma: minimal gap antigap} that $H_n$ is almost surely non-degenerate. In this case, elements of $\caP_n$ do not have repeated elements. Therefore, we use the terminology of ``sets" and ``multisets" for elements of $\caP_n$ interchangeably.}

\begin{figure}[h]
\centering
\begin{tikzpicture}[scale=1.2]
    \draw[->] (-0.5,0) -- (9.1,0);

    \node[left] at (-0.5,0) {$\sigma(H_n)$};

    \foreach \x in {1.7,8.7} {
        \filldraw (\x,0) circle (0.05);
    }

    \draw[thick] (0,-0.3) rectangle (0.6,0.3);
    \foreach \x in {0, 0.3, 0.6} {
        \filldraw (\x,0) circle (0.05);
    }

    \draw[->] (0.3,0.8) -- (0.3,0.3);
    \node[above] at (0.3,0.8) {res. patch};

    \draw[<->] (0.6,-0.5) -- (1.7,-0.5);
    \node[below] at (1.1,-0.5) {\footnotesize gap $>\alpha^{\theta n}$};

    \draw[thick] (2.8,-0.3) rectangle (3.3,0.3);
    \foreach \x in {2.8, 3.0, 3.2, 3.3} {
        \filldraw (\x,0) circle (0.05);
    }

    \draw[thick] (4.8,-0.3) rectangle (5.4,0.3);
    \foreach \x in {4.8, 5.0, 5.2, 5.4} {
        \filldraw (\x,0) circle (0.05);
    }

    \draw[<->] (4.8,0.45) -- (5.4,0.45);
    \node[above] at (5.1,0.45) {$s_n$};

    \draw[thick] (6.9,-0.3) rectangle (7.6,0.3);
    \foreach \x in {6.9,7.2,7.35, 7.6} {
        \filldraw (\x,0) circle (0.05);
    }

 \end{tikzpicture}
\caption{Partition of the spectrum into resonant patches. The dots on the horizontal line are eigenvalues. The rectangular boxes group eigenvalues belonging to a resonant patch. In this picture, $\caP_{n,1}$ consists of two singletons, $\caP_{n,2}$ is empty, $|\caP_{n,3}|=1$ and $|\caP_{n,4}|=3 $, where we recall that $\mathcal{P}_{n,k}$ is the multiset of resonant patches with $k$ levels.}
\label{fig: partition resonant patches}
\end{figure}
The rationale for such a definition goes as follows. The perturbation that couples the spin at site $n$ directly to other spins has norm bounded by $C \alpha^n$. Therefore, the effect of different resonant patches on each other can be described by spectral perturbation theory.  The most complete understanding of our model is obtained when $\caP_n$ consists only of singletons, and we suspect that this becomes true almost surely as $n\to \infty$. Indeed, note that the mean level spacing around zero, which was called $\sss_n$ in \eqref{sec: stat sn}, is of order $2^{-n}$, which is much larger than $C \alpha^{\theta n}$ if $\alpha$ is small enough and $n$ is large enough. 

We have not been able to establish that $\caP_n$ consists only of singletons, but we can prove that the fraction of eigenvalues that are in a non-singleton resonant patch vanishes almost surely, as we state now.  
For any $k \in \mathbb{N}$, we write $\caP_{n,k} \subset \caP_n$ for the multiset of resonant patches that have $k$ eigenvalues. Since the total number of eigenvalues is $2^n$, we have
$$
\sum_{k>0} k|\caP_{n,k}| =2^n.
$$

Our result reads 


\begin{theorem}\label{thm: dissolution of patches}
 Let Assumption \ref{ass: nondegeneracy} hold, then, there is $\alpha_0$  depending on $\theta$, such that for all $0\leq \alpha \leq \alpha_0$,  we have
 $$ \frac{|\caP_{n,1}|}{2^n} \to 1, \qquad \text{almost surely}. $$
\end{theorem}
This theorem is a crucial ingredient in the proof of Theorem \ref{thm: main theorem}. Crudely speaking, since the effect of the interactions is perturbative for levels in singleton patches, Theorem \ref{thm: dissolution of patches} allows us to compare the spectral statistics of $H_n$ to the spectral statistics of $H_n^{(\mathrm{free}')}$, as defined in Section \ref{sec: level statistics}. 

\subsection{Remarks} \label{rmkL intro}

We conclude the presentation of the results with a few brief remarks.

\begin{enumerate}

        \item  Our result holds for a very wide class of bath Hamiltonian; the many-body localization result, Theorem \ref{thm: fock space localization}, only assumes $H_B$  is self-adjoint and bounded, and acts on finitely many spins. The Poisson spectral statistics result, Theorem \ref{thm: main theorem},  only requires an additional mild assumption on $H_B$; see Assumption \ref{ass: nondegeneracy}. In all instances where this model has been studied in the literature (see the above-mentioned references), the bath Hamiltonian considered satisfy the above assumptions.
        \item In our model \eqref{eq: main model1}, the bath interacts with spin $i>n_B$ via the term $g_i\alpha^{i-n_B} X_1X_i$. The choice of the interaction term $X_1X_i$ is merely for convenience. Our results hold if this term is replaced by any interaction of the form $V_{b}V_i$ where 
        \begin{enumerate}
        \item $V_b$  acts only on the bath spins $i=1,\dots,n_B$, 
        \item $V_i$ acts on the spin $i$, and
        \item  $\|V_{b}V_i \| \leq C$ for an absolute constant $C>0$.
        \end{enumerate}
        \item The choice of a  uniform distribution for $h_i,g_i$ is made for convenience. Theorem \ref{thm: fock space localization}, as mentioned before, holds with deterministic $g_i$  that are bounded in absolute value by one. For this theorem, we may take the distribution of $h_i$ to have a compactly supported and smooth density with mean zero. Such a choice only affects the localization threshold in the theorem: localization can be proven for $\alpha< \alpha_0$ with a new value of $\alpha_0$ instead of $1/8$.

        In Theorems \ref{thm: main theorem} and \ref{thm: dissolution of patches}, we believe that our results are still true and our proof method works with small modifications if one assumes that the variables $h_i$ and $g_i$ have   compactly supported and 
        smooth densities, with $h_i$ having mean zero. In this case, the value of $\alpha_0$ should be modified depending on the distributions of $h_i$ and $g_i$ (see also the end of Section \ref{sec: dissolution technical} for a comment in this direction).

        On the other hand, our current method relies on the smoothness of the densities, and singular distributions are out of reach of our current tools.  
    \end{enumerate}

\subsection{Short outline  of the proofs}
\label{sec: outline}

In this section, we present a detailed outline of the proof. The proof of Theorem \ref{thm: fock space localization} is straightforward; it uses perturbation theory and simple probabilistic estimates.  Theorem  \ref{thm: dissolution of patches} is the first significant novelty of this work. We are not aware of any results in the spirit of Minami estimates (see \cite{minami1996local}) for disordered quantum many-body systems. This theorem provides a first step in this direction. The proof relies on perturbation theory and certain inequalities from potential theory: the Cartan and Remez inequalities. Such inequalities have been used in the context of one-body localization \cite{bourgain2009Cartan,kruger2012localization, DeRoeck2023LongPersistence}. However, we apply them in a quite different  manner compared to the above-mentioned works. We believe our methods can be used for a more general class of random operators. 
Proof of Theorem \ref{thm: main theorem} is the most technical part of the paper. It uses various probabilistic and combinatorial arguments, as well as the Brascamp-Lieb inequality. Some parts of our methods here can be compared to those of \cite{borgs2009proof}. We believe that our methods can also be used to give a different and shorter proof of the Poisson spectral statistics in classical random  systems, such as the one considered in \cite{borgs2009proof}.

\subsubsection{``Old'' and ``new'' patches} \label{sec: intro proof idea I}
We track the spectrum as we move from scale $n$ to $n+1$. 
To this end, we introduce an intermediate Hamiltonian $H_{n+1/2}$: 
$$H_{n+1/2}= H_n+h_{n+1}Z_{n+1}, \quad H_{n+1}= H_{n+1/2}+ g_{n+1}\alpha^{n+1-n_B}X_1X_{n+1}.$$

The spectrum of $H_{n+1/2}$ is easily obtained from  $H_n$. Indeed, let us write  $\mathrm{spec}(\cdot)$ for the spectrum. Then 
\begin{equation}\label{eq: shifted copies}
    \mathrm{spec}(H_{n+1/2} )  =  (\mathrm{spec}(H_n)+ h_{n+1} )  \cup   (\mathrm{spec}(H_n)-h_{n+1} ),
    \end{equation}
    where, for any $A \subset \mathbb{R}$, and $x \in \mathbb{R}$, the notation $x+A$ denotes the  set $\{x+a|a\in A \}$.
In Figure \ref{fig: spin addition}, we present the above process pictorially, where we make two shifted copies of $\mathrm{spec}(H_n)$ and then \textit{fuse} them. Note that here, in the definition of the spectrum, we consider the possible multiplicity of the eigenvalues, and above union is a multiset union (see Section \ref{sec: non res event}).

Recall that given the resonant threshold $\alpha^{\theta n}$, we constructed a partition $\mathcal{P}_n$ of $\mathrm{spec}(H_n)$, consisting of resonant patches (the name will become clear in Section \ref{sec: term}). We use the same threshold ($\alpha^{\theta n}$) to partition $\mathrm{spec}(H_{n+1/2})$ into  patches, and we denote this partition by $\caP_{n+1/2}$. 




\begin{figure}
\centering

\begin{subfigure}{0.75\textwidth} 
\centering
\begin{tikzpicture}[x=1cm,y=1cm,scale=1.2]

\def\xmin{-2.6} \def\xmax{2.6}
\def\shift{1.5}
\def\tickh{0.20}

\def\axisth{0.75pt}
\def\tickth{1.10pt}
\def\guideth{0.50pt}
\def\dotsize{2.25pt}

\def\linecol{black!55}
\def\guidecol{black!30}

\def\dotsep{0.25}
\def\yA{1.2} \def\yB{0.4} \def\yC{-0.4} \def\yD{-1.2}
\def\Ymin{-1.8} \def\Ymax{1.6}

\def\tick#1#2{%
  \draw[line width=\tickth, draw=\linecol] (#1, #2-\tickh) -- (#1, #2+\tickh);
}

\draw[->, line width=\axisth, draw=\linecol] (\xmin,\yA) -- (\xmax,\yA)
  node[right,text=black]{$\mathrm{spec}(H_{n+1/2})$};

\draw[->, line width=\axisth, draw=\linecol] (\xmin,\yB) -- (\xmax,\yB)
  node[right,text=black]{$\mathrm{spec}(H_n)+h_{n+1}$};

\draw[->, line width=\axisth, draw=\linecol] (\xmin,\yC) -- (\xmax,\yC)
  node[right,text=black]{$\mathrm{spec}(H_n)-h_{n+1}$};

\draw[->, line width=\axisth, draw=\linecol] (\xmin,\yD) -- (\xmax,\yD)
  node[right,text=black]{$\mathrm{spec}(H_n)$};

\tick{0}{\yA}
\tick{\shift}{\yB}
\tick{-\shift}{\yC}
\tick{0}{\yD}

\draw[dashed, line width=\guideth, draw=\guidecol] (-\shift,\Ymin) -- (-\shift,\Ymax);
\draw[dashed, line width=\guideth, draw=\guidecol] (0,\Ymin)       -- (0,\Ymax);
\draw[dashed, line width=\guideth, draw=\guidecol] (\shift,\Ymin)  -- (\shift,\Ymax);

\def\dotsonly#1#2#3{%
  \fill[#3] (#1-0.25,#2) circle (\dotsize);
  \foreach \i in {0,1,2}{
    \fill[#3] (#1+0.25+\i*\dotsep,#2) circle (\dotsize);
  }
}

\dotsonly{\shift}{\yB}{red}
\dotsonly{-\shift}{\yC}{blue}
\dotsonly{0}{\yD}{black}
\dotsonly{\shift}{\yA}{red}
\dotsonly{-\shift}{\yA}{blue}

\def\yArrow{-1.6}
\draw[<->, line width=\axisth, draw=\linecol] (-\shift,\yArrow) -- (0,\yArrow)
  node[midway,below,text=black]{$h_{n+1}$};

\draw[<->, line width=\axisth, draw=\linecol] (0,\yArrow) -- (\shift,\yArrow)
  node[midway,below,text=black]{$h_{n+1}$};

\end{tikzpicture}
\caption{}
\end{subfigure}

\caption{
Pictorial procedure of obtaining $\mathrm{spec}(H_{n+1/2})$ from $\mathrm{spec}(H_n)$.
From below: the first line represents the spectrum of $H_n$.  Black dots are the eigenvalues.
The second and third lines are two shifted copies of the spectrum by $\pm h_{n+1}$.
Blue and red dots show the shifted spectra. The fourth line shows the spectrum of
$H_{n+1/2}$, where the spectrum is the union of blue and red dots.
}
\label{fig: spin addition}
\end{figure}


\begin{figure}
\centering


\begin{subfigure}{\textwidth}
\centering
\begin{tikzpicture}[x=1cm,y=1cm,scale=1.3]

\def\xmin{-2.6} \def\xmax{2.6}
\def\shift{1.5}
\def\tickh{0.20}

\def\axisth{0.75pt}
\def\tickth{1.10pt}
\def\guideth{0.50pt}
\def\rectth{1.05pt}
\def\dotsize{2.25pt}

\def\linecol{black!55}
\def\guidecol{black!30}

\def\dotsep{0.25}
\def\yA{1.2} \def\yB{0.4} \def\yC{-0.4} \def\yD{-1.2}
\def\Ymin{-1.8} \def\Ymax{1.6}

\def\tick#1#2{%
  \draw[line width=\tickth, draw=\linecol] (#1, #2-\tickh) -- (#1, #2+\tickh);
}

\draw[->, line width=\axisth, draw=\linecol] (\xmin,\yA) -- (\xmax,\yA) node[right,text=black]{$\mathrm{spec}(H_{n+1/2})$};
\draw[->, line width=\axisth, draw=\linecol] (\xmin,\yB) -- (\xmax,\yB) node[right,text=black]{$\mathrm{spec}(H_{n})+h_{n+1}$};
\draw[->, line width=\axisth, draw=\linecol] (\xmin,\yC) -- (\xmax,\yC) node[right,text=black]{$\mathrm{spec}(H_{n})-h_{n+1}$};
\draw[->, line width=\axisth, draw=\linecol] (\xmin,\yD) -- (\xmax,\yD) node[right,text=black]{$\mathrm{spec}(H_{n})$};

\tick{0}{\yA}
\tick{\shift}{\yB}
\tick{-\shift}{\yC}
\tick{0}{\yD}

\draw[dashed, line width=\guideth, draw=\guidecol] (-\shift,\Ymin) -- (-\shift,\Ymax);
\draw[dashed, line width=\guideth, draw=\guidecol] (0,\Ymin)       -- (0,\Ymax);
\draw[dashed, line width=\guideth, draw=\guidecol] (\shift,\Ymin)  -- (\shift,\Ymax);

\def\dotsrect#1#2#3{%
  \fill[#3] (#1-0.25,#2) circle (\dotsize);
  \foreach \i in {0,1,2}{
    \fill[#3] (#1+0.25+\i*\dotsep,#2) circle (\dotsize);
  }
  \pgfmathsetmacro{\xstart}{#1+0.25}
  \pgfmathsetmacro{\xend}{#1+0.25+2*\dotsep}
  \draw[#3, line width=\rectth] (\xstart,#2-0.22) rectangle (\xend,#2+0.22);
}

\dotsrect{\shift}{\yB}{red}
\dotsrect{-\shift}{\yC}{blue}
\dotsrect{0}{\yD}{black}
\dotsrect{\shift}{\yA}{red}
\dotsrect{-\shift}{\yA}{blue}

\def\yArrow{-1.6}
\draw[<->, line width=\axisth, draw=\linecol] (-\shift,\yArrow) -- (0,\yArrow)
  node[midway,below,text=black]{$h_{n+1}$};

\draw[<->, line width=\axisth, draw=\linecol] (0,\yArrow) -- (\shift,\yArrow)
  node[midway,below,text=black]{$h_{n+1}$};

\end{tikzpicture}
\caption{}
\end{subfigure}
\hfill
\vspace*{6mm}
\begin{subfigure}{\textwidth}
\centering
\begin{tikzpicture}[x=1cm,y=1cm,scale=1.3]

\def\xmin{-2}
\def\xmax{2}
\def\shift{0.6}
\def\tickh{0.20}

\def\axisth{0.75pt}
\def\tickth{1.10pt}
\def\guideth{0.50pt}
\def\rectth{1.05pt}
\def\dotsize{2.25pt}

\def\linecol{black!55}
\def\guidecol{black!30}

\def\dotsep{0.25}

\def\yA{1.2}
\def\yB{0.4}
\def\yC{-0.4}
\def\yD{-1.2}

\def\Ymin{-1.8}
\def\Ymax{1.6}

\def\tick#1#2{%
  \draw[line width=\tickth, draw=\linecol] (#1, #2-\tickh) -- (#1, #2+\tickh);
}

\draw[->, line width=\axisth, draw=\linecol] (\xmin,\yA) -- (\xmax,\yA)
  node[right,text=black]{$\mathrm{spec}(H_{n+1/2})$};

\draw[->, line width=\axisth, draw=\linecol] (\xmin,\yB) -- (\xmax,\yB)
  node[right,text=black]{$\mathrm{spec}(H_n)+h_{n+1}$};

\draw[->, line width=\axisth, draw=\linecol] (\xmin,\yC) -- (\xmax,\yC)
  node[right,text=black]{$\mathrm{spec}(H_n)-h_{n+1}$};

\draw[->, line width=\axisth, draw=\linecol] (\xmin,\yD) -- (\xmax,\yD)
  node[right,text=black]{$\mathrm{spec}(H_n)$};

\tick{0}{\yA}
\tick{\shift}{\yB}
\tick{-\shift}{\yC}
\tick{0}{\yD}

\draw[dashed, line width=\guideth, draw=\guidecol] (-\shift,\Ymin) -- (-\shift,\Ymax);
\draw[dashed, line width=\guideth, draw=\guidecol] (0,\Ymin)       -- (0,\Ymax);
\draw[dashed, line width=\guideth, draw=\guidecol] (\shift,\Ymin)  -- (\shift,\Ymax);

\def\dotsrect#1#2#3{%
  \fill[#3] (#1-0.25,#2) circle (\dotsize);
  \foreach \i in {0,1,2}{
    \fill[#3] (#1+0.25+\i*\dotsep,#2) circle (\dotsize);
  }
  \pgfmathsetmacro{\xstart}{#1+0.25}
  \pgfmathsetmacro{\xend}{#1+0.25+2*\dotsep}
  \draw[#3, line width=\rectth] (\xstart,#2-0.22) rectangle (\xend,#2+0.22);
}

\def\dotsonly#1#2#3{%
  \fill[#3] (#1-0.25,#2) circle (\dotsize);
  \foreach \i in {0,1,2}{
    \fill[#3] (#1+0.25+\i*\dotsep,#2) circle (\dotsize);
  }
}

\dotsrect{\shift}{\yB}{red}
\dotsrect{-\shift}{\yC}{blue}
\dotsrect{0}{\yD}{black}

\dotsrect{\shift}{\yA}{red}
\dotsonly{-\shift}{\yA}{blue}

\pgfmathsetmacro{\xBlueStart}{-\shift+0.25}
\pgfmathsetmacro{\xRedLeft}{\shift-0.25}

\filldraw[
  fill=purple!20,
  draw=purple,
  line width=\rectth
]
(\xBlueStart,\yA-0.22) rectangle (\xRedLeft,\yA+0.22);

\dotsonly{-\shift}{\yA}{blue}
\fill[red] (\shift-0.25,\yA) circle (\dotsize);

\draw[-, line width=\axisth, draw=\linecol] (\xBlueStart,\yA) -- (\xRedLeft,\yA);

\def\yArrow{-1.6}
\draw[<->, line width=\axisth, draw=\linecol] (-\shift,\yArrow) -- (0,\yArrow)
  node[midway,below,text=black]{$h_{n+1}$};

\draw[<->, line width=\axisth, draw=\linecol] (0,\yArrow) -- (\shift,\yArrow)
  node[midway,below,text=black]{$h_{n+1}$};

\end{tikzpicture}
\caption{}
\end{subfigure}

\caption{These two panels show a similar procedure as in Figure \ref{fig: spin addition}. 
In sub-window (a), boxes show the non-singleton resonant patches.
This is an example where no new resonance occurs.
In sub-window (b), the same procedure with the same $\mathrm{spec}(H_n)$ and a different $h_{n+1}$ is shown.
This is an example of a resonance event. The blue resonant block and the single red
eigenvalue create a new resonant block, shown by a filled purple rectangle.
}
\label{fig: res event}
\end{figure}

$\mathcal{P}_{n+1/2}$ has two types of resonant patches: 
\begin{enumerate}

    \item \textit{Old patches:} 
    Patches that are contained entirely in one of the shifted spectra and do not intersect with the other shifted spectrum. 
   These are shifted copies $p\pm h_{n+1}$ of the patches $p$ from the previous scale. See Figure \ref{fig: res event} sub-window (a): solid blue and red rectangles show the old patches.
    \item  \emph{New patches}:  These are created in the \textit{fusion} process above, i.e.\, they contain eigenvalues from both $\mathrm{spec}(H_n)\pm h_{n+1}$. See Figure \ref{fig: res event} sub-window (b): the filled purple rectangle shows the new patch.
\end{enumerate}
The event that no new patches appear in  $\mathcal{P}_{n+1/2}$ is denoted by $\mathcal{A}_{n,n+1}$, see the definition in  Section \ref{sec: non res event}. The event that no new patch will ever appear at higher scales is denoted by  $\caA_{n,\infty}$,  i.e.\ $\caA_{n,\infty}:=\cap_{i\geq n} \caA_{i,i+1}$.  

We first observe that the probability of having a new resonance after scale ${n}$, $\mathbb{P}(\caA_{n,\infty}^c)$, decays exponentially in $n$ (see Lemma \ref{lem: prob event caa}).   Let us give  a brief explanation: at scale $n$, denoting  the eigenvalues of $H_{n}$ by $\{E_a\}_a$ we have a ``new resonance" with threshold $\epsilon$ if and only if for some $a,b$, we have: $|E_a-E _b-2h_{n+1}| \leq \epsilon$. Given a realization of the disorder $\{ h_i,g_i\}_{i\leq n}$, for  fixed $a$ and $b$, such an event has probability  $\epsilon$ w.r.t. the random variable $h_{n+1}$. Then a union bound gives us the desired result for $\mathcal{A}_{n,n+1}^c$ by noting that we have at most $4^n$ pairs of eigenvalues. Another union bound over different scales allows us to bound the probability of $\mathcal{A}_{n,\infty}^c$. 

The above observation, which is proven in Lemma \ref{lem: prob event caa}, is the first part of Theorem \ref{thm: fock space localization}. Moreover, it serves as the backbone of the proof of Theorems \ref{thm: main theorem} and \ref{thm: dissolution of patches}. Thanks to this bound, in the rest of the present section  `Short outline of the proofs', we condition on the event $\mathcal{A}_{\ell,\infty}$ for a fixed $\ell$, and we always assume $n>\ell$, i.e., we do not deal with events in which  a new resonance occurs after scale $\ell$. We do not discuss the main technical difficulties of such ``conditioning" in the outline, except for a brief comment at the end of Section \ref{sec: error terms intro}.

\subsubsection{From $n$ to $n+1/2$ by tensoring, and  from $n+1/2 $ to $  n+1$ by perturbation theory} \label{sec: intro proof idea II}

Let $p \in \caP_n$
 be a patch of $H_n$ and let us assume that $\caP_{n+1/2}$ contains only ``old" patches, i.e.\ patches of the form $p\pm h_{n+1}$, as explained above.  Let us write these patches as $(p, \mu)$ with $\mu\in \{\pm 1\}$. The spectral projector corresponding to a patch is
 $$
 P_{(p, \mu)} = 
P_{p} \otimes\delta_{Z_{n+1}-\mu},$$
where we slightly abuse the notation in the above expression (see Section \ref{sec: conventions} Item 3).

We view $H_{n+1}$ as a perturbation of $H_{n+1/2}.$ For large $n$,  $\alpha^{\theta n} \gg \alpha^{n+1-n_B}$ i.e.\ the gaps between different resonant patches $(p,\mu)$ of $H_{n+1/2}$,  are much larger than the norm of the perturbation.
Therefore, each resonant patch $(p,\mu) \in \caP_{n+1/2}$ gives rise to a spectral patch of $H_{n+1} $ that we call $S_{p,\mu}$, with 
$$
|S_{p,\mu}|= |(p,\mu)| =|p|,
$$
where $|\cdot|$ denotes the number of elements of a multiset (i.e., taking into account the multiplicity). In fact, the above observation allows us to relate $S_{p,\mu}$  to $(p,\mu)$ via analytic perturbation theory.

Moreover, the convex hulls of different $S_{p,\mu}$ do not overlap (see the second line of Figure \ref{fig:perturbation}), and the distances between these convex hulls are not smaller than the resonance threshold at the next scale, i.e.\ $\alpha^{\theta(n+1)}$. 
This yields the following crucial observation:  Even though $\{ S_{p,\mu} \}_{p \in \caP_{n}, \mu \in \{\pm 1\}}$ is a partition of $\mathrm{spec}(H_{n+1})$, it is not necessarily equal to the partition $\caP_{n+1}$. 
Indeed,  for any $(p,\mu)$, either 
\begin{enumerate}
    \item $S_{p,\mu}  \in \caP_{n+1}$
\item or $S_{p,\mu} $  can be partitioned into (at least two) sets in  $\caP_{n+1}$.
\end{enumerate}
In the first case, we say that the patch $(p,\mu)$ \emph{remains a resonant patch} whereas in the second case, we say that it is  \textit{dissolved}. In Figure \ref{fig:perturbation}, the red patch is an example of the first possibility, and the blue patch is an example of the second case.

\begin{figure}
\centering
\begin{tikzpicture}[x=1cm,y=1cm,scale=1.2]

\def\xmin{-4}
\def\xmax{4}
\def\shift{2.5}
\def\tickh{0.20}

\def\axisth{0.75pt}
\def\tickth{1.10pt}
\def\guideth{0.50pt}
\def\rectth{1.05pt}
\def\dotsize{2.25pt}

\def\linecol{black!55}
\def\guidecol{black!30}

\def\yA{1.3}
\def\yB{0.0}
\def\yC{-1.3}

\def\Ymin{-1.4}
\def\Ymax{1.2}

\def\tick#1#2{%
  \draw[line width=\tickth, draw=\linecol] (#1, #2-\tickh) -- (#1, #2+\tickh);
}

\draw[->, line width=\axisth, draw=\linecol] (\xmin,\yA) -- (\xmax,\yA)node[right,black] {$\mathrm{spec}(H_{n+1})$};
\draw[->, line width=\axisth, draw=\linecol] (\xmin,\yB) -- (\xmax,\yB)node[right,black] {$\mathrm{spec}(H_{n+1})$};
\draw[->, line width=\axisth, draw=\linecol] (\xmin,\yC) -- (\xmax,\yC)node[right,black] {$\mathrm{spec}(H_{n+1/2})$};

\tick{0}{\yA}
\tick{0}{\yB}
\tick{0}{\yC}

\draw[dashed, line width=\guideth, draw=\guidecol] (-\shift,\Ymin) -- (-\shift,\Ymax);
\draw[dashed, line width=\guideth, draw=\guidecol] (0,\Ymin) -- (0,\Ymax);
\draw[dashed, line width=\guideth, draw=\guidecol] (\shift,\Ymin) -- (\shift,\Ymax);

%
%

\dotsrectarray{\yA}
  {-3.10,-2.20,-1.50,-1.30, 0.80,2.20,2.45,2.65}
{-1.50,-1.30, 2.20,2.65}
  [solid]

\dotsrectarray{\yB}
  {-3.10,-2.20,-1.50,-1.30, 0.80,2.20,2.45,2.65}
  {-2.20,-1.30, 2.20,2.65}
  [dashed]

\dotsrectarray{\yC}
  {-3.00,-2.00,-1.60,-1.20, 1.00,2.00,2.40,2.80}
  {-2.00,-1.20, 2.00,2.80}
  [solid]

\end{tikzpicture}
\caption{
Effect of the perturbation. From below, the first line is the spectrum of
$H_{n+1/2}$, where resonant patches are grouped by solid rectangles.
The second line is the spectrum of $H_{n+1}$: the  dashed rectangle encloses each patch after perturbation. The third line is the spectrum of $H_{n+1}$, similar to the second line: the solid rectangles denote the resonant patch of $H_{n+1}$. The blue patch is dissolved, and the red patch remains a resonant patch after the perturbation. The perturbation {decreases} the inner level distance of the red patch and the remaining levels of the blue patch.
}
\label{fig:perturbation}
\end{figure}

\subsubsection{Proof of Theorem \ref{thm: fock space localization}}

We have already argued that the spectral patch $S_{p,\mu}$ of $H_{n+1}$ is a deformed version of the resonant patch $(p,\mu) \in \caP_{n+1/2}$, described by analytic perturbation theory.  In fact, we have the following bound; see Lemma \ref{lem: projectors PT theory}, 
\begin{equation} \label{eq: projection bound intro}
    \|P_{(p,\mu)}-P_{S_{p,\mu}} \| \leq C(2\alpha^{(1-\theta)})^{n+1},
\end{equation}
 where we denoted the \textit{spectral projectors} associated with $(p,\mu)$ and $S_{p,\mu}$ by $P_{(p,\mu)}$, and $P_{S_{p,\mu}}$, respectively. 
 This bound tells us that each eigenstate $\psi$ belonging to the spectral patch $S_{p,\mu}$ satisfies
 $$
 \langle \psi, Z_{n+1} \psi\rangle =\mu + O((2\alpha^{(1-\theta)})^{n+1}),
 $$
 with the error term exponentially small in $n$. This bound shows that eigenvectors are localized as far as the $(n+1)$-th spin is concerned. 
One can continue this reasoning at higher scales, and obtain a more general statement about the joint localization of all spins $j>n$, as  stated in Theorem  \ref{thm: fock space localization}. Finally, we recall that the reasoning outlined above is only valid provided the event $\caA_{n,\infty}$ holds; this event has  probability tending to one exponentially fast in $n$, as explained in Section \ref{sec: intro proof idea I}.

\subsubsection{Dissolution: conceptual aspects} \label{sec: intro dissolution}

 The main conceptual tool to prove Theorem \ref{thm: dissolution of patches} is the notion of ``\textit{patch dissolution}." Here we elaborate on this concept. 
In Section \ref{sec: intro proof idea II}, we introduced the notion of a resonant patch $(p,\mu) \in \caP_{n+1/2}$, with $p\in \caP_{n},\mu \in \{\pm1\}$, being dissolved or remaining a resonant patch. 
We start with a trivial observation that can be summarized as: 
 ``turning off the interaction renders dissolution inevitable".

If the interaction is turned off, i.e.\ $g_{i}=0$ for $i>n$, then the patch $S_{p,\mu}$ is equal to the patch $(p,\mu) \in \caP_{n+1/2}$, hence equal to the set $p+\mu h_{n+1}$, with $p\in \caP_n$.  In particular, the maximal distance between levels inside the patch is the same for $S_{p,\mu}$ as for $p$. 
The crux, however, is that the resonance threshold has shrunk from $\alpha^{\theta n}$ at scale $n$ to $\alpha^{\theta (n+1)}$ at scale $n+1$. Hence it is quite possible that the maximal distance is large enough so that it exceeds the resonance threshold at scale $n+1$, in which case the patch is dissolved since at least two of its adjacent levels are no longer resonant.  Moreover, this maximal distance is nonzero almost surely, because our Hamiltonians are non-degenerate almost surely, as explained in Appendix \ref{sec: app Non-degeneracy}.   Therefore, even if the patch did not dissolve at scale $n+1$, it is bound to dissolve eventually, at some larger scale.



Therefore, the only obstacle to dissolution is the possibility that the interaction pulls the levels inside the patch towards each other, reducing the distance between them by a factor $\alpha^{\theta}$ at each scale. This is the fate of the red resonant patch in Figure \ref{fig:perturbation}.

Since the patches are well-separated from each other, because we are on the event $\caA_{n,\infty}$, we can study the levels inside the patch $S_{p,\mu}$ independently of the rest of the spectrum of $H_{n+1}$, in the sense that we explain now.

Let us introduce  shorthand notations for $P_{(p,\mu)}$ and $P_{S_{p,\mu}}$. Let $P_0$ be the spectral projector of $H_{n+1/2}$ on the patch $(p,\mu)$ and let $P_{g}$, with $g\equiv g_{n+1}$, be the spectral projector of $H_{n+1}$ on the patch $S_{p,\mu}$. The set $S_{p,\mu}$ equals the spectrum of $P_gH_{n+1}P_g$ restricted to $\mathrm{Ran}\:  P_g$. This matrix is unitarily equivalent, up to a shift, to an effective Hamiltonian $H_{P}$, a $|p|\times |p|$ matrix acting on $\mathrm{Ran} P_0$ and of the form  (see Section \ref{sec: effective Ham derivation})
$$
H_{P}= D+V, \qquad D=P_0 H_{n+1/2} P_0, \quad  ||V||\leq C(4 \alpha^{1-\theta})^{n+1}.
$$ We also denote the \textbf{spectral width} of $H_{P}$ (i.e.\ the difference between the largest and the smallest eigenvalues of $H_P$),  by $s_{n+1}(g)$ (see Section \ref{sec: control on patch shrink}). In the following, this is the main quantity of interest. \\


  As a simple illustrative situation,
 let us assume that the patch consists of two eigenvalues $E_1,E_2$, i.e.\ $|p|=2$.  By the  definition of a resonant patch, we have $|E_1-E_2|\leq \alpha^{\theta n}$.
 We  hence see that the patch only fails to dissolve if
 \begin{equation} \label{eq: d plus v}
  s_{n+1}(g)=  |E_2(D+V)-E_1(D+V) |\leq \alpha^{\theta (n+1)} .
 \end{equation}
 Keeping in mind that $\theta$ can be chosen close to $1$ and $\alpha$ is a small parameter, the condition \eqref{eq: d plus v} suggests a remarkable fine-tuning. One is tempted to conjecture that this happens for a set of small measure in the possible values of $g$ and that it is therefore unlikely. We sketch a proof of this conjecture in the upcoming section  \ref{sec: dissolution technical}. 
 Although the above discussion was specific for $|p|=2$, it has an obvious generalization for $|p|>2$. If the distances between neighboring levels are not larger than $\alpha^{(n+1)\theta}$, then the spectral width $s_{n+1}(g)$ is necessarily smaller than $(|p|-1)\alpha^{(n+1)\theta}$.

 
 
 \subsubsection{Dissolution: technical aspects} \label{sec: dissolution technical}
 The above-mentioned conjecture is proved using the  Remez and Cartan inequalities, see Section \ref{sec: Remez and cartan} for the precise statements. The gist of these inequalities can be summarized as follows: if a \textit{nice} (complex analytic on the unit ball/polynomial)   function $f$ is \textit{not too small} at one point on the interval $[-1,1]$,  then the Lebesgue measure of the set of points in $[-1,1]$ that $f$ is \textit{too small}  cannot be \textit{too large}. 

Given a  patch $(p,\mu) \in \mathcal{P}_{n+1/2}$ with $p \in \mathcal{P}_n,\mu \in \{\pm 1 \}$, let us first pretend that the above inequalities can be applied directly to the spectral width of $S_{p,\mu}$, $s_{n+1}$.  Viewed as a function of $g=g_{n+1}$, we use $g=0$ as the \textit{good point}. For $g_{n+1}=0$, the patch does not shrink at all, so these inequalities tell us that for most values of $g \in [-1/2,1/2]$ the patch cannot shrink too much. By a change of variable, this is tantamount to the fact that the variable $$v_{n+1}(g):=\ln(s_{n+1}(g))/(n+1) \ln(4\alpha^{1-\theta})$$ does not grow too much  (see \eqref{eq:mainp}); i.e.  
\begin{equation} \label{intro: eq: patch shrink}
    \left|\{ g \in [-1/2,1/2] \: | \: v_{n+1}(g) \geq x v_{n+1}(0) \}\right| \leq p_{\alpha,n+1}(x),
\end{equation}
for a function $p_{\alpha,n+1}$ that only depends on $n,\alpha$ and that satisfies
 $p_{\alpha,n+1}(x) \to 0$ as $x \to \infty$. {Crucially, $p_{\alpha,n+1}$ does not depend on $v_{n+1}$.}


In case that the patch did not dissolve at scale $n+1$, we repeat  the above construction at subsequent scales, leading us to consider $$
\frac{v_{n+m}}{v_n}= \frac{v_{n+m}}{v_{n+m-1}} \ldots    \frac{v_{n+1}}{v_{n}} ,
$$ 
where we notice that from the definition, one can see  $v_{n+1}(0)=v_n$. 
We apply the bound \eqref{intro: eq: patch shrink} to each of the factors $ \frac{v_{j+1}}{v_{j}}$  in the product above, since $v_j$ is related to $v_{j+1}(0)$ in a trivial manner. The question whether the patch eventually dissolves is then translated to a question about the product of \textit{independent} random variables $Y_{n+1},\dots Y_{n+m}$ whose tail probability is $p_{\alpha,j}$, see\  \eqref{intro: eq: patch shrink}.  {Here the independence is deduced from the fact that $p_{\alpha,j}$ does not depend on $v_j$.}

Having outlined the main point of the proof of the dissolution of patches, let us conclude with a few remarks:  
\begin{enumerate}
  \item Notice that the convex hull of $S_{p,\mu}$ is well separated from the rest of the spectrum for all values of $g \in [-1/2,1/2]$ and hence $s_{n+1}(g)$ is well defined, even if $S_{p,\mu}$ is dissolved for some values of $g$.
    
    \item   The above reasoning suffers from the following flaw: the spectral width is neither a polynomial nor analytic in $g=g_{n+1}$. Therefore, neither the Remez nor the Cartan inequality is immediately applicable.  This is solved by bounding $s_{n+1}(g)$ from below by a function $|\tilde \varphi(g)|$, where $\tilde\varphi(g)$ is an analytic function of the 
 matrix elements of $H_P(g)$ and $\tilde{\varphi}(0)=s_{n+1}(0)$ (see the equation \eqref{eq:diag}).
    

    \item   One may be rightfully confused by the use of the Remez inequality since this inequality only applies to polynomials.  In fact, one can apply the Cartan Lemma directly to the function $\tilde \varphi(g)$ referred to in the previous remark.  However, this bound is too weak for our purposes. The remedy is that for sufficiently large values of $\lambda$, namely  $\lambda \geq s_n(0)^{n}$,  we cut the Taylor series of $\tilde \varphi(g)$  at a certain order $n_*$, use the fact that the remainder is small, and apply the Remez inequality to the remaining polynomial. This provides a stronger bound compared to using the Cartan Lemma.
   In contrast,  for extremely small values of $\lambda$, the effect of the tail of the series becomes important, and the use of the Remez inequality becomes less efficient compared to directly applying the Cartan Lemma. Hence, we need both of these inequalities for different regimes of $\lambda$. 
\end{enumerate}

At first sight, one may think that using the above tools makes our result heavily dependent on the assumption that $g_i$ are uniformly distributed. However, adapting the above tools to the case where $g_i$ has a smooth density is rather straightforward (see Section 11 of \cite{kruger2012localization} as an instance of this adaptation). We choose not to pursue this direction in order to keep the setup transparent. 
Let us emphasize that this is the only ``non-straightforward" instance of adapting our proof to more general distributions with smooth densities.
 \subsubsection{Dissolution of multiple patches}

 So far, we have illustrated the proof of the fact that a single resonant patch dissolves with high probability.  This was reflected in the fact that we fixed a value of $\mu\in\{\pm 1\}$.  The descendants of $p \in \caP_n$ at scale $n+m$ are labeled by $\underline{\mu} \in \{\pm 1\}^m$, though they are not necessarily all resonant patches, as some of them might have dissolved. 
 
 In Section \ref{sec: dissolution technical}, we considered multiple scales, and then we implicitly imagined $\underline{\mu} \in \{\pm 1\}^m$  to be fixed. 
 We then established that dissolution is very likely to happen, as $m$ grows,  along a fixed branch of descendants. 
Also, our bounds were uniform in the choice of the branch, i.e.\ in the choice of $\underline{\mu} \in \{\pm 1\}^m$. 
 However,  the number of descendants at scale $m$ is $2^m$ whereas our bounds are not exponential in $m$. Therefore, we certainly cannot conclude that all descendants will dissolve with high probability. 

 This is ultimately also the reason that we cannot derive any bound on the smallest level spacing in our system. For example, we cannot prove the \emph{limited level attraction} principle for our model, see \cite{Imbrie2016a}.

 


 On the other hand, to prove Theorem \ref{thm: dissolution of patches},  it suffices to prove that the majority of descendants dissolve, or, more precisely, that the ratio of levels belonging to non-dissolved descendants of a patch tends to zero.

 To achieve this, we relate the expectation of the ratio of levels belonging to undissolved (descendants of) patches to the \textit{average} over the probability of a \textit{single} descendant not being dissolved; see \eqref{eq: patchdissolutionsum}.  Since our bounds are uniform in the choice of the initial patch $p$ and the choice of the descendant, we can indeed prove that this average probability tends to zero. 
 
Let us mention a last technical nuance: so far, we discussed the event that each patch is dissolved only once. However, a patch may contain more than two levels, and we are interested in the event that the patch is eventually dissolved into singletons. To fill the remaining gap, we may repeat the above argument inductively (i.e., each patch will dissolve several times with high probability). This works because of the following observation: for any scale $n+m>n$, given the event $\mathcal{A}_{n,\infty}$, the maximal patch size at scale $n+m$ is a non-increasing function of $m$ (see Section \ref{sec: dissolution of most}). This means that the argument just has to be run for a finite number of times.

  
 \subsubsection{Proof Idea: Poisson statistics}
The proof of {Theorem \ref{thm: main theorem}} uses {Theorem \ref{thm: dissolution of patches}} as a crucial input, but it still makes up the bulk of our paper. 
We prove it essentially by moment methods. 
First, we recall that the eigenvalues of $H_n^{(\mathrm{free})}$ are labeled by configurations, i.e.\ elements of $\Sigma_n=\{\pm 1\}^n$. By Weyl's inequality (see Lemma \ref{lemma : labeling lemma}), we can use this labeling also for the interacting model, although this of course does not imply anything about the eigenvectors. 
We denote hence the eigenvalues of $H_n$ by $E_{\eta}$ with $\eta \in \Sigma_n$.  Given a $k$-tuple $(\eta_1,\dots,\eta_k) \in \Sigma_n^k$, the main building block of the proof is the ``\textit{factorization property}" of $(E_{\eta_1},\dots,E_{\eta_k})$: 
The joint probability density function (pdf) of $(E_{\eta_1},\dots,E_{\eta_k})$ is a product of pdfs of $E_{\eta_i}$, up to some error that is small when $n$ is large and $k$ is  $O(1)$, and only for \emph{most} of such $k-$tuples.    
Once such a factorization property is established, the rest of the proof concerns identifying and bounding all possible error terms. We elaborate on this in Section \ref{sec: error terms intro}.

In the following, we describe the main conceptual tools required to prove the factorization Lemma. To keep the picture transparent, we simplify certain parts and hide some technical issues in this introduction.  
\subsubsection{Factorization for the free model $H_n^{(\text{free})}$}




Let us first assume for simplicity that the $h_i$ variables are Gaussian with mean zero and unit variance. Then for any $\sigma \in \Sigma_{n}$, $E_{\sigma}^o:={\sigma}.h=\sum_{x=1}^n\sigma(x)h_x $ is a Gaussian random variable with mean zero and variance $n$. Consequently, by a simple Gaussian computation we observe that for $\sigma_1,\sigma_2 \in \Sigma_n$, the pdf of $(E_{\sigma_1}^o,E_{\sigma_2}^o)$ approximately factorizes if the following holds:
\begin{equation}\label{eq: covariance vanishes}
   \frac{1}{n} \text{Cov}(E_{\sigma_1}^o,E_{\sigma_2}^o)=\frac1{n} \sum_x \sigma_1(x) \sigma_2(x)=o(1). 
\end{equation}
This is the manifestation (in an approximate way) of the 
well-known fact that Gaussian random variables are independent iff their covariance vanishes. Similarly, for $(E_{\sigma_1}^o,\dots,E_{\sigma_k}^o) $ the factorization  holds if for every pair  of indices $1\leq i\neq j\leq k$ the estimate \eqref{eq: covariance vanishes} holds with $\sigma_1,\sigma_2$ replaced by $\sigma_i,\sigma_j$. 

Standard large deviations estimates teach us that for  most  $k$-tuples $(\sigma_1,\dots,\sigma_k)$ 
the estimate \eqref{eq: covariance vanishes} indeed holds for every pair $1\leq i\neq j\leq k$, and this implies that for most $k$-tuples, factorization indeed holds. In Section \ref{sec: typical def}, we introduce the notion of a \emph{typical} $k$-tuple, formalizing the above considerations. In Lemma \ref{lem: counting 00}, we establish that the number of \textit{atypical} $k$-tuples, divided by the total number of possible $k$-tuples, is bounded by $C\exp(-n^a)$ with $a>0$. 


Let us now discuss how the Poisson property relates to this factorization.
Let $I$ be an interval of length $\lambda \sss_n$ with $\lambda=O(1)$ such that $I\ni 0$ (recall $\sss_n$ from \eqref{sec: stat sn} as the mean level spacing). 
One possible approach to prove that ``the eigenvalue process converges to a Poisson process" begins  by showing  that the number of eigenvalues in $I$, denoted by $\mathcal{N}_I$, has a Poisson distribution with rate $\lambda$ as $n\to\infty$.
In particular, the $k$-th moment $\mathbb{E}(\caN_I^k)$ should tend to $m_k(\lambda)$, the $k$-th moment of this Poisson distribution, which is a polynomial in $\lambda$ whose highest-degree term is  $\lambda^k$. Using the shorthanded notation $\underline{\sigma}=(\sigma_1,\dots,\sigma_k) \in \Sigma_n^k$, we compute 
\begin{equation} \label{eq: moment first encounter}
\mathbb{E}(\caN_I^k)=   \sum_{\underline{\sigma}} \mathbb{P}(\bigcap_i {E}^o_{\sigma_i} \in I).
\end{equation}
If the approximate factorization were to hold for all $k$-tuples $\underline{\sigma}$, then \eqref{eq: moment first encounter}  would be well-approximated by 
\begin{equation}\label{eq: wrong result from fact}
2^{nk} (\frac{1}{\mathsf{s}\sqrt{2 \pi n}} |I|)^k =  \lambda^k,
\end{equation}
because the  pdf of $E^o_{\sigma}$ around zero is approximately constant and equal to $\frac{1}{\mathsf{s} \cdot\sqrt{2 \pi n}}$, and because $\sss_n= \mathsf{s} \cdot\sqrt{2 \pi n}2^{-n}$. 
Obviously, this computation leading to \eqref{eq: wrong result from fact} has to be flawed because $m_k(\lambda)\neq \lambda^k$.
The point is this: even though the fraction of $k$-tuples for which approximate factorization fails is small, some of these atypical $k$-tuples \emph{do yield an important contribution to \eqref{eq: moment first encounter}}. Indeed, assume that the tuple $(\sigma_1,\ldots,\sigma_k)$ consists of only $k'<k$ different configurations $(\sigma'_1,\ldots,\sigma'_{k'})$ (which is hence an atypical case); then we obviously have 
$$
\mathbb{P}(\bigcap_i {E}^o_{\sigma_i} \in I)= \mathbb{P}(\bigcap_{i=1,\ldots,k'} {E}^o_{\sigma'_i} \in I) \approx (\frac{|I|}{\mathsf{s}\sqrt{2 \pi n}})^{k'}
$$
where we used factorization for the second intersection of $k'$ events, not for the first one of $k$ events.  This expression gets multiplied by the number of $k'$-tuples, and it then gives a contribution of the form $\lambda^{k'}$ to \eqref{eq: moment first encounter}. In this way, the polynomial $m_k(\lambda)$ is reconstructed taking combinatorial factors into account. Then, the convergence to $m_k(\lambda)$ can be proven by controlling the remaining ``error terms", i.e., atypical terms in the sum with vanishing contributions: (e.g. see Section \ref{sec: error terms intro} on how to control the error terms for our model).

\subsubsection{Beyond the Gaussian case}

In the previous section, we discussed the factorization property for typical $k$-tuples, in case the random variables $h_x$ are Gaussian. Let us briefly comment on how to treat the non-Gaussian case.  

First, given a typical $k$-tuple $(\sigma_1,\dots,\sigma_k)$, we construct  in Lemma \ref{lem: indep r.v. contruction}, $2^{k-1}$ \textit{independent} random variables $\{ y_{\tau}\}_{\tau \in \Sigma_{k-1}}$ with the following properties
\begin{enumerate}
    \item The pdf of each $y_{\tau}$ is given by an $m_{\tau}$-fold convolution of the pdf of $h_x$, with $m_{\tau}=n(1/2^{k-1}+o(1))$.
    \item For all $1 \leq i\leq k$, $E_{\sigma_i}^o= h.\sigma_i$ is written as a linear combination of $y_{\tau}$, see \eqref{eq:changeofvariable}.
\end{enumerate}
Item 1 allows us to approximate the pdf of $y_{\tau}$ with the pdf of a Gaussian with variance proportional to $n/2^{k-1}$. This follows from local central limit  (LCLT) estimates \eqref{eq: LCLT1}.  
Item 2 implies that to obtain the pdf of $(E_{\sigma_1}^o,\dots,E_{\sigma_k}^o)$ from $y_{\tau}$, we only need to integrate over the remaining {orthogonal} $2^{k-1}-k$ directions.  To ensure that the error terms stemming from LCLT estimates remain under control after the integration, we use the Brascamp-Lieb inequality stated in Lemma \ref{BL ineq lemma}.

Finally, we comment on the main idea of the construction of $\{ y_{\tau}\}_{\tau \in \Sigma_{k-1}}$. 
We start with the case $k=2$, in which case $\tau=+,-$.  Let  $\mathfrak{J}_+$,$\mathfrak{J}_-$ be the set of all indices $x \in [n]=: \{1,2,\dots,n\}$ such that $\sigma_1(x)=\pm \sigma_2(x)$ (plus and minus for $\mathfrak{J}_{+}$ and $\mathfrak{J}_-$ respectively). Then $y_\tau$ is given by $\sum_{x \in \mathfrak{J}_{\tau}} \sigma_1(x) h_x$  for $\tau = \pm$.

For $k=3$, we further partition $\mathfrak{J}_+$ into $\mathfrak{J}_{++}$ and $\mathfrak{J}_{+-}$ as follows: $\mathfrak{J}_{+\tau}$ is the set of $x \in \mathfrak{J}_+$ such that $\sigma_1(x) = \tau \sigma_3(x)$ for $\tau = \pm 1$. In this case, $y_{++} = \sum_{x \in \mathfrak{J}_{++}}\sigma_1(x) h_x$. Other combinations such as  $-+$ are treated analogously. In case $k=3$ this gives us $2^2$ independent random variables.  Since $(\sigma_1,\sigma_2,\sigma_3)$ is a typical tuple, we can deduce that $|\mathfrak{J}_{\tau_1\tau_2}|=n (1/4+o(1))$ for all $\tau_1,\tau_2$ (see Lemma \ref{lem: indep r.v. contruction}). Then, since the pdf of $h_x$ is symmetric around zero, we deduce item 1 above. Item 2 is also straightforward from the construction. We may continue this process inductively and obtain the promised set of $2^{k-1}$ independent random variables for any $k$.


 \subsubsection{Semi-perturbed process: definition}
 The second step in the proof of the factorization is to introduce a  \textit{semi-perturbed} process. We decompose $n=n_0+n_1$, such that $n_0= \varrho n$, with a judicious choice of $0<\varrho <1$. The semi-perturbed process consists of the  eigenvalues of $\widetilde{H}_n:= H_{n_0} +\sum_{i=n_0+1}^n h_iZ_i$ (see\ \eqref{eq: tilde def}), and these eigenvalues are denoted by $\widetilde{E}_{\eta}$, $\eta \in \Sigma_n$. We also decompose any $\eta \in \Sigma_n$ as $\eta=(\mu,\sigma)\in \Sigma_{n_0}\times \Sigma_{n_1}$. Notice that  we have $\widetilde{E}_{\eta}={E}_{\mu}+E^o_{\sigma}=: E_{\mu}+\sum_{x=n_0+1}^n\sigma(x)h_x$. Also, notice the slight difference in the definition of $E_{\sigma}^o$ here, compared to the previous Section, i.e., the shifted indices.

 Let us explain the choice of $n_0=\varrho n$ and what this achieves: recall the scaling parameter $\sss_n\sim 2^{-n}$ \eqref{sec: stat sn}. Recall that $\sss_n$ is the typical level spacing at scale $n$. We choose $n_0$ such that: 
 $$\alpha^{n_0} \ll \sss_n \ll \alpha^{\theta n_0}.$$
 \begin{enumerate}
     \item $\alpha^{n_0} \ll \sss_n$ ensures the fact that $\widetilde{E}_{\eta}$ is a good approximation of $E_{\eta}$, since $\|H_n -\widetilde{H}_n \| \sim \alpha^{n_0}$. In the  proof of Theorem \ref{thm: main theorem}, we look at scales of order $\sss_n$; therefore, moving the levels by an exponentially smaller value will not change the Poisson property. From now on, we will focus on proving the factorization property and Poisson statistics for the semi-perturbed process. 
     \item $\sss_n \ll \alpha^{\theta n_0}$ means non-resonant levels of $H_{n_0}$ are sufficiently far from each other: we will comment on this more later.
 \end{enumerate}
\subsubsection{Semi-perturbed process: factorization and beyond}

Recall  $I$  the interval of length $\lambda \sss_n$ with $\lambda=O(1)$ and $0 \in I$. Similar to the case of the free Hamiltonian, the main workhorse to prove the Poisson statistics is the convergence of $\mathbb{E}(\mathcal{N}_I^k) \to m_k(\lambda)$. This means  ($(\underline{\mu},\underline{\sigma})\in \Sigma_{n_0}^k \times \Sigma_{n_1}^k $): 
\begin{equation} \label{eq: factor caricature}
    \sum_{\underline{\mu},\underline{\sigma}} \mathbb{P}\left(\bigcap_{i=1}^k {E}_{\mu_i}+ \sigma_i.h \in I\right) =  m_k(\lambda)(1+o(1)).
\end{equation}
Of course, reality is much more complicated, as addressed in Sections \ref{sec: factor 2} and \ref{sec:kth moment main}: one should take into account different intervals. However, the above statement contains the main idea of the proof.   \\

We group the terms in the above sum into three main categories: 
\begin{enumerate}
    \item \textit{Leading terms.} These are terms in the above sum such that both $\underline{\mu}$ and $\underline{\sigma}$ are \textbf{typical} $k$-tuples. In the limit, such terms will create the term $\lambda^k$ appearing in $m_k(\lambda)$.
    \item \textit{Subleading terms.} Recall $\eta_i=(\mu_i,\sigma_i)$. These are terms such that $(\eta_1,\dots,\eta_k)$ has only $k'<k$ distinct elements. Such terms are responsible for creating the expression $m_k(\lambda)-\lambda^k$.  
    \item \textit{Error terms.} Any other terms in the sum not mentioned above. The main remaining task is to prove that the contribution of these terms is negligible. 
\end{enumerate}
In the following, we outline the general approach to treat each of the above cases. \\

The \textbf{subleading} terms:  assuming that the above claims hold for the first and third cases, the \textit{subleading} terms can be treated inductively via a simple combinatorial construction. This is done in Section \ref{sec: computation of moments} and it is similar to the free case as explained before.  \\

 The \textbf{leading terms}: consider a $k$-tuple $(\eta_1,\dots,\eta_k)$ with $\eta_i=(\mu_i,\sigma_i)$, belonging to this set as defined earlier. If we turn off the interaction and neglect the bath, factorization is explained in the case of the free model $H_n^{(\text{free})}$. Now, if we add the interaction and consider the effect of the bath, each energy level may vary by an amount of order one, i.e., $|\widetilde{E}_{\eta_i}-E^o_{\eta_i}| \leq C=O(1)$. Such variations may ruin the factorization property, especially since we are looking at intervals of order $\sss_n \sim 2^{-n}$. The remedy is to compare $\widetilde{E}_{\eta_i}$ and $E_{\eta_i}^o$ in the following way. For the sake of clarity, let us do this comparison for a single spin configuration $\eta_1\equiv \eta=(\mu,\sigma)$. \\

We know that $\widetilde{E}_{\eta}=E_{\mu}+ E^o_{\sigma}$ and ${E}_{\eta}^o= E_{\mu}^o+E_{\sigma}^o=:\sum_{x=1}^{n_0}\mu(x)h_x+\sum_{x=n_0+1}^{n}\sigma(x)h_x$, i.e., the  pdfs of $\widetilde{E}_{\eta}$ and ${E}^o_{\eta}$ are given by the convolution of the  pdf of $ E_{\sigma}^o$ with the pdf of $E_{\mu}$ and $E_{\mu}^o$ respectively.  Now the point is the following: The pdfs of $E_{\sigma}^o$ and $E_{\mu}^o$ are well approximated by  Gaussian pdfs with variances $n_1$ and $n_0$, respectively.  Moreover, $E_{\mu}$ and $E_{\mu}^o$ differ by a perturbation of order one. Since the density of a Gaussian with variance $n_0$ (i.e., $E_{\mu}^o$) is approximately constant on scales of order $\sqrt{n_0}$, we have good control over the mentioned perturbation (which is of order one) \textit{after the convolution with the density of $E_{\sigma}^o$}. This means that the pdf of $E_{\eta}^o$ and $\widetilde{E}_{\eta}$ remain close enough. This approximation can also be done for $k>1$ thanks to the technical tools developed for the factorization of the free model, as explained above. \\

The above approximation allows us to carry over the factorization of the free model to the semi-perturbed process. We formalize the above intuition in Section \ref{sec: factor 2}.

The fate of the third category, the \textbf{error terms}, is discussed in the next section.

\subsubsection{Semi-perturbed process: error terms} \label{sec: error terms intro}

Here we give a general simplified picture of the possible error terms in \eqref{eq: factor caricature} and the idea to treat them. Section \ref{sec:kth moment main} is devoted to this task.
Consider $\underline{\sigma}=(\sigma_1,\dots,\sigma_k)$ as a collection of $k$ vectors in $\mathbb{R}^{n_1}$, whose spanned  subspace  has dimension  denoted by $\text{dim}(\underline{\sigma})$. Then, the error terms in \eqref{eq: factor caricature} belong to either of the following cases: 
\begin{enumerate}

\item $\text{dim}(\underline{\sigma})=k$: In this case, first, we argue that the contribution of $\mathbb{P}(\cdots)$ in \eqref{eq: factor caricature}  is bounded by $Cn^k |I|^k$ uniformly in $\underline{\mu}$: the mentioned probability is trivially bounded by 
\begin{equation} \label{intro: eq: bound linf}
\sup_{a_i} \mathbb{P}(\cap_i \sigma_i.h \in I^{a_i}),
\end{equation} where $I^{a_i}$ is the interval $I$ shifted by the real number $a_i$.  For any fixed set of $a_i$, this is simply an integration over a  $ k$-dimensional subspace of $\mathbb{R}^{n_1}$.  Brascamp-Lieb inequality tells us that this integration will at most create a prefactor of order $n_1^k$ compared to the volume of the $k$-cube $|I|^k$ (see Lemma \ref{lem:carrot}), hence the mentioned bound holds. \\
On the other hand, either $\underline{\sigma}$ or $\underline{\mu}$ should be \textit{atypical} (if not, we are in the first case of the previous section). This means the number of such configurations is bounded by $2^{nk}\times \exp(-n^a)$, with some $a>0$, thanks to the bound on the number of atypical configurations (see Lemma \ref{lem: counting 00}). Combining this combinatorial factor and the previous probability bound, we deduce that the contribution of such terms vanishes in the limit, recalling that $|I|\sim \sqrt{n}2^{-n}$.

    \item  {$\text{dim}(\underline{\sigma})=\ell<k$}. In this case, several ``pathologies" may occur that require treatment. Instead of trying to be exhaustive, we give simple examples of each of these pathologies and the tool to treat them. \\
    In this case, among $\underline{\sigma}$, we can find $\ell$ linearly independent  spin configurations (as vectors of $\mathbb{R}^{n_1}$). We denote the set of their indices by $F\subset [k]=: \{1,2,\dots,k\}$, and $D:=[k]\setminus F$. By $\sigma_{F/D}$, we mean $\{\sigma_i \}_{i \in F/D}$ respectively.
    For any $j \in D$, $\sigma_j$ is written as a linear combination of $\sigma_F$. In other words, there is a $|D| \times |F|$ matrix, $v$, such that $\sigma_D$ is given by multiplication of the matrix $v$ with the vector $\sigma_F$.   \\
    The input of Lemma \ref{lem: coeff in module group} is that the number of ``\textit{admissible}" matrices $v$ only depends on $k$, and not on $n$. Therefore, it is sufficient to consider a fixed, arbitrary partition of $k$ into $F,D$, and a fixed,  arbitrary admissible $v$, i.e., from now on we fix $F,D,v$ as above and look at the sum in \eqref{eq: factor caricature} such that for all $\underline{\sigma}$, $\sigma_F$ is related to $\sigma_D$ by $v$.    
    \\
    Here, 
    for any realization of $\{h_x\}_x$, a given $\{\sigma_i.h\}_{i \in F}$ fully determines $\{\sigma_i.h \}_{i \in D}$. Hence, if we are unlucky, the best we can do is to bound $$\mathbb{P}(\cap_i 
  E_{\mu_i}+\sigma_i.h \in I) \leq  \mathbb{P}(\cap_{i \in F} E_{\mu_i}+\sigma_i .h \in I).$$ Now, if we bound the latter similarly to  \eqref{intro: eq: bound linf}, we get a factor of order $2^{-n_1|F|}$ with polynomial corrections. Notice that now we have $|F|$ instead of $k$ free variables,  which gives us the volume of an $|F|$-cube as the leading order. The first problem here is that now we sum this contribution over $\underline{\sigma}$ and $\underline{\mu}$. Recall that we fixed $F,D,v$. Then, the sum over $\underline{\sigma}$ will only involve a sum over $\sigma_F$, which gives us a factor $2^{n_1|F|}$ . However, we have not restricted the sum over $\underline{\mu}$. A priori, this creates a deadly factor of order $2^{n_o|D|}$. Considering all the factors,  we get a term of order $2^{n_o|D|}$ instead of $o_n(1)$. \\

  To refine the above argument, we argue that summing over ${\mu}_D \equiv\{ \mu_j\}_{j\in D}$ creates only $O_n(1)$ non-zero terms instead of $2^{n_o|D|}$, with high probability. This can be achieved thanks to the following two observations: 
  \begin{enumerate}
      \item For $j \in D$,  $\mu_j\neq \mu_{j}'$, 
      $$\{E_{\mu_j}+\sigma_j.h \in I \bigcap_{i\in F} E_{\mu_i}+\sigma_i.h  \in I\} \quad \text{and}  \quad \{E_{\mu_j'}+\sigma_j.h \in I \bigcap_{i\in F} E_{\mu_i}+\sigma_i.h  \in I \} $$ implies $\mu_j$ and $\mu_{j}'$ belong to the same resonant patch i.e., same member of $\caP_{n_0}$, where we recall the definition of $\caP_{n_0}$ in Section \ref{sec: intro dissolution statement}.  
  \item By ``maximal patch", we mean the resonant patch(es) whose number of levels (size) is largest. The second observation is that the probability of the maximal patch's size being larger than $K$ decays to zero as $K \to \infty$ uniformly in $n$. 
  
  \end{enumerate}
   The above observations mean that we can replace the factor of $2^{n_0|D|}$ by $K^{|D|}$, assuming that the maximal patch size is $K$.  In other words, these two observations imply the following bound: 
   \begin{equation} \label{intro: eq: bound K}
       \sum_{\underline{\mu}, \sigma_F}\mathbb{P}_A(\cap_{i} E_{\mu_i}+\sigma_i.h \in I) \lesssim  K^{|D|} \sum_{\mu_F,\sigma_F} \mathbb{P}_A(\cap_{i \in F} E_{\mu_i}+\sigma_i.h \in I),
   \end{equation}
   where $\mathbb{P}_A(\cdot) = \mathbb{P}(\cdot \cap A)$ with $A$ being  the event where maximal patch  size is $K$; moreover, in the first expression, $\sigma_D$ is determined by  $\sigma_F$ and $v$. Before proceeding, let us give brief comments about the above observations:\\
   Observation (a) is a result of the choice of $n_0,n_1$, and Lemma \ref{lem: coeff in module group}. This lemma tells us that all the non-zero elements of $v$ are lower and upper bounded in absolute value by functions of $k$ (independent of $n$). Combining this with  the fact that $|I| \sim \sss_n \ll \alpha^{\theta n_0}$,  and a simple computation, gives us the  bound $|E_{\mu_j}-E_{\mu_j'}|\leq C\sss_n \ll\alpha^{\theta n_0}$ i.e., as desired in observation (a) for $n$ large enough. \\
   Recall the event $\mathcal{A}_{m,\infty}$: the event where there is no new resonance event after scale $m$. Given this event, the maximal patch size does not grow after scale $m$. Therefore, observation (b)  is tantamount to a bound on $\mathbb{P}(\mathcal{A}_{m,\infty})$. This has been discussed in the previous sections. \\
    In \eqref{intro: eq: bound K}, we replaced $\mathbb{P}$ with $\mathbb{P}_A(\cdot) = \mathbb{P}(\cdot \cap A)$, where $A=A_K$ is the event that the maximal patch size is at most $K$. In particular, $\mathbb{P}(A_K) \to 1$ as $K \to \infty$. We comment later on why such replacements are benign. \\

   Although we succeeded in reducing the factor of $2^{n_0|D|}$ to $K^{|D|}$, we are still far from getting what we wanted: if we bound the right-hand side of \eqref{intro: eq: bound K} similar to \eqref{intro: eq: bound linf}, we will get $K^{|D|}2^{n|F|} \times 2^{-n|F|} $ up to a polynomial correction which can be large in $n$.  Further refinement of the above bound depends on the structure of $v$. Here we encounter two cases, the second of which splits into two subcases:
    \begin{enumerate}
        \item  $v$ contains a non-trivial relation: for each $i \in D$, $\sigma_i$ is a linear combination of $\sigma_F$ with coefficients given by ${v}_i:=\{v_{i,j}\}_{j \in F}$.  If we find one ${v}_i$ with at least two non-zero elements $v_{i,j},v_{i,j'}$, we say that $v$ has a non-trivial relation.
        \textit{Not having a non-trivial relation} is more intuitive: it means we can partition $\underline{\sigma}$ into $\ell$ different sets, say $A_1,\dots,A_{\ell}$, such that for any $A_i$, and any of its elements $\sigma_j,\sigma_{j'}$ we have either $\sigma_j=\sigma_{j'}$ or $\sigma_j=-\sigma_{j'}$.   
        
        In the above analysis, we got a combinatorial factor $2^{n_1|F|}$ from the sum over $\underline{\sigma}$. Lemma \ref{lem: counting} tells us that if $v$ has a non-trivial relation, this combinatorial factor can be replaced by $(2^{|F|}-1)^{n_1}$. This provides an additional factor of order $\exp(-cn)$ compared to the above analysis, which allows us to conclude the desired smallness in this case. 
        \item The only remaining case is when $v$ does not have any non-trivial relation. In this case, we rely essentially on Theorem \ref{thm: dissolution of patches}. In this scenario, the main conceptual tool can be explained for $k=2$. Here again, we have two cases: 
        \begin{enumerate}
            \item \textit{Resonant case:} this is the case where $\sigma_1=\sigma_2\equiv \sigma $ and $\mu_1\neq \mu_2$ (we exclude $\mu_1=\mu_2$, because this means we are in the subleading case). Here, the crucial observation is that $E_{\mu_1}+\sigma.h\in I$ and $E_{\mu_2}+\sigma.h\in I$ can only happen if $E_{\mu_1},E_{\mu_2}$ are in the same resonant patch, i.e., same member of $\caP_{n_0}$. We denote this event by $\chi((\mu_1,\mu_2) \text{ are in res.})$. 
            
            The above observation is a consequence of the fact that $|I|\sim \sss_n \ll \alpha^{\theta n_0}$. If we denote $\caA=\caA^{K,\epsilon}_m$ to be the event where maximal patch size is $K$, and the ratio of resonant levels  are smaller than $\epsilon$ after certain scale $m<n_0$, then the above observation means:
    \begin{align} 
                \sum_{\mu_1 \neq \mu_2,\sigma} \mathbb{P}_{\caA}(\cap_{i=1}^2E_{\mu_i}+\sigma.h \in I) &\leq \sum_{\mu_1\neq \mu_2}\mathbb{E}_{\caA}(\chi((\mu_1,\mu_2)\text{ are in res.})) \times \sum_{\sigma}\sup_a \mathbb{P}(\sigma.h \in I^a) \nonumber \\
                & \leq  \epsilon K^22^{n_{0}} \times C 2^{n_1} |I|/\sqrt{n} \leq   C\epsilon K^2
                \label{eq: intro: res},
     \end{align}
            where $\mathbb{P}_{\caA}(\cdot) = \mathbb{P}(\cdot \cap \caA)$. On the RHS of the first line: the first sum is bounded by $\epsilon K^2 2^{n_0}$. This is a direct consequence of the definition of event $\caA$, which bounds the number of in resonant pairs; 
            the probability appearing in the second sum is bounded by $C|I|/\sqrt{n}$ as we mentioned before.
            The last bound is obtained by recalling that $|I|\sim \sqrt{n}2^{-n}$. Taking $\epsilon$ sufficiently small allows us to control the terms belonging to this category (resonant case). \\  
            On the LHS of the first line, we replaced $\mathbb{P}(\cdot)$ with $\mathbb{P}_{\caA}(\cdot)$.
            This replacement can  be done because $\caA$ is a ``high probability" event:  $\mathbb{P}(\caA^{K,\epsilon}_m)=1-o_m(1)-o_K(1) $ for any fixed $\epsilon$ as proved in Lemma \ref{lem: bound on caa}: this is an application of Theorem \ref{thm: dissolution of patches}. We comment more later on why such a replacement is allowed for our purposes.

        
        \item \textit{Anti-resonant case:} this is the case where $\sigma_1=-\sigma_2$. To control sums appearing in the LHS of \eqref{eq: factor caricature}, in such cases it is essential to have a bound on the following quantity: the number of pairs $(\mu_1,\mu_2) \in \Sigma_{n_0}\times \Sigma_{n_0}$, such that $|E_{\mu_1}+E_{\mu_2} |\ll \sss_n$. Indeed, if such a quantity grows as fast as $2^{n_0}$, then the right-hand side of \eqref{eq: factor caricature} cannot be controlled: 
        assume $I$ is symmetric around zero, if $|E_{\mu_1}+E_{\mu_2}| \ll \sss_n$  then by a simple calculation one can see that   
        $\mathbb{P}(E_{\mu_1}+\sigma.h \in I \cap E_{\mu_2}-\sigma.h \in I ) \approx \mathbb{P}(E_{\mu_1}+\sigma.h \in I)$. Then, if we sum the LHS of \eqref{eq: factor caricature}  over all pairs of $(\sigma,-\sigma)$ and all pairs of $(\mu_1,\mu_2)$ such that $|E_{\mu_1}+E_{\mu_2}| \ll \sss_n$, we get a factor of order one, provided that the number of such pairs grows as $O(2^{n_0})$; whereas, such terms should have a contribution of $o(1)$.

        This observation leads us to define the notion of \textit{anti-resonant} pairs. We say that  $\mu$ and $\mu'$ are anti-resonant if $|E_{\mu}+E_{\mu}'| < \alpha^{\theta n_0}/4$. In Section \ref{sec: antiresonances}, we prove that the number of such pairs divided by $2^{n_0}$ goes to zero almost surely. The proof of this theorem follows the same line as the proof of  Theorem \ref{thm: dissolution of patches}. Thanks to this theorem, treatment of the anti-resonant terms follows the same steps as the resonant terms discussed before. \\

        At first sight, the notion of \textit{anti-resonant} pairs, and the fact that we need to control them, may seem mysterious. One may try to demystify this notion by realizing the following fact: if we turn off both the interaction and the bath, i.e., considering $H_{n_0}^{\mathrm{(free)}}$, then for each $\mu$, $(\mu,-\mu)$ are anti-resonant since in this case $E_{\mu}=-E_{-\mu}$. Then, the number of anti-resonant pairs will be at least  $2^{n_0}$. On the other hand, in this case, the semi-perturbed process coincides with the eigenvalues of $H_{n}^{(\textrm{free})}$. We recall that $H_{n}^{\mathrm{(free)}}$ does not have the Poisson property around the origin because of the reflection symmetry.  We already explained that adding the bath with Assumption \ref{ass: nondegeneracy} removes this symmetry. The role of Theorem \ref{thm: antires vanish} is to ensure that after adding the interaction this symmetry is not revived, not even in an approximate way.

        \end{enumerate}

    \end{enumerate}
      \end{enumerate}
    Let us finish this section with the following comment. We wanted to prove the convergence of moments as in \eqref{eq: factor caricature}. However, in \eqref{intro: eq: bound K} and \eqref{eq: intro: res}, error terms are shown to be small, assuming a certain high-probability event $\caA$. Of course, this is not sufficient to prove the convergence of moments: in those small-probability events, our observable can attain abnormally large values, preventing the mentioned convergence. However, strictly speaking, to prove Theorem \ref{thm: main theorem} we do not need the convergence of moments.  This becomes apparent by looking at the expression \eqref{sec: stat Laplace convergence}. Since the test function $\varphi$ is taken to be positive, $\exp(-\xi_n(\varphi))$ is a bounded observable; hence restricting ourselves to high-probability events at this stage is harmless. We deal with such subtleties in Section \ref{sec: proof of them 2}.

\subsection{Plan of the paper}

Sections \ref{sec: sec pre+proof loc} and \ref{sec: dissolution} are devoted to the proofs of Theorem \ref{thm: fock space localization} and \ref{thm: dissolution of patches}. In particular, Section \ref{sec: labeling} and Section \ref{sec: term} contain basic preliminaries;  Section \ref{sec: proof of thm loc} contains the proof of Theorem \ref{thm: fock space localization}. Section \ref{sec: dissolution} contains the proof of Theorem \ref{thm: dissolution of patches}.  Section \ref{sec: antiresonances}  shows that the number of anti-resonant pairs is small.  Section \ref{sec: factor1} proves the factorization property discussed above for the free Hamiltonian. Section \ref{sec: factor 2} is devoted to the factorization property for the semi-perturbed process.  Section \ref{sec:kth moment main} combines the machinery of the previous sections to prove the ``full" factorization property. Section \ref{sec: proof of them 2} deduces Theorem \ref{thm: main theorem} from the full factorization property.  In Appendix \ref{sec: app Non-degeneracy}, we prove that $H_n$ is almost surely non-degenerate and non-anti-degenerate; these are properties that are used in our analysis. Appendix \ref{sec: LCLT} recalls basic results concerning the local central limit theorem from the literature.  Appendix \ref{sec: app  LD estimate} contains a simple combinatorial argument which helps us to deal with non-typical cases. Appendix \ref{sec:PTtheory} recalls basic tools of analytic perturbation theory. Appendix \ref{sec: app Gaussians} gathers some preliminary estimates.

\subsection{Conventions} \label{sec: conventions}
We use the following conventions throughout the paper. 
\begin{enumerate}
\item Symbols  $C,c,c',...$ are reserved for constants. These constants may depend on various parameters such as $\alpha,\theta, \mathsf{s}$,$H_B$,$n_B$; they may also depend on $k,K$; but they never depend on $n,n_0,n_1$. We may make this dependence explicit from time to time by writing symbols such as $C(K,k),C(\alpha, \theta),\cdots$. Such precision may be dropped for brevity purposes. Let us emphasize that dependence on $K,k$ is more subtle compared to other parameters since they will also be sent to infinity.  
\item We use the small $o$-notation in an unconventional way:   We will have three large parameters $n,k,K$. We define:  
\begin{enumerate}
    \item $f(n)=o_n(1)$ means $f(n) \to 0$ as $n \to \infty$ \textit{pointwise in $k,K$}. 
    \item $g(K)={o}_K(1)$ means $g \to 0$ as $K \to \infty$;  \textit{pointwise  in $k$;  and uniformly in $n$}.
    \item   $h(k)={o}_k(1)$ means $h(k) \to 0$ as $k\to \infty$  \textit{uniformly in $n,K$}. 
\end{enumerate}
\item By abusing the notation, we denote both of the following  operators by the same symbol $Z_i$: 
\begin{enumerate}
    \item $Z_i\otimes_{j\neq i}I_j: \otimes_n\mathbb{C}^2 \to \otimes_n \mathbb{C}^2$, where $I_j$ is the identity acting on site $j$. 
    \item $Z_i:\mathbb{C}_i^2 \to \mathbb{C}_i^2$, where $\mathbb{C}_i$ refers to the $i$th copy in the product $\otimes_n\mathbb{C}^2$.
\end{enumerate}
Almost always, the distinction between these two is clear from the context.
\item We set $\mathbb{N}=\{1,2,\dots\}$ and $\mathbb{N}_0=\{0,1,2,\dots\}$.
\item For any $n \in \mathbb{N}$, we denote:
$[n]:= \{1,\dots,n\},\:\Sigma_n:= \{ \pm1\}^n.$
\end{enumerate}

\section{Preliminaries and proof of localization}  \label{sec: sec pre+proof loc}

\subsection{ Labeling }\label{sec: labeling}
We first construct the (disorder-dependent) labeling of the spectrum of $H_n$ by $\Sigma_n\equiv \{ \pm1\}^n$ for $n>n_B$. By spectrum, we mean the multiset of eigenvalues considering the algebraic multiplicity.   The labeling is done inductively. For $n=n_B$, we do an arbitrary labeling.  
 For $n>n_B$, we assume that eigenvalues of $H_{n-1}$ are labeled by $\Sigma_{n-1}$ and we write these eigenvalues as $E_\sigma $ with $\sigma \in \Sigma_{n-1}$. Consider the operator 
$$
H_{n-1/2}:=H_{n-1}+h_n Z_n.
$$
Since this operator does not couple spin $n$ to the first $n-1$ spins, its eigenvalues are of the form $E_\sigma+h_n \mu$ with $\mu \in \{\pm 1\}$ and thus they are naturally labeled by $\Sigma_n=\Sigma_{n-1}\times \{\pm 1\}$, i.e., $E_{\sigma}+\mu h_n$ is labeled by $(\sigma,\mu) \in \Sigma_{n-1} \times \{\pm 1 \}$. 
We now lift this labeling to the spectrum of $H_n$ as follows: 
Let $j=j(\sigma,\mu)$ be the index of the eigenvalue of $H_{n-1/2}$, when enumerated in increasing order, i.e.\ $j=1$ corresponds to the smallest eigenvalue. Then the label $(\sigma,\mu)$ is given to the $H_n$-eigenvalue with index $j(\sigma,\mu)$. We make an arbitrary choice if $H_n$ has a degeneracy. In particular, this arbitrary choice means different copies of a repeated eigenvalue receive different labels.  Almost surely, we do not have any degeneracy, thanks to Lemma \ref{lemma: minimal gap antigap} if one assumes Assumption \ref{ass: nondegeneracy}. In fact, from Section \ref{sec: dissolution} onward such subtleties do not occur since we assume Assumption \ref{ass: nondegeneracy}.

This construction has the following useful property: 
\begin{lemma}
\label{lemma : labeling lemma}
For $n>n_B$, the eigenvalues of $H_n$   satisfy
$$
\left|E_{(\sigma,\mu)}-(E_\sigma +h_n \mu ) \right|  \leq C\alpha^n,
$$
for any $\sigma \in \Sigma_{n-1}$ and $\mu \in \{\pm 1\}=\Sigma_1$. More generally, eigenvalues of $H_{n-1+m}$
satisfy
$$
\left|E_{(\sigma,{\mu})}-(E_\sigma +h .{\mu} ) \right|  \leq C\alpha^n
,$$
for any $\sigma \in \Sigma_{n-1}$, and ${\mu} \in \Sigma_{m}$, where $h. {\mu}=\sum_{x=1}^{m}h_{x+n-1} \mu(x)$, and $C$ only depends on the bath and $\alpha$, and is uniform in the spin configuration.
\end{lemma}

\begin{proof}
The first bound follows by Weyl's inequality since $\| H_{n-1/2}-H_{n} \| \leq \alpha^{n-n_B} $, it is enough to  take $C=\alpha^{-n_B}$. To get the second bound, we apply the first bound inductively and estimate $\alpha^n+\alpha^{n+1}+\ldots$ by $2\alpha^n$, using that $\alpha<1/2$, and adjust $C$ by a factor of two.  
\end{proof}

 \subsection{Patches} \label{sec: term}

\subsubsection{The filtration $\caF_n$}
We also define the sub-algebras  for $n \geq n_B$: 
$\caF_n$ is the $\sigma-$algebra generated by $\{g_{\leq n}, h_{\leq n}\}$, and 
$\caF_{n+1/2}$ is the sigma-algebra generated by $\{g_{\leq n}, h_{\leq n+1}\}.
$
From the previous section, recall $H_n$, and also recall
$$
H_{n+{1}/{2}}= H_n+ h_{n+1} Z_{n+1}.
$$
Thus, in particular,
$$H_n=H_{n-1/2}+g_n\alpha^{n-n_B} X_1X_n. $$
We note that $H_n$ is $\caF_n$-measurable and $H_{n+1/2}$ is $\caF_{n+1/2}$-measurable.
\subsubsection{The events $\caA_{n_0,n_1}$} \label{sec: non res event}
Fix $\theta \in(2/3,1)$, and $\alpha\in(0,1)$ such that the  assumptions of Theorem \ref{thm: fock space localization} hold; then recall from Section \ref{sec: intro dissolution statement} that the spectrum of $H_n$ is partitioned into resonant patches. The partition is called $\caP_n$, and we write $\caP_{n,k}$ for the set of resonant patches with $k$ levels.
For $n \geq n_B$, the spectrum of $H_{n+1/2}$ is determined in a simple way from the spectrum of $H_n$, namely
$$
\mathrm{spec}(H_{n+1/2}) = \bigsqcup_{p \in \caP_n} ((p+ h_{n+1}) \sqcup   (p-h_{n+1}))
$$
where $p+x=\{E+x \, |\, E \in p\}$, and $\sqcup$ denotes the multiset union (e.g. $\{1,1,2 \} \sqcup \{1,2 \}=\{1,1,1,2,2\}$). 
This means that the spectrum of $H_{n+1/2}$ can be viewed as the multiset union of $p+\mu h_{n+1}$ with $(p,\mu) \in\caP\times \{\pm 1\}$.
Let us emphasize that from Section \ref{sec: dissolution} onward, given assumption \ref{eq: assumption 1}, in Lemma \ref{lemma: minimal gap antigap}
 we prove the non-degeneracy of $H_n$. This means that all the multisets mentioned above will become sets.

For $n\geq n_B$, we let $\caA_{n,n+1}$ be the event that all these multisets are at a larger distance than $\alpha^{\theta n}$  from each other, i.e.\ 
$$
\mathrm{dist}(p+\mu h_{n+1}, p'+\mu' h_{n+1}) > \alpha^{\theta n}, \qquad \forall (p,\mu) \neq  (p',\mu'), 
$$
where, for any two multisets  $X,Y$, $\mathrm{dist}(X,Y):= \inf_{x\in X,y \in Y} 
|x-y|$. Remark that in the above event, if $\mu=\mu'$, then $p \neq p'$. However, in this case $\mathrm{dist}(p+\mu h_{n+1},p'+\mu h_{n+1})=\mathrm{dist}(p,p')$ which itself is lower bounded by $\alpha^{\theta n}$ thanks to the definition of patches  of $H_n$.  \\

{For later use, we note that the above event also implies that 
$$
\mathrm{dist}(\mathrm{Conv}(p+\mu h_{n+1}), \mathrm{Conv}({p'+\mu' h_{n+1}})) > \alpha^{\theta n}
$$
where, for any multiset $S$, the interval $\mathrm{Conv}(S)=[\inf S,\sup S]$  is the convex hull of  $S$. This implication follows from the fact that none of the sets $p+\mu h_{n+1}$ can be partitioned into two subsets such that these subsets are at a distance larger than $\alpha^{\theta n}$ from each other;
 recall that this was the defining property of the collection $\caP_n$.}

We will also write, for $n_1>n_0\geq n_B$; 
$$
\caA_{n_0,n_1} :=  \caA_{n_0,n_0+1} \cap    \caA_{n_0+1,n_0+2} \cap \ldots \cap  \caA_{n_1-1,n_1}  
$$
and we note that this remains also meaningful for $n_1=\infty$. 
We note that 
 $$
 \caA_{n,n+1} \in \caF_{n+1/2}.
 $$
The event $ \caA_{n,n+1}$ has high probability, uniformly in $\caF_n$, namely: 
\begin{lemma}\label{lem: prob event caa} For any $n\geq n_B$, and any $m$, including $m=\infty$:
$$\mathbb{P}(\caA_{n,n+m}| \caF_n) \geq 1-C_{\alpha,\theta}(4\alpha^{\theta})^{n}, \qquad \mathbb{P}-\text{almost surely.}$$
\end{lemma}
\begin{proof}
  For a finite multiset $X$, with real elements, the set
  $$
  \{ h \in \mathbb{R} \,  |  \,   \mathrm{dist}(X-h,X+h) \leq \epsilon \}
  $$
  has Lebesgue measure bounded by $\epsilon |X|^2$, where $|X|$ denotes the number of elements of $X$.   Taking $X=\mathrm{spec}(H_n)$, it follows that  $\mathbb{P}(\caA_{n,n+1}^c| \caF_n)$ is bounded by $4^n \times \alpha^{\theta n}$ almost surely, since for this choice of $X$,  $\caA_{n,n+1}^c$ is a subset of the above-mentioned set, thanks to the remark following the definition of $\caA_{n,n+1}$. This yields the claim for $m=1$. Notice that the same bound holds for $\mathbb{P}(\mathcal{A}^c_{n,n+1}|\mathcal{F}_{n'})$ for all $n'<n$, thanks to the tower property of the conditional expectation. A  union bound and proper use of the mentioned tower property yields the claim for any $m$, including $m=\infty$, where we used the fact that $4\alpha^{\theta}<1$; hence, we set $C_{\alpha,\theta}=1/(1-(4\alpha^{\theta}))$. 
\end{proof}

\subsubsection{Spectral patches and resonant patches} \label{sec: spectral patches and resonant patches}
Any resonant patch $p\in \caP_{n-1}$ is in particular a multiset of eigenvalues of $H_{n-1}$ and so we can consider the spectral projector $P_p$ associated with that patch.  Recall $\delta_{x}$ as the Kronecker delta. 
Then, for $n>n_B$, 
$$
P_p \otimes \delta_{Z_n-\mu},\qquad \mu=\pm 1
$$
 are spectral projectors of $H_{n-1/2}$, associated with the sets $p+h_n\mu$; where in the above expression we abused the notation $Z_n$ as explained in the third item of Section \ref{sec: conventions}. This follows from the fact that $H_{n-1/2}$ does not contain any interaction of the spin at site $n$ with the other spins.

Let us now assume that the event $\caA_{n-1,n}$ is satisfied. 
We will study the spectrum of $H_n$ as a perturbation of $H_{n-1/2}$. 
Since $\|H_n-H_{n-1/2}\| \leq \alpha^{n-n_B}=C\alpha^n$ is smaller than half of the smallest distance between sets $p+\mu h_n$ with $p \in \caP_{n-1}, \mu \in \{\pm 1\}$, for large $n$, we can treat these patches by analytic perturbation theory. 
\begin{lemma} \label{lem: projectors PT theory} 
 For any $n>C$ ($C$ depends on the bath, $\alpha,\theta$) assuming that the event $\caA_{n-1,n}$ holds,  for any $p\in \caP_{n-1}$ and $\mu\in \{\pm 1\}$, there is a \emph{spectral patch} $S_{p,\mu}$ of $H_n$ (i.e.\ a multiset of eigenvalues with algebraic multiplicity), emanating from the spectral patch 
$p+\mu h_n$ of $H_{n-1/2}$,  such that
\begin{enumerate}
\item  Its cardinality is unchanged: $|S_{p,\mu}|=|p+\mu h_n|=|p|$, where $|\cdot|$ denotes the cardinality of a multiset where the multiplicity is taken into account. 
    \item The intervals $\mathrm{Conv}(S_{p,\mu})$ for different $(p,\mu)$ are disjoint. Moreover, $S_{p,\mu}$ and $S_{p',\mu'}$ are gapped for $(p,\mu)\neq (p',\mu')$ i.e. $\mathrm{dist}(S_{p,\mu},S_{p',\mu'})> \alpha^{\theta n}$.
    \item The endpoints of $\mathrm{Conv}(S_{p,\mu})$ and $\mathrm{Conv}(p+\mu h_n)$ coincide up to corrections at most $\alpha^{n-n_B}$.
    \item The corresponding spectral projector $P_{S_{p,\mu}}$ satisfies
\begin{equation}\label{eq: analytic pt}
|| P_{S_{p,\mu}}- P_p \otimes \delta_{Z_n-\mu}  || \leq C \left(2\alpha^{(1-\theta)}\right)^n.
\end{equation}
\end{enumerate}
\end{lemma}

\begin{proof}
These claims follow by basic analytic perturbation theory (see \cite{kato1995perturbation} Sections 1 and 4, and see Appendix \ref{sec:PTtheory}). In particular, Item 2) follows by Weyl's inequality, given the event $\caA_{n-1,n}$ and knowing that for $n$ sufficiently large  we have $\alpha^{\theta (n-1)}>\alpha^{\theta n}+ 2 \alpha^{n-n_B}$ since we have taken $\alpha^{(1-\theta)}<1/2$.  Item 3) follows by Weyl's inequality, and Item 4) is deduced by taking $n$ sufficiently large (e.g. such that $(2\alpha^{1-\theta})^n<1/2C$ with $C$ appearing in \eqref{eq:expansion bound0}), summing over $m$ in \eqref{eq:expansion bound0},  and adjusting the constant.
\end{proof}
\subsection{Proof of Theorem \ref{thm: fock space localization} }
\label{sec: proof of thm loc}

We now have sufficient notation to give the short proof of Theorem \ref{thm: fock space localization}. 
\begin{proof}
The labeling constructed in Section \ref{sec: labeling} imposes a bijection $\Sigma_n \to \{1,2,\dots,2^n \}$. The bijection promised in Theorem \ref{thm: fock space localization} is this bijection. In the following, we observe that this bijection satisfies the properties claimed in Theorem \ref{thm: fock space localization}.
We start with the following observation: For $i$ sufficiently large ($i>C_{B,\alpha,\theta}$), let $\caA_{i-1,i}$ hold. Then given $p \in \caP_{i-1}$ and $\mu \in \{\pm1\}$, we consider the patch $S_{p,\mu}$ as described in Lemma \ref{lem: projectors PT theory}. For any eigenvalue  $E\in S_{p,\mu}$, it holds that its label $\sigma \in \Sigma_i$ (as constructed in Section \ref{sec: labeling}) satisfies 
$\sigma(i)=\mu$,  thanks to Lemma \ref{lemma : labeling lemma} and definition of $\caA_{i-1,i}$.  Therefore, writing $\psi_\sigma$ for the corresponding eigenvector,  we have for $i$ sufficiently large: 
$$
\langle \psi_\sigma, \delta_{Z_i-\sigma(i)}\psi_\sigma\rangle  \geq 
\langle \psi_\sigma, P_p\otimes \delta_{Z_i-\sigma(i)}  \psi_\sigma \rangle
\geq \langle \psi_\sigma, P_{S_{p,\mu}}  \psi_\sigma \rangle -C (2\alpha^{(1-\theta)})^i=    1 -C (2\alpha^{(1-\theta)})^i
$$
where we abused the notation $Z_i$ in the second expression (see Item 3 of Section \ref{sec: conventions}). For the first bound, we used the fact that both $I-P_{p}$ and $\delta_{Z_i-\sigma(i)}$ are nonnegative operators; for the second bound, by taking $i$ large enough, we used 
\eqref{eq: analytic pt}. Hence, by partition of unity:
\begin{equation}\label{eq: bound on delta z}
 |\langle \psi_\sigma, \delta_{{Z}_i+\sigma(i)}\psi_\sigma\rangle|= \langle \psi_\sigma, \delta_{{Z}_i+\sigma(i)}\psi_\sigma\rangle \leq  C (2\alpha^{(1-\theta)})^i.  
\end{equation}
 Let $\ell^*$ be the smallest natural number such that $\caA_{\ell_*,\infty}$ holds, and such that the above bound holds (i.e. $\sigma(\ell_*)=\mu$, and \eqref{eq: analytic pt} is applicable). By Lemma \ref{lem: prob event caa}, this has indeed the claimed exponential tail (upon taking $\ell$ large and fixing $C_{\alpha,\theta,B}$).
We recall the projector $P_{\sigma_{\geq \ell}}$ from Section 
\ref{sec: fock space localization} and we note that 
$$
1-P_{\sigma_{\geq\ell}} \leq \sum_{i\geq\ell} \delta_{{Z}_i+\sigma(i)},
$$
in the operator sense, 
and hence, for $\ell>\ell_*$ 
\begin{equation}
    1- \langle \psi_\sigma, P_{\sigma_{\geq\ell}} \psi_\sigma \rangle \leq   \sum_{i\geq\ell}   \langle \psi_\sigma, \delta_{{Z}_i+\sigma(i)} \psi_\sigma \rangle \leq C\sum_{i \geq \ell} (2\alpha^{(1-\theta)})^i,
\end{equation}
where $\sum_{i\geq \ell}$ means sum over $i=\ell$ to $n$. 
The claim of Theorem \ref{thm: fock space localization}  follows by recalling that  $\alpha^{1-\theta}<1/2$, and fixing $C$ (it is sufficient to multiply the above constant by $ 1/(1-2\alpha^{1-\theta})$). 
          
 \end{proof}

\section{Dissolution of resonant patches}  \label{sec: dissolution}
From now on, we always assume $\alpha>0$. The proof of Theorem \ref{thm: dissolution of patches} in case $\alpha=0$ is trivial thanks to Appendix \ref{sec: app Non-degeneracy}. The proof of Theorem \ref{thm: main theorem} can be adapted to the case $\alpha=0$ in a straightforward way. We also assume that Assumption \ref{ass: nondegeneracy} holds. This means that $H_n$ is almost surely non-degenerate thanks to Lemma \ref{lemma: minimal gap antigap}. From now on, we take advantage of this non-degeneracy implicitly.

\subsection{Resonant patches} \label{sec: res patch def}
Recall the setup of Section \ref{sec: spectral patches and resonant patches}: let us assume that the event $\caA_{n-1,n}$ holds. Given  $p \in \caP_{n-1}$ and $\mu \in \{\pm1 \}$ recall the spectral patch $S_{p,\mu}$ of $H_n$ from Lemma \ref{lem: projectors PT theory}. Thanks to  Lemma \ref{lem: projectors PT theory}, given  a spectral patch $S_{p,\mu}$ of $H_{n}$,  there are two alternatives:
\begin{enumerate}
    \item  either $S_{p,\mu} \in \caP_n$, i.e.\ the spectral patch $S_{p,\mu}$ is also  a resonant patch,
    \item or $S_{p,\mu}$ can be partitioned into elements (more than one) of $\caP_n$. In this case, we say that \textit{the resonant patch has dissolved}. 
\end{enumerate}
Which of these alternatives occurs is the central question studied in this section.

If the spectral patch does not dissolve, i.e.\ if $S_{p,\mu} \in \caP_n$, then we can ask similar questions at higher scales.  
Let us introduce some notation to systematize these considerations: 
Let us assume that the event $\caA_{n-1,n-1+m}$ holds, for some $m>1$. 
We write now $$\underline{\mu}=(\mu_n,\ldots,\mu_{n+m-1}) \in \{\pm 1\}^m $$ and we let $\underline{\mu}_{n-1+\ell}$, for $\ell=1,\ldots,m$ to be the restriction of $\underline{\mu}$ to its first $\ell$ components.  Note that previously $\mu_n$ was simply called $\mu$, but now we need to keep track of the sites. 

   Let $p\in \caP_{n-1}$.  The event $S_{p,\mu_n
   } \in \caP_{n} $ is well-defined because the event $  \caA_{n-1,n-1+m} \subset \caA_{n-1,n} $ holds.
   Let us now assume that the event $S_{p,\mu_n
   } \in \caP_{n} $ holds true and let us rename $p'=S_{p,\mu_n
   } \in \caP_n$. 
   Therefore, for any $\mu_{n+1}$,  the set $p'+\mu_{n+1} h_{n+1}$ is a spectral patch of $H_{n+1/2}$, whose convex hull is sufficiently far away from the convex hull of other such spectral patches. Just as above, we can then consider the perturbed spectral patch $S_{p',\mu_{n+1}}$ that we also denote by  $S_{p,\underline{\mu}_{n+1}}$, referring back to the original patch $p$ at scale $n-1$, and recalling that  $\underline{\mu}_{n+1}=(\mu_n,\mu_{n+1})$.
   We can now again ask whether $S_{p,\underline{\mu}_{n+1}} \in \caP_{n+1}$. 
   If the answer is ``yes", then we continue the procedure. If, for a given $\underline{\mu}$ the answer was ``yes" $\ell-1$-times, then we conclude that 
$S_{p,\underline{\mu}_{n-1+j}} \in \caP_{n-1+j}$ for all $j=1,\ldots,\ell-1$ and we can  meaningfully ask whether    $S_{p,\underline{\mu}_{n-1+\ell}} \in \caP_{n-1+\ell}$. 

The event $S_{p,\underline{\mu}_{n-1+\ell}} \in \caP_{n-1+\ell}$ is actually only unambiguously defined provided that $S_{p,\underline{\mu}_{n-1+j}} \in \caP_{n-1+j}$ for all $j=1,\ldots,\ell-1$ (and provided that the event $\caA_{n-1,n-1+\ell}$ holds, but this is always assumed).   For that reason, and to avoid repetition, we \emph{define} the event $S_{p,\underline{\mu}_{n-1+\ell}} \in \caP_{n-1+\ell}$ as implying also the events $S_{p,\underline{\mu}_{n-1+j}} \in \caP_{n-1+j}$ for all $j=1,\ldots,\ell-1$.

The whole point will be that this sequence of consecutive ``yes"  is highly unlikely as $\ell$ grows. This will be the main point of Section \ref{sec: dissolution of a single patch}.

\subsection{Controlling the spectral width of a resonant patch} \label{sec: control on patch shrink}

For $n> n_B$, fix a patch $p\in \mathcal{P}_{n-1,k}$ with $k>1$ (see Section \ref{sec: intro dissolution statement}), and $\mu \in \{\pm 1 \}$. We assume that the event
$\mathcal{A}_{n-1,n}$ is true, so that we consider the spectral patch $S_{p,\mu}$ of $H_n$, see\ previous sections.  Let us make the dependence of this patch on the disorder variable $g \equiv g_n$ explicit: $S_{p,\mu}(g)$. Recall the convex hull of a set  $\mathrm{Conv}(\cdot)$. We define the spectral width of the patch $S_{p,\mu}$ as: 
\begin{equation} \label{def:s}
    s_n(g):= |\mathrm{Conv}(S_{p,\mu}(g))|. 
\end{equation}
Notice that since $H_n$ is almost surely non-degenerate (see Lemma \ref{lemma: minimal gap antigap}), $s_n(g)>0$,  $\mathbb{P}$- almost surely, since $k>1$. 
Then, we may define the following variable:
\begin{equation} \label{def:p}
  \tilde{\alpha}:=4\alpha^{1-\theta} \implies   v_n(g):= \frac{\ln(s_n(g))}{n \ln(\tilde{\alpha})}.
\end{equation}
 Notice that since we are investigating the patch $S_{p,\mu}$, the number $s_n(0)$ is the spectral width of the resonant patch  $p \in \caP_{n-1}$ at the scale $n-1$; then it is natural to also set: 
\begin{equation} \label{eq: def v_{n-1}}
v_{n-1}:= \frac{\ln(s_n(0))}{(n-1) 
\ln(\tilde{\alpha})}.
\end{equation}
As explained, $S_{p,\mu}(g)$ does not necessarily belong to $\mathcal{P}_{n,k}$. However, if it does, then we have for all $k$:
\begin{equation} \label{eq: condition on sn}
   s_n(g) \leq 2^{n-1} \alpha^{\theta n},  
\end{equation}
  which means for $\alpha$ sufficiently small and $n>\max\{n_B,4 \}$, we have: 
\begin{equation} \label{eq: patch bound}
v_n, v_{n-1} \geq C(\theta)=\frac{\theta}{2(1-\theta)}>0.
\end{equation}
We also take $\theta \in (2/3,1)$ so that $C(\theta) \geq 1$. This is only for later convenience. Note that to have the above bound, it is sufficient to take $ 1/16 > \alpha^{1-\theta}$. \\

The above bound means that if $v_n<C(\theta)$, then the patch has already been broken, i.e.\ $S_{p,\mu} \notin \caP_{n,k}$.  Therefore, to show that ``most patches" will be broken after certain scales, it is sufficient to upper bound moments  of the random variable  $v_n/v_{n-1}$ properly. This motivates the main proposition of this section:   

\begin{proposition}\label{prop: main moment bound}
Fix $b>4$, then there is $n_*:=n_*(\alpha,\theta,B,b)< \infty$ such that for all $n>n_*$, we have the following: 
 let $p \in \mathcal{P}_{n-1,k}$ with $k>1$ and let $\mu \in \{\pm 1 \}$. Recall $v_n,v_{n-1}$ as defined above on the event $\mathcal{A}_{n-1,n}$. 
     Let $0<\alpha <\alpha_*$ with $\alpha_*(\theta)$ be sufficiently small.  Then  
\begin{equation} \label{eq: main moment bound}
     \chi(\mathcal{A}_{n-1,n})\mathbb{E}((\frac{v_n}{v_{n-1}})^b | \caF_{n-1/2})  \leq \exp(-\frac{b/2-1}{n}), \qquad  \mathbb{P}-\text{almost surely}.
 \end{equation}
\end{proposition}
In the above proposition, $n_*$ depends on $\alpha, \theta,b$ and the bath Hamiltonian through $n_B$ and $\|H_B\|$. 
 Fixing any $\theta \in (2/3,1)$, the exact numerical value of $\alpha_*$ can be computed; however, this value is not sharp in any respect! Also note that the right-hand side is independent of $k$.   We devote the next two sections, namely \ref{sec: proof of dissol} and \ref{sec: proof of lemma cor: main p}, to the proof of the above proposition.
\subsection{Proof of Proposition \ref{prop: main moment bound}} \label{sec: proof of dissol}

\subsubsection{Tail estimate}
The main ingredient of the proof is the following lemma, controlling the tail of $v_n/v_{n-1}$ given a configuration at scale $n-1$, and given the event $\mathcal{A}_{n-1,n}$.  For any $y \geq 0$ we set: 
$$x=\frac{n-1}{n}\left(1+ \frac{y}{n}\right), \quad \text{also set:} \quad 
a:= - \ln(\tilde{\alpha}). $$
Then we have: 
\begin{lemma}  \label{cor: main p}
    For any $y \geq 0$, and for  $n$ sufficiently large ($n>  n(\alpha,\theta,B)$) we have: 
\begin{equation} \label{eq:mainp}
   \chi(\mathcal{A}_{n-1,n}) \mathbb{P}(v_{n} \geq x v_{n-1}|\mathcal{F}_{n-1/2}) \leq \min\{32 \cdot\exp\left(-a\left(\frac{y}{1+y/n}\right)\right)                           , C\exp(-cy/n) \}.
\end{equation}
\end{lemma}
The above bound is far from being optimal. In particular, the constants are not optimal at all. Proof of the above lemma requires more setup, and we postpone it to the next sections. First, we deduce Proposition \ref{prop: main moment bound} from Lemma \ref{cor: main p}.
\subsubsection{Deducing Proposition \ref{prop: main moment bound} from Lemma \ref{cor: main p} }
\begin{proof}[Proof of Proposition \ref{prop: main moment bound}]  For $n>n_*$ with $n_*$ given in Lemma \ref{cor: main p}, we abbreviate
$$
v(x)=\chi(\mathcal{A}_{n-1,n})\mathbb{P}((\frac{v_n}{v_{n-1}}) \geq x | \caF_{n-1/2} ).
$$
Then since by definition $v_n/v_{n-1}$ is non-negative, we have
\begin{align} \label{eq: moment decompostion}
     &\chi(\caA_{n-1,n})\mathbb{E}((\frac{v_n}{v_{n-1}})^b | \caF_{n-1/2}) \leq   \int_{0}^{\infty} v(x) b x^{b-1} dx  \nonumber 
     =\sum_{i=1}^4 \underbrace{\int_{a_{i-1}}^{a_i} v(x) x^{b-1} bdx}_{I_i}, \: \mathbb{P}-a.s.,
     \end{align}
  
      with  $a_0=0,a_1=(n-1)/n, a_2=2(n-1)/n, a_3=n^2,a_4=\infty$. Let us bound each term in the above sum separately:

    \begin{enumerate}
        \item $I_1 \leq ((n-1)/n))^b$ almost surely, by trivially bounding $v(x) \leq 1$.  
        \item Bounding $v(x)$ by the second expression appeared in \eqref{eq:mainp}, we get for $n$ large enough: 
        \begin{equation}
            I_4 \leq C \int_{n^2}^{\infty} b x^{b-1} e^{c}\exp(-cnx/(n-1)) \leq \exp(-cn),
        \end{equation}
        where we used the fact that $bx^{b-1}<e^{cx/2}$ for all $x\geq n^2 $ for $n$ sufficiently large; we then   integrated the resulting expression and used the fact that $n$ is large enough.
        \item To treat $I_3$, for $n$ sufficiently large, we bound $v(x)$ by the first expression appearing in \eqref{eq:mainp}: 
        \begin{equation} \label{eq: I3 pbound}
            I_3 \leq 32\int_{\frac{2(n-1)}{n}}^{n^2}
            \underbrace{\exp(-a(nx-(n-1))/x)}_{=:g(x)} bx^{b-1} \leq 32e^{-an/2} bn^{2b} \leq \exp(-an/4),
        \end{equation}
         where since $g(x)$ (abbreviated in the above expression) is decreasing  for $x>0$, we bounded it  by $\exp(-an/2)$ for $x \geq 2(n-1)/n$, the last bound holds for $n$ sufficiently large. 
        \item For $I_2$, again we bound $v(x)$ by the first expression of \eqref{eq:mainp}. Moreover, for $(n-1)/n\leq x\leq 2(n-1)/n$ we bound $g(x)$ (appeared in \eqref{eq: I3 pbound}) as $g(x)\leq \exp(-a(nx-(n-1))/2)$.  We get: 
        \begin{equation} \label{eq: I2pbound}
        I_2 \leq 32\cdot be^{a(n-1)/2} \int_{\frac{n-1}{n}}^{\frac{2(n-1)}{n}} \exp(-anx/2) x^{b-1} dx \leq 32\cdot be^{a(n-1)/2}\left(2/an\right)^b \Gamma(b,a(n-1)/2),
        \end{equation}
    where $\Gamma(b,m)=\int_{m}^{\infty} e^{-z}z^{b-1} dz$  is the incomplete Gamma function, obtained by changing  variables and changing the 
 integration bound. Now, we use  a well-known bound concerning $\Gamma$ (see \cite{Gamma2000inequalities} Section 3.2), namely $\Gamma(b,m) < 2 \exp(-m)m^{b-1}$ valid for $m>2(b-1)$. Hence, for $n$ sufficiently large, we have: 
\begin{equation}
    I_2 \leq \frac{128b}{an} \left( \frac{n-1}{n}\right)^{b-1}.
\end{equation}
  \end{enumerate}
  Combining bounds in the above four steps, assuming $n$ is large enough (to get $I_4+I_3<I_2/2)$, and taking $a>384$, the desired result \eqref{eq: main moment bound} follows thanks to the inequality: $1+x\leq e^x$ for all $x \in \mathbb{R}$.  
\end{proof}

\subsection{Proof of Lemma \ref{cor: main p}} \label{sec: proof of lemma cor: main p}
Before proceeding, we recall two rather classical inequalities for the reader's convenience. These inequalities are at the heart of the proof. 

\subsubsection{Preliminaries: Cartan and Remez inequalities} \label{sec: Remez and cartan}

The first is the Remez inequality, where we refer to \cite{lubinsky1997small}, Corollary 5.2: 
 
 \begin{lemma}[Remez] \label{lemma:Remez}
    Let $P$ be a polynomial of degree at most $n$, with real coefficients, normalized by 
    $\|P \|_{L^{\infty}[-1,1]} =1$. Let 
    $\varepsilon \in (0,1]$. Then 
    \begin{equation} \label{eq:Remez}
        | \{x \in [-1,1] : |P(x)| \leq \varepsilon \}  | < 8 \varepsilon^{\frac{1}{n}},
    \end{equation}
    where $| \cdot|$ denotes the one-dimensional Lebesgue measure. 
    \end{lemma}
   The above bound is not sharp (the constant $8$). But it is sufficient for our purposes. Notice that although the mentioned estimate is ``meaningful" for $\varepsilon \in (0,1]$, strictly speaking, it is true for all $\varepsilon>0$. \\
   
   The second inequality is the Cartan bound, for which we refer to \cite{bourgain2009Cartan} Lemma 1 (see also \cite{kruger2012localization} Section 10):
\begin{lemma} [Cartan] \label{lemma:Cartan1d}
    Let $F$ be a real analytic function on $[-\frac12, \frac 12] $ which extends to an analytic function on $D= \{z \in \mathbb{C}; |z|<1 \}$, with $|F| \leq 1$ on $D$. If 
    $|F(0)| > \epsilon$, where $0<\epsilon< \frac 12$, then for $\delta>0$ we have: 
    \begin{equation} \label{eq:Cartan}
        \left|\left\{ x \in [-1/2,1/2] \: \Big| \: 
        |F(x)| < \delta\right\} \right| \leq C \delta^{\frac{c}{\ln(1/\epsilon)}},
    \end{equation}
    where, as before, $|\cdot|$ denotes the one-dimensional Lebesgue measure. In particular, for $0<\lambda<1$ , taking $\delta=\lambda \epsilon$ a straightforward computation gives: 
    \begin{equation} \label{eq:Cartan2}
         \left|\left\{ x \in [-1/2,1/2] \: \Big| \: 
        |F(x)| < \lambda \epsilon \right\} \right| \leq C \lambda^{\frac{{c}}{\ln(1/\epsilon)}}.
    \end{equation}
    
\end{lemma}

\subsubsection{Perturbative Expansion} \label{sec: effective Ham derivation} 

As we discussed before, given the event $\mathcal{A}_{n-1,n}$, $S_{p,\mu}(g)$ can be treated by spectral perturbation theory: by expanding around $g\equiv g_n=0$ for $n$ large enough. The associated spectral projection is denoted by $P_g$. By classical results in perturbation theory (see \cite{kato1995perturbation} Chapter 2,  p.67, 99; also see Appendix \ref{sec:PTtheory}), there exists an invertible operator-valued function $U_g$--as a function of $g$-- such that $U_gP_0U_g^{-1} =P_g$.  Define
\begin{equation} \label{eq: effective patch}
   H_P(g) := U_g^{-1} P_g H_gP_g  U_g,
 \end{equation}
where we denoted $H_n= H_{n-1/2}+g X_1X_n$
 by $H_g$. Notice that by $U_gP_0U_g^{-1} =P_g$, we have $H_P(g)= P_0U_g^{-1}H_g U_g P_0$. 
More importantly, notice that the spectrum of $H_P(g)$ is identical to  $P_gH_gP_g$. Recall the spectrum of $P_gH_gP_g$ on $\mathrm{Ran} P$ is the patch $S_{p,\mu}(g)$.\\
Appealing to results from perturbation theory (see \cite{kato1995perturbation}, also see Section \ref{sec:PTtheory}, \eqref{eq:seriesexpansion3}, \eqref{eq:boundPT}), we express the main expansion of this section.  
 Let 
 \begin{equation} \label{eq: alpha tilde def}
     \tilde{\alpha}:= 4 \alpha^{1-\theta}.
 \end{equation} 
  Then, we can write
\begin{equation}\label{eq:expansion}
H_P(g)=H_{P}(0)+ B(g), \qquad 
B(g)=\sum_{k=1}^{\infty}g^{k}\tilde{\alpha}^{nk}B_k,
\end{equation}
where we have the following  bound for $n$ sufficiently large:   
\begin{equation} \label{eq:boundexpansion}
\|B_k\| \leq 1.
\end{equation}

The above considerations are rather general (i.e., they are valid for any other self-adjoint perturbation $V$ with $\|V \|=1$). Since the above series is convergent for any complex $g$ in a disk of radius $\tilde{\alpha}^{-n}$, $H_P(g)$ is analytic in this disk. Moreover, since $H_{n-1/2}$ and the interaction term are self-adjoint, we have that for any \textit{real} $g$ in the above disk, $U_g$ is unitary and $U_g^{-1}=U_g^{\dagger}$. In particular, this means that $B_k$ is self-adjoint (see Appendix \ref{sec:PTtheory}).


\subsubsection{Main part of the proof} 
 
Throughout this section, we always assume $n>n_B$; also, $n_*>n_B$. Lemma \ref{cor: main p} is stated in terms of $v_n, v_{n-1}$. However, there is a natural correspondence between $v_n$ and $s_n$. For the purpose of the proof, it is more convenient to rewrite \eqref{eq:mainp} in terms of $s_n$. To this end, let us introduce some abbreviations and notations. These symbols will only be used in this section. 
\\

\textbf{Abbreviations.} We abbreviate: 
$$v_n(0)= :v_0, \quad s_n(0)=:s_0.$$
For any $\lambda \in (0,1)$ we define $\nu>0$ as: 
$$\lambda=: (\tilde{\alpha}^n)^{\nu}.$$
\textbf{Recast of Lemma \ref{cor: main p}.} Then,  \eqref{eq:mainp} can be recast as    
\begin{equation} \label{eq: main bound s}
    \chi(\mathcal{A}_{n-1,n})\mathbb{P}(s_n(g) \leq \lambda s_n(0) |\mathcal{F}_{n-1/2}) \leq 
    \min\{32 \lambda^{1/(v_0+\nu)}, C \lambda^{c/(nv_0 a)} \}, \quad \mathbb{P}-a.s,
\end{equation}
where we recall $a=-\ln(\tilde{\alpha})$. Notice that the above expression is identical to \eqref{eq:mainp} by taking $\lambda=s_0^{y/n}$ i.e. $y=\nu n /v_0$. 
To prove Lemma \ref{cor: main p}, it is sufficient to prove the above bound, as we do in the following: 
\begin{proof}[Proof of Lemma \ref{cor: main p}] \textbf{Step 1.}
Notice that thanks to the expression $\chi(\mathcal{A}_{n-1,n})$ we assume throughout the proof that the event $\mathcal{A}_{n-1,n}$ holds. For $n$ large enough, recall the patch $S_{p,\mu}(g)$, and its effective Hamiltonian $H_P(g)$ on $\mathrm{Ran} P$.  Also recall the expansion of $H_P(g)$ \eqref{eq:expansion}. The above assumption means this expansion and  $v_n(g), s_n(g)$ are well-defined for all values of the  random variable $g$. \\

Denote the smallest and largest eigenvalues of $H_{P}(g)$ by $E_1(g)$, $E_k(g)$. The corresponding eigenvectors are denoted by ${\psi_1(g)},{\psi_k(g)}$. Define: 
\begin{equation} \label{eq:diag}
    \tilde{\varphi}(g):= \langle{\psi_k(0)},H_P(g) {\psi_k(0)} \rangle - \langle{\psi_1(0)},H_P(g) {\psi_1(0)}\rangle. 
\end{equation}
First, we observe that: 
\begin{equation} \label{eq: s to phi bound}
   \chi(\mathcal{A}_{n-1,n}) \mathbb{P}(s_n(g)\leq \lambda s_n(0)| \mathcal{F}_{n-1/2}   ) \leq  
      \chi(\mathcal{A}_{n-1,n})\mathbb{P}(|\tilde{\varphi}(g)| \leq \lambda \tilde{\varphi}(0)| \mathcal{F}_{n-1/2}).  
\end{equation}
The above bound is deduced by the following facts: $H_P(g)$ has the same spectrum as $P_gH_gP_g$; moreover, notice that $\tilde{\varphi}(0)=s_n(0)$ by definition. Finally, we use the bound $|\tilde{\varphi}(g)| \leq s_n(g)$, which is a direct consequence of the {Schur-Horn inequality}.\\

\textbf{Step 2.} In this step, we show for any $\lambda \in (0,1)$, and $n$ large enough: 
\begin{align} \label{eq:Cartanestimate}
    \mathfrak{p}(\lambda,n):=\chi(\mathcal{A}_{n-1,n})\mathbb{P}(|\tilde{\varphi}(g)| \leq \lambda \tilde{\varphi}(0)| \mathcal{F}_{n-1/2})  \leq C \lambda^{{c}/(nv_n(0)a)}, 
 \mathbb{P}-a.s.
\end{align}

For any $r>0$, we denote $D_{r}:=\{z \in \mathbb{C}| |z|<r \}.$ We also identify $D_1=:D$. Thanks to the expansion \eqref{eq:expansion},  and the bound 
\eqref{eq:boundexpansion} we  deduce that $\tilde{\varphi}$ is analytic in the disk 
$D_{\tilde{\alpha}^{-n}}$.  Moreover, $\tilde{\varphi}$ is analytic on $D$ and  for all $z\in D$
$$|\tilde{\varphi}(z) - s_n(0)|  \leq  3\tilde{\alpha}^n.$$\
On the other hand, recalling the choice $\theta \in (2/3,1)$, thanks to \eqref{eq: patch bound} we observe that $s_n(0) \leq \tilde{\alpha}^n$ for $n$ large enough.   This means for all $z \in D$: 
$$\tilde{\varphi}(z) \leq 4 \tilde{\alpha}^n.$$
In light of the above bound, we define: 
$$\phi(z):= \frac{\tilde{\varphi}(z)}{4 \tilde{\alpha}^n}.$$
Thanks to the construction, $\phi$ is analytic on $D$ and $|\phi(z) |\leq 1$ for all $z\in D$. Moreover, we have: 
$$\phi(0)= \frac{s_n(0)}{4 \tilde{\alpha}^n}=: \epsilon. $$
Given the above properties, we may apply Cartan Lemma \ref{lemma:Cartan1d} for $\phi$, with $\epsilon$ given as above and deduce: 
\begin{equation} \label{eq: Cartan estimate patch}
   \mathfrak{p}(\lambda,n)= \left| \left\{ g \in [-1/2,1/2] \: \big| \: |\phi(g)| \leq \lambda \epsilon \right\} \right| \leq  C \lambda^{c/nv_n(0)a} , 
\end{equation}
for $n>n(\alpha,\theta,B)$ sufficiently large. Notice that in the Cartan Lemma, it is required that $F(0)>\epsilon$; whereas here $F(0)=\epsilon$. This yields the same bound, by changing $\epsilon$ to $\epsilon/2$ and modifying the constant. Here we also used the fact that for $n$ large enough $\ln(4)<na$. This concludes the proof of  this step \eqref{eq:Cartanestimate}.

\textbf{Step 3.} In this step, we observe that for any $\lambda \in (0,1)$, 
 defining $\nu>0$ such that $\lambda=:(\tilde{\alpha}^n)^{\nu}$, we have: 
\begin{equation} \label{eq:bound phi poly}
\mathfrak{p}(n,\lambda)\leq 32\cdot\lambda^{1/(v_0+\nu)}, \quad \mathbb{P}-a.s.. 
\end{equation}
Let $\lfloor v_0 +\nu \rfloor$ be the largest integer not larger  than $v_0+\nu$. Thanks to \eqref{eq: patch bound}, we have $\lfloor v_0+\nu \rfloor \geq 1$. Recalling the expansion \eqref{eq:expansion}, we define for any $g \in [-1,1]$: 
\begin{align}
   \tilde{\varphi}_<(g):= s_n(0)+ \sum_{j=1}^{\lfloor v_0+\nu \rfloor } 
    g^j \tilde{\alpha}^{nj}  \big( \langle{\psi_k(0)},B_j {\psi_k(0)}\rangle - \langle{\psi_1(0)},B_j{\psi_1(0)}\rangle \big). \label{eq: phi poly dep 1}
\end{align}
Notice that thanks to the above definition and in light of \eqref{eq:expansion}, \eqref{eq:boundexpansion} for $n$ large enough ($n>n(\|H_B \|)$), we have for any $g \in [-1,1]$, 
$$ |\tilde{\varphi}(g)-\tilde{\varphi}_<(g)| \leq 3 \lambda s_0. \implies \mathfrak{p}(n,\lambda) \leq \mathbb{P}(|\tilde{\varphi}_{<}(g)| \leq 4 \lambda s_0 |\mathcal{F}_{n-1/2}).$$ Having the above estimate, the result of this step \eqref{eq:bound phi poly}, follows by applying the Remez inequality  \eqref{eq:Remez} for the polynomial 
$$\varphi(g):= \frac{ \tilde{\varphi}_<(g)} { \|\tilde{\varphi}_<(g)\|_{L^{\infty}([-1,1])}}.$$
Notice that in the above argument, we used the following facts:  $0<s_0\leq \sup_{g}|\tilde{\varphi}_<(g)|$; $|\varphi|$ is bounded by one and has degree at most $\lfloor v_0+\nu \rfloor \geq 1$, with real coefficients (since $B_j$ in the expansion are self-adjoint). We also bounded $4^{1/[v_0+\nu]}$ by $4$.\\
 Estimate  \eqref{eq: main bound s} and consequently Lemma \ref{cor: main p} follows by combining \eqref{eq:bound phi poly} and \eqref{eq:Cartanestimate}.

\end{proof}

\subsection{Dissolution of a single patch} \label{sec: dissolution of a single patch}

For $n\geq n_B$, we abbreviate 
$$
\mathbb{P}_n=   \mathbb{P}\left(\cdot | \caA_{n,\infty}\right).
$$
We recall the notation and conventions of Section \ref{sec: 
spectral patches and resonant patches}. 
The following proposition is crucial.
\begin{proposition}
\label{prop: dissolution single patch} There exists a finite $n(\alpha,\theta,B)$ such that for any $n_1> n(\alpha,\theta,B)$, and 
for any $p\in \caP_{n_1}$ such that the patch size $|p|>1$,  
$$ 
\sum_{n=n_1+1}^\infty\sup_{\underline{\mu} \in \Sigma_{n-n_1}}{\mathbb{P}_{n_1}(S_{p,\underline{\mu}} \in \caP_n | \caF_{n_1})}  <\infty, \quad   {\mathbb{P}-a.s.} .
$$

\end{proposition}

The rest of this section is devoted to the proof of Proposition \ref{prop: dissolution single patch}. We implicitly assume $|p|>1$ in the rest of the section.   
Let, for $n_1<i\leq n$, 
\begin{equation} \label{eq: dis sol X def}
\hat{X}_i := \frac{v_i}{v_{i-1}} \chi(S_{p,\underline{\mu}_{i-1}} \in \caP_{i-1} )\chi(\caA_{n_1,i}),
\end{equation}  
where $\underline{\mu}_{i-1}$ is the restriction of $\underline{\mu} \in \{\pm 1\}^{n-n_1}$ to its first $i-1-n_1$ components, as defined in Section \ref{sec: res patch def}.  
On the event 
$$\{ S_{p,\underline{\mu}_{i-1}}  \in \caP_{i-1}  \} \cap \caA_{n_1,i},$$
the variable $\frac{v_i}{v_{i-1}}$ is indeed well-defined, hence $\hat X_i$ is well-defined.  Also note that  $0<v_{i-1},v_i<\infty$ a.s. in this event since $s_i>0$ a.s.. 
\begin{lemma}\label{eq: resonant implies bound on v}
For $n_1 >C(\alpha,\theta,B)$,
 there is a random variable $f_{n_1}$, measurable with respect to $\caF_{ n_1}$, such that $f_{n_1} >0$, $\mathbb{P}_{n_1}$-almost surely, and such that for all $\underline{\mu}\in \{\pm 1\}^{n-n_1}$:
$$
  \{ \hat{X}_{n_1+1}\ldots \hat{X}_{n} \geq f_{n_1} \} \supset  \{S_{p,\underline{\mu}} \in \caP_{n} \}   \cap \caA_{n_1,n}.
$$
\end{lemma}

\begin{proof}
For $n_1$ large enough, by definition, we have 
$$
\hat{X}_{n_1+1}\ldots \hat{X}_{n} = \frac{v_n}{v_{n_1}}\chi(\{S_{p,\underline{\mu}_{n-1}} \in \caP_{n-1}  \} ) .\chi(\caA_{n_1,n}),
$$
notice the following  subtlety in the above identity: $v_{n-1}$ in the denominator, defined in \eqref{eq: def v_{n-1}} coincides with $v_{n-1}$ in the numerator thanks to the indicator function above. 
 Let $k>1$ be the size of the patch $p$, i.e.\ $p \in \caP_{n_1,k}$.  
If $S_{p,\underline{\mu}}\in \caP_{n}$ and $\caA_{n_1,n}$ holds,  then necessarily $S_{p,\underline{\mu}} \in \caP_{n,k}$. This means 
that  $v_n \geq C(\theta)>0$ thanks to \eqref{eq: patch bound}. We know by definition that $v_{n_1}>0$, and, on the other hand $v_{n_1}<\infty$ almost surely because of Lemma \ref{lemma: minimal gap antigap} (since $k>1$).  Therefore, taking $f_{n_1}:= C(\theta)/v_{n_1} \leq v_n/v_{n_1} $ finishes the proof.


\end{proof}
Fix $b=\mathsf{b}>4$.
Now we consider:
\begin{align}
 \mathbb{P}_{n_1}(    S_{p,\underline{\mu}} \in \caP_{n}   | \caF_{n_1}) & \leq \frac{1}{\mathbb{P}(\mathcal{A}_{n_1,\infty}|\caF_{n_1})} \mathbb{P}( 
 \{S_{p,\underline{\mu}} \in \caP_{n} \} \cap \caA_{n_1,n} | \caF_{n_1}) \nonumber \\
 &\leq \frac{1}{\mathbb{P}(\mathcal{A}_{n_1,\infty}|\caF_{n_1})} \mathbb{P}( \hat{X}_{n_1+1}\ldots \hat{X}_{n}  \geq f_{n_1} | \caF_{n_1})   \nonumber \\ &\leq \frac{1}{\mathbb{P}(\mathcal{A}_{n_1,\infty}|\caF_{n_1})} 
 \frac{1}{f_{n_1}^{\mathsf{b}}}   \mathbb{E}(  \hat{X}^\mathsf{b}_{n_1+1}\ldots \hat{X}^{\mathsf{b}}_{n}  | \caF_{n_1})  \: \: \mathbb{P} -a.s.. \label{eq: bound by x prod}
\end{align}
The first bound is a consequence of the definition of conditional expectation and an inclusion bound ($\mathcal{A}_{n_1,\infty} \subset \mathcal{A}_{n_1,n}$); the second bound is deduced by Lemma \ref{eq: resonant implies bound on v}, and the last bound is a Markov inequality.  \\
We will bound
\begin{align}
  \mathbb{E}(  \hat{X}^{\mathsf{b}}_{n_1+1}\ldots \hat{X}^{\mathsf{b}}_{n}  | \caF_{n_1}) & =     \mathbb{E} \left(  \hat{X}^{\mathsf{b}}_{n_1+1}\ldots \hat{X}^{\mathsf{b}}_{n-1}    \mathbb{E}( \hat{X}^{\mathsf{b}}_n |\mathcal F_{n-1})    \middle| \caF_{n_1} \right) \nonumber\\
  & \leq     \mathbb{E} \left(  \hat{X}^{\mathsf{b}}_{n_1+1}\ldots \hat{X}^{\mathsf{b}}_{n-1})  )   \middle| \caF_{n_1} \right) 
  {||}\mathbb{E}( \hat{X}^{\mathsf{b}}_n |\mathcal F_{n-1}) {||}_{L^\infty(\mathbb{P})}. 
\end{align}

The first equality follows from the
 tower property $\mathcal{F}_{n_1} \subset \mathcal{F}_{n-1}$ and  because $\hat{X}_{n'}$ is $\caF_n$-measurable for any $n'\leq n$. The second inequality is the $L^1-L^{\infty}$ bound.

Iterating this bound, we get 

\begin{equation}
  \mathbb{E}(  \hat{X}^{\mathsf{b}}_{n_1+1}\ldots \hat{X}^{\mathsf{b}}_{n}  | \caF_{n_1})
     \leq    \prod_{i=n_1+1}^n
  {||}\mathbb{E}( \hat{X}^{\mathsf{b}}_i |\mathcal F_{i-1}) {||}_{L^\infty(\mathbb{P})} \:. \label{eq: iteration ultimate} 
\end{equation}

To continue, we establish
\begin{lemma}\label{lem: bound on single x} For $i$ large enough ($i>i(\alpha,\theta,B)$):
 $$   \mathbb{E}( \hat{X}^{\mathsf{b}}_i |\mathcal F_{i-1})   \leq e^{-\frac{{\mathsf{b}}/2-1}{i}}, \qquad  \text{ $\mathbb{P}$-a.s.} \: .  $$
\end{lemma}
\begin{proof}
Thanks  to the tower property, since $\mathcal{A}_{i-1,i} \in \mathcal{F}_{i-1/2}$, we write
$
        \mathbb{E}( \hat{X}^{\mathsf{b}}_i |\mathcal F_{i-1})   \leq  \mathbb{E}\left(
           Y^{\mathsf{b}}_i  \middle|\mathcal F_{i-1}  \right) $
           where 
           $$Y^{\mathsf{b}}_i=  \chi(\caA_{i-1,i})   \chi(S_{p,\underline{\mu}_{i-1}} \in \caP_{i-1})   \mathbb{E}(  ( \frac{v_i}{v_{i-1}})^{\mathsf{b}} |\mathcal F_{i-1/2} ),$$
           where, after applying the tower property, the inequality is deduced by  the fact that $\mathcal{A}_{n_1,i} \subset \mathcal{A}_{i-1,i}$.
Proposition \ref{prop: main moment bound} tells us that, almost surely, for $i$ large enough
$$
 Y^{\mathsf{b}}_i  \leq    e^{-\frac{{\mathsf{b}}/2-1}{i}}.   
$$
Notice that thanks to the second indicator function in the definition of $Y_i^{\mathsf{b}}$, we are in the realm of Proposition \ref{prop: main moment bound}, i.e., there exists a patch at scale $i-1$.
Therefore, the claim follows.


\end{proof}
By Lemma \ref{lem: bound on single x}, we  bound \eqref{eq: iteration ultimate}  for $n_1$ large enough  ($n_1>n(\alpha,\theta,B)$, with deterministic $n(\alpha,\theta,B)$: notice that $n(\alpha,\theta)$ also depends on $\mathsf{b}$, we drop this since $\mathsf{b}>4$ is fixed) as 
\begin{equation}\label{eq: products of xs}
\mathbb{E}(  \hat{X}^{\mathsf{b}}_{n_1+1}\ldots \hat{X}^{\mathsf{b}}_{n}  | \caF_{n_1}) \leq   \prod_{i=n_1+1}^n   e^{-\frac{{\mathsf{b}}/2-1}{i} }  . 
\end{equation}
From \cite{GUO2011991} equation (8), we have the following bound on harmonic sums,
\begin{equation}
    \ln(n) + \gamma + \frac{1}{2n+1} \leq \sum_{i=1}^n \frac{1}{i} \leq \ln(n) +\gamma +\frac{1}{2n-1}, 
\end{equation}
with a universal constant $\gamma$. Therefore, we get 
\begin{equation} \label{eq: LDestimate}
\mathbb{E}(  \hat{X}^b_{n_1+1}\ldots \hat{X}^{\mathsf{b}}_{n}  | \caF_{n_1}) \leq C (n/n_1)^{-({\mathsf{b}}/2-1)}.
\end{equation}
Plugging these bounds into \eqref{eq: bound by x prod}, we have obtained the almost sure bound for $n_1> n(\alpha,\theta,B)$
\begin{align}
\mathbb{P}_{n_1}(    S_{p,\underline{\mu}} \in \caP_{n}   | \caF_{n_1}) & \leq  \frac{1}{f_{n_1}^{\mathsf{b}}}  \frac{1}{\mathbb{P}(\mathcal{A}_{n_1,\infty}|\caF_{n_1})} C (n/n_1)^{-({\mathsf{b}}/2-1)}.
\end{align}
By Lemma \ref{lem: prob event caa}, we can bound the denominator from below by $c>0$ (uniformly in $n,n_1$). Since  $f_{n_1} >0$ almost surely, we  conclude the proof of Proposition \ref{prop: dissolution single patch} by recalling that we fixed  $\mathsf{b}>4$.

\subsection{Dissolution of most patches: Proof of Theorem \ref{thm: dissolution of patches}} \label{sec: dissolution of most}
In this section, we will finally prove Theorem \ref{thm: dissolution of patches}.
To save writing, let us introduce the fraction $r_n(k)$ of levels that are in a resonant patch of size  $k$:
$$
r_n(k)= 2^{-n} k|\caP_{n,k}|.
$$
Our goal is to prove $r_n(k) \to 0$, for any $k>1$, and also $\sum_{k>1} r_n(k)\to 0$,  almost surely.

We first introduce slightly modified ratios at scale $n$ that depend on a fiducial scale $n_1<n$ and that are well-defined on the event $\caA_{n_1,n}$.  Namely
$$
r_{n_1,n}(k)= k2^{-n}\sum_{\underline{\mu} \in \Sigma_{n-n_1} } \sum_{p \in \caP_{n_1,k}} \chi( S_{p,\underline{\mu}} \in \caP_{n}).
$$
Note that $r_{n_1,n}(k) \leq r_{n}(k)$, the ratio $r_{n_1,n}(k) $ considers only the resonant patches of size $k$ that descend from resonant patches of size $k$ at scale $n_1$.

\begin{lemma}\label{prop: weird ratios vanish}
 For any $ n_0 >C(\alpha,\theta,B)$, $k>1$, and any $n_1>n_0$ we have
$\lim_{n\to\infty} r_{n_1,n}(k)=0$, $\mathbb{P}_{n_0}$-almost surely.
\end{lemma}
\begin{proof}
First, we prove that $r_{n_1,n}(k) \to 0$, $\mathbb{P}_{n_1}$- almost surely . To this end, we prove that
\begin{equation}\label{eq: summable expectation}
  \sum_{n > n_1} \mathbb{E}_{n_1}(r_{n_1,n}(k) | \caF_{n_1}) <\infty,  \qquad  \mathbb{P}_{n_1}-\text{a.s.}. 
\end{equation}
By Markov's inequality and the Borel-Cantelli  Lemma, this implies that $\lim_{n\to\infty} r_{n_1,n}(k)=0$, $\mathbb{P}_{n_1}$- almost surely. 
The application of  the Borel-Cantelli Lemma is straightforward in the above conditional setup, since the above bound holds $\mathbb{P}_{n_1}-$a.s.
We now start with the proof of \eqref{eq: summable expectation}.
By definition of the fraction $r_{n_1,n}(k)$, we have
        \begin{equation} \label{eq: patchdissolutionsum}
            \mathbb{E}_{n_1}(r_{n_1,n}({k})  | \caF_{n_1})=\frac{k}{2^{n}} \sum_{p \in \mathcal{P}_{n_1,k}} \sum_{\underline{\mu}} {\mathbb{P}_{n_1}(S_{p,\underline{\mu}} \in \caP_{n}
   | \caF_{n_1})}.
        \end{equation}
        Note that  $\mathcal {P}_{n_1,k}$ is measurable w.r.t.\ $\caF_{n_1}$, and the event $\caA_{n_1,n}$ is satisfied $\mathbb{P}_{n_1}$-a.s., 
        so the above formula makes sense.   By Proposition \ref{prop: dissolution single patch}, for $n_1>C$ we deduce  almost surely
        $$ \sum_{n=n_1+1}^\infty 2^{n_1-n}\sum_{\underline{\mu}} \mathbb{P}_{n_1}(S_{p,\underline{\mu}} \in \caP_{n}|\caF_{n_1} ) <\infty,$$
        
        then bounding $k\sum_{p \in \mathcal{P}_{n_1},k}1$, by $2^{n_1}$  yields \eqref{eq: summable expectation}.\\ 
        
        So far we proved that $r_{n_1,n} \to 0$, $\mathbb{P}_{n_1}$ a.s.. To prove the same limit  $\mathbb{P}_{n_0}$-almost surely, we argue as follows:
   Since $\caA_{n_0,\infty} \subset \caA_{n_1,\infty}$ (because $n_0<n_1$) and $\mathbb{P}(\caA_{n_1,\infty}) >0, \mathbb{P}(\caA_{n_0,\infty}) >0 $ (thanks to Lemma \ref{lem: prob event caa}, for any $n_0$ large enough and $\alpha$ small enough) any event that happens $\mathbb{P}_{n_1}$-almost surely, also happens $\mathbb{P}_{n_0}$-almost surely. This finishes the proof of this lemma.

\end{proof}

As a consequence of the above lemma, we derive:


  

  
     

\begin{lemma}\label{lem: vanishing fraction conditional}
For any $n_0> C(\alpha,\theta,B) $, 
$$\lim_{n\to\infty} \sum_{k>1} r_n(k)=0, \qquad  \mathbb{P}_{n_0}-\text{almost surely}. $$
\end{lemma}
\begin{proof}
At any $n$, there is a maximal patch size $K(n)=\max \{ k | \caP_{n,k} \neq \emptyset \}$. 
   Note first that for $n>n_0$, $K(n)$ is non-increasing $\mathbb{P}_{n_0}$-almost surely, because of the conditioning on $\caA_{n_0,\infty}$  thanks to Lemma \ref{lem: projectors PT theory}. 
   Similarly the quantity 
   $$
  r_n^{\geq}(\ell)= \sum_{i=\ell}^{K(n_0)} r_n(i)
   $$
   is non-increasing in $n$,
    for $n \geq n_0$, for every $1\leq \ell \leq K(n_0)$.
   Therefore, we deduce that $r_n^{\geq}(\ell)$ converges to a (random) limit that we call $r^{\geq}(\ell)$.  
   Let $\ell_*=\max\{\ell \geq 1 | g(\ell)>0\}$. Since $r_n^{\geq}(1)=1$, we have that $1\leq \ell_* \leq K(n_0)$. 
   Let us now force a contradiction by assuming that $\ell_*>1$ with positive $\mathbb{P}_{n_0}$ probability. 
 We now pick a time $n'$  large enough (larger than $C(\alpha,\theta,B)$ appeared in Lemma \ref{prop: weird ratios vanish}) such that 
   $$
   r_n^{\geq}(\ell_*) > r^{\geq}(\ell_*)/2,\qquad \forall n\geq n',
   $$
   and
   $$
 r_n^{\geq}(\ell_*+1)  \leq  r^{\geq}(\ell_*)/4,\qquad \forall n\geq n',
   $$
  with the convention $r_n^{\geq}(K(n_0)+1)=0$. 
It follows that $r_{n',n}(\ell_*)$ does not converge to zero, with positive $\mathbb{P}_{n_0}$ probability and with $n'\geq n_0$ and $\ell_*$ random. In fact the liminf of $r_{n',n}$ is lower bounded by $r^{\geq}(\ell_*)/4$: with positive $\mathbb{P}_{n_0}$ probability. This is because for $n>n'>n_0$ given the event $\mathcal{A}_{n_{0},\infty}$, $r_n^{\geq}(\ell_*)-r_n^{\geq}(\ell_*+1)$ represents either levels emanating from a patch of size $\ell_*$ at scale $n'$ which remained unbroken,  or levels emanating from patches of size $\ell>\ell_*$ at scale $n'$ which have been broken before $n$. This is a contradiction with Lemma \ref{prop: weird ratios vanish}. Notice that by a countable intersection, Lemma \ref{prop: weird ratios vanish} holds for all deterministic $n',k$. Hence we can use this Lemma for our random choice of $\ell_*,n'$.

  Therefore, we conclude that, $\mathbb{P}_{n_0}$-almost surely,  $r^{\geq}(\ell)=0$ for all $\ell>1$, which proves the claim.
\end{proof}

Finally, we need this conclusion to hold $\mathbb{P}$-almost surely, instead of only $\mathbb{P}_{n_0}$-almost surely. By Bayes' formula, we get for any event $E$,
$$ \mathbb{P}(E) \geq  \mathbb{P}(\caA_{n_0,\infty})  \mathbb{P}_{n_0}(E).   $$
By Lemma \ref{lem: prob event caa}, $\mathbb{P}(\caA_{n_0,\infty})\to 1$ as $n_0\to \infty$. Taking now $E$
 to be the event that $\lim_{n\to\infty} \sum_{k>1} r_n(k)=0$, the desired conclusion follows from the above lemma and taking the $\limsup$ of the above expression. This proves Theorem \ref{thm: dissolution of patches}.



\section{Anti-resonances} \label{sec: antiresonances}

This section is very analogous to Section \ref{sec: dissolution}. Just as in that section, we proved that the relative number of resonances decreases with scale; we will now prove the analogous statement for anti-resonances. The analogy is not perfect; whereas the discussion of resonances proceeded via a partition $\caP_n$ of $\Sigma_n$, we discuss anti-resonances by considering the set of anti-resonant 2-tuples:
\begin{equation} \label{eq: def Z_n}
\caZ_{n} := \{  ( \sigma,\sigma') \in \Sigma^2_n  \, \big|  \, \left| {E}_{\sigma} + {E}_{\sigma'}  \right| <  \alpha^{\theta n}/4 \},    
\end{equation}
for $n>n_B$. Let us note that 
\begin{enumerate}
    \item We allow for the case where $\sigma=\sigma'$.
    \item If $( \sigma,\sigma'), ( \sigma,\sigma'')  \in \caZ_n$, then $E_{\sigma'}, E_{\sigma''}$ are resonant. In particular, for any $\sigma\in \Sigma_n$, there is at most one resonant patch in $\caP_n$ that contains levels that are anti-resonant with $\sigma$.
\end{enumerate}
Let us now state the main result of this section.
\begin{theorem} \label{thm: antires vanish}
Let Assumption \ref{ass: nondegeneracy} hold, and assume $\alpha>0$ is sufficiently small. 
    Let $\widehat{\caZ}_n \subset \Sigma_n$ be the set of $\sigma$ such that, for some $\sigma'$, $(\sigma,\sigma') \in \caZ_n$. 
    Then, almost surely,
    $$
    \frac{|\widehat{\caZ}_n |}{2^n}  \rightarrow 0.
    $$
\end{theorem}

\subsection{Preliminaries}\label{sec: prelim anti}
One of the first steps in Section \ref{sec: dissolution} was the definition of events $\caA_{n_0,n_1}$. On these events, it was guaranteed that all resonances at scale $n_1$ could be related to resonances at scale $n_0$ for $n_0>C(\alpha,\theta,B)$.  
The analogous event for anti-resonances for $n>n_B$ is  
$$
\mathcal G_{n, n+1} =  \{ \forall \sigma,\sigma' \in \Sigma_n:  \min_{\nu=\pm 1} |{E}_{\sigma} + {E}_{\sigma'}  +2\nu h_{n+1}| > \alpha^{\theta(n+1)} \}.
$$

We will also write, for $n_1>n_0 >n_B$; 
$$
\caG_{n_0,n_1} =  \caG_{n_0,n_0+1} \cap    \caG_{n_0+1,n_0+2} \cap \ldots \cap  \caG_{n_1-1,n_1}  
$$
and we note that this remains also meaningful for $n_1=\infty$. 
We note that, just as it was the case for the event  $ \caA_{n,n+1}$, we have 
 $$
 \caG_{n,n+1} \in \caF_{n+1/2}.
 $$
The event $ \caG_{n,n+1}$ has high probability, uniformly in $\caF_n$, namely: 
\begin{lemma}\label{lem: prob event cag} For any $n>n_B$, and any  $m$, including $m=\infty$:
$$\mathbb{P}(\caG_{n,n+m}| \caF_n) \geq 1-C\alpha^{c_{\theta}n}, \qquad \mathbb{P}-\text{almost surely}.$$
\end{lemma}
\begin{proof}
Analogous to the proof of Lemma \ref{lem: prob event caa}. 
\end{proof}

As already mentioned, the key property of the event $\caG_{n_0,n_1} \cap   \caA_{n_0,n_1}$ is that it allows us to relate an anti-resonant pair at scale $n_1$ to an anti-resonant pair at scale $n_0$ for $n_0>C$, as we state in the upcoming lemma.
Also, we can decouple the problem of anti-resonances from the problem of resonances. 

Let us say that $\sigma \in \Sigma_n$ is isolated if $\{E_\sigma\} \in \caP_{n,1}$. 
We define the set of resonance labels as: 
\begin{equation} \label{eq: res def}
    \caR_n:= \{\sigma \in \Sigma_n | \{E_\sigma \} \notin \caP_{n,1} \}.
\end{equation}
and let us define
 
$$
\caZ_{n}^{\text{f}} = \caZ_{n} \cap \{ (\sigma,\sigma') \in \Sigma_n^2 \, | \,  \sigma,\sigma' \notin \caR_n \}.
$$
Moreover, given $n>n_1$ we write for any $\sigma \in \Sigma_n$, $\sigma=(\sigma_1,\mu)$ such that $\sigma_1 \in \Sigma_{n_1}$ and $\mu\in \Sigma_{n-n_1}$; then we define: 
\begin{equation} \label{eq: Z ff def}
    \mathcal{Z}_{n_1,n}^{\mathrm{ff}}:= \mathcal{Z}_n \cap  \{(\sigma,\sigma') \in \caR_n^c\times \caR_{n}^c \: \big| \sigma_1,\sigma_1' \notin \caR_{n_1}  \}. 
\end{equation}

\begin{lemma}\label{lem: significance of set cag}
For $\alpha$ sufficiently small, let $n_1>n_0>C(\alpha,\theta,B)$,  and let the event $\caG_{n_0,n_1}\cap \caA_{n_0,n_1}$ be true. Given the above setup,
\begin{enumerate} 
\item If 
 $$(\sigma_1,\sigma_1') \in \caZ_{n_1}$$ then $\sigma_{1} \neq \sigma'_1 $ and there is a unique  2-tuple $(\sigma_{0},\sigma'_0 )\in \caZ_{n_0}$ and a choice of ${\mu} \in \Sigma_{n_1-n_0}$
such that 
$$
\sigma_1=(\sigma_0,{\mu}),\qquad  \sigma'_1=(\sigma'_0,-{\mu}).
$$
Moreover, for any $\ell=1,\ldots, n_1-n_0$,  we then have
$$
( (\sigma_0,\underline{\mu}_{\ell}),  (\sigma'_0,-\underline{\mu}_\ell)  ) \in \caZ_{n_0+\ell},
$$
where here $\underline{\mu}_{\ell}$ denotes $\underline{\mu}$ restricted to its first $\ell$ components.

\item Additionally,  if we assume
$$
(\sigma_{1},\sigma'_1 )\in \caZ^{\mathrm{ff}}_{n_0,n_1},
$$
then in addition to the above conclusions, we have: 
$(\sigma_{0},\sigma'_0 )\in \caZ^{\mathrm{f}}_{n_0}$ 
and  for any $\ell=1,\ldots, n_1-n_0$,  we  have
$$
( (\sigma_0,\underline{\mu}_{\ell}),  (\sigma'_0,-\underline{\mu}_\ell)  ) \in \caZ^{\mathrm{ff}}_{n_0,n_0+\ell}.
$$
\end{enumerate}
\end{lemma}
Notice that in the above lemma, the notation $\underline{\mu}_{\ell}$ is defined in a slightly different manner compared to Section \ref{sec: res patch def}. The same object is denoted by $\underline{\mu}_{n_0-1+\ell}$ in Section \ref{sec: res patch def}. We make this choice to lighten the notation.

\begin{proof}
  From the assumption $(\sigma_{1},\sigma'_1 )\in \caZ_{n_1}$, we get 
 $$
 |E_{\sigma_{1}}+E_{\sigma'_{1}}  | \leq \alpha^{n_1 \theta}/4.
 $$
 Let us write $\sigma_1=(\sigma,\nu), \sigma_1'=(\sigma',\nu')$ with $\nu', \nu \in \{\pm 1\}$, $\sigma,\sigma' \in \Sigma_{n_1-1}$.
 Combining the above bound with the definition of the event $\caG_{n_1-1,n_1}$ and the bound from Lemma \ref{lemma : labeling lemma}, we get that necessarily $\nu'=-\nu$  and also that $\sigma_1\neq \sigma_1'$ for $n_0$ sufficiently large (such that the bound in Lemma \ref{lemma : labeling lemma} becomes less than $\alpha^{\theta n_1}/4$). Uniqueness of $(\sigma,\sigma')$ is almost surely ensured by the labeling process.  Therefore, we have 
 $$
  |E_{\sigma}+E_{\sigma'}  | \leq \alpha^{n_1 \theta}/4 + C\alpha^{n_1}.
 $$
For sufficiently small $\alpha$, and $n_0$ large enough, this is bounded by $\alpha^{(n_1-1) \theta}/4$ and so $(\sigma,\sigma') \in \caZ_{n_1-1}$. We repeat this argument inductively, using the fact that $\caG_{n-1,n}$ holds for any $n_0<n\leq n_1$, and deduce that there exists $\sigma_0,\sigma_0'\in \Sigma_{n_0}$, and $\mu \in \Sigma_{n_1-n_0}$ such that 
$\sigma_1=(\sigma_0,\mu)$, $\sigma_1'=(\sigma_0',-\mu)$ and such that $((\sigma_0,\underline{\mu}_{\ell}),(\sigma_0',-\underline{\mu}_{\ell})) \in \mathcal{Z}_{n_0+\ell}$ for all $\ell \in \{ 1,\dots,n_1-n_0\}$. Repeating the same argument allows us to deduce that $(\sigma_0,\sigma_0') \in \caZ_{n_0}$. This finishes the proof of the first claim. \\

To prove the second claim, thanks to the definition of $\caZ_{n_0,n_1}^{\mathrm{ff}}$, and the fact that $(\sigma_1,\sigma_1')\in \caZ_{n_0,n_1}^{\mathrm{ff}}$ we deduce that $\{{\sigma_0}\},\{{\sigma_0'}\} \in \caR_{n_0}^c $
 i.e., they are isolated; this concludes the proof of the fact that $(\sigma_0,\sigma_0') \in \caZ_{n_0}^{\mathrm{f}}$. 
 Now, using the definition of the event $\caA_{n_0,n_1}$, we can deduce that any descendant of $\sigma_0$ and any descendant of $\sigma_0'$ are isolated. This means $\{(\sigma_0, \underline{\mu}_{\ell})\}, \{(\sigma_0', -\underline{\mu}_{\ell})\} \in \caR_{n_0+\ell}^c$ for any $\ell \in \{1,\dots,n_1-n_0\}$. Combining this with the above conclusion, this finishes the proof of the second claim. 
\end{proof}

\subsection{Dissolution of a single anti-resonant 2-tuple}

We now fix a 2-tuple $p \in \caZ_{n-1}^{\mathrm{f}}$, for $n>C$, say 
$p=(\sigma,\sigma')$, 
and  $\mu \in \{\pm 1\}$.
In particular, $\sigma,\sigma'$ correspond to isolated levels of $H_{n-1}$ (we do not exclude the possibility that $\sigma=\sigma'$). 
We assume that the event $\caA_{n-1,n} \cap  \caG_{n-1,n} $ holds, which implies in particular that also the levels corresponding to $(\sigma,\mu),(\sigma',-\mu)$  are isolated levels of $H_n$.
To mimic the notation of Section \ref{sec: dissolution}, let us  write
$$
A_{p,\mu}= ((\sigma,\mu),(\sigma',-\mu) ),
$$
where we use the symbol $A$ instead of $S$ to underscore anti-resonances.
Now, the crucial question is whether 
$$
A_{p,\mu}  \in \caZ_n,
$$
if it is, then it is also an element of $\caZ_n^{\mathrm{f}}$ since we assumed $\caA_{n-1,n}$ for $n>C(\alpha,\theta,B)$; furthermore,  for any $n-1>n_0>C(\alpha,\theta,B)$, if we assume $\caA_{n_0,n}\cap \caG_{n_0,n}$, taking  $p \in \caZ_{n_0,n-1}^{\mathrm{ff}}$, if we establish  $A_{p,\mu} \in \caZ_n$, we deduce that $A_{p,\mu} \in \caZ_{n_0,n}^{\mathrm{ff}}$.\\

In the case of resonances, we defined random variables $s_n$ and $v_n/v_{n-1}$ that were helpful in diagnosing if $A_{p,\mu} \in \caZ_n$ or not, see \eqref{eq: condition on sn} and \eqref{eq: patch bound}. We will now introduce the analogue of $s_n$ for $n>C$; 
\begin{equation}
    \mathrm{a}_n:= {E}_{(\sigma,
    \mu)}+{E}_{(\sigma',
    -\mu)}.
\end{equation}
If $A_{p,\mu} \in \caZ^{\mathrm{f}}_n$, then  
\begin{equation}
    |\mathrm{a}_n|<  \alpha^{\theta n}/4.
\end{equation}
Just as for $s_n$, we know that $|\mathrm{a}_n|>0$, almost surely, thanks to Lemma \ref{lemma: minimal gap antigap}.
Since $E_{\sigma}+\mu h_n$
and ${E}_{\sigma'}-\mu h_n$ are sufficiently isolated, which is ensured by the event $\caA_{n-1,n} $, one can do perturbation theory in the parameter $g=g_n$ exactly as in \eqref{eq:expansion}. Therefore, there is a convergent expansion for $\mathrm{a}_n(g)$ for $n>C$:
\begin{equation} \label{eq:expansion anti}
    \mathrm{a}_n(g)= \mathrm{a}_n(0)+ \sum_{k=1}^{\infty} g^k \tilde{\alpha}^{nk} b_k, 
\end{equation}
where $\tilde{\alpha}=4\alpha^{1-\theta}$ as before and $|b_k|\leq 2$.   We also define the analogue of $v_n$: 
\begin{equation} \label{eq: p def anti}
   \mathrm{v}_n(g):= \frac{\ln(|\mathrm{a}_n(g)|)}{n \ln (\tilde{\alpha})}.
\end{equation}
We also denote as before, $\mathrm{a}_0:= \mathrm{a}_n(0)$, and $\mathrm{v}_0:=\mathrm{v}_n(0)$. As before, we also set:
$$\mathrm{v}_{n-1}:=\frac{\ln(|\mathrm{a}_{n}(0)|)}{(n-1)\ln(\tilde{\alpha})}. $$ 
For $n$ large enough, if 
\begin{equation}
A_{p,\mu}  \in \caZ_n, \quad 
\text{ then } 
\quad 
    \frac{1}{\mathrm{v}_n(g)} \leq \frac{2(1-\theta)}{\theta}. 
\end{equation}



With these quantities in place, we study the probability that $A_{p,\mu} \in \caZ_n$, for a fixed $p \in \caZ_{n-1}^{\mathrm{f}}$ and $\mu \in \{\pm 1\}$. The result is

\begin{proposition}\label{prop: main moment bound anti}
We let $\alpha$ be sufficiently small, and we fix $b>4$.
We fix $n>C(\alpha,\theta,B)$ and we assume the event $\mathcal{A}_{n-1,n} \cap \mathcal{G}_{n-1,n}$ to hold.  We fix  $p \in \caZ_{n-1}^{\mathrm{f}}$ and $\mu \in \{\pm 1 \}$, and we define 
 the  corresponding variables $\mathrm{v}_n,\mathrm{v}_{n-1}$, as explained above. 
Then, there is $n_*=n_*(\alpha,b,\theta,B)$ such that, for $n\geq n_* $, the following holds:
\begin{equation} \label{eq: main moment bound anti}
     \chi(\mathcal{A}_{n-1,n} \cap  \mathcal{G}_{n-1,n} ) \mathbb{E}((\frac{\mathrm{v}_n}{\mathrm{v}_{n-1}})^b | \caF_{n-1/2})  \leq \exp(-\frac{b/2-1}{n}), \qquad  \mathbb{P}-\text{almost surely}.
 \end{equation}
\end{proposition}
This proposition is the analogue of Proposition \ref{prop: main moment bound} and its proof follows  the proof of Proposition \ref{prop: main moment bound} closely, as given in 
 Section \ref{sec: proof of dissol}. One difference is that we use $|\mathrm{a}_0|$ instead of $s_0$, as the former can be negative. However, this does not present any complications. We omit the almost analogous proof. However, we highlight the fact that in the proof of Proposition \ref{prop: main moment bound anti} our main workhorse is the expansion \eqref{eq:expansion anti} instead of \eqref{eq:expansion}.

Then, we formulate the analogue of Proposition \ref{prop: dissolution single patch}.
Let us abbreviate ($n>n_B)$)
$$
\mathbb{P}_n(\cdot)=   \mathbb{P}(\cdot| \caA_{n,\infty}\cap \caG_{n,\infty})
$$
Then, 
\begin{proposition}
\label{prop: dissolution single pair} 
For any $n_1>C(\alpha,\theta,B)$, and 
 for any $p=(\sigma,\sigma')\in \caZ^{\mathrm{f}}_{n_1}$ 
$$ 
\sum_{n=n_1+1}^\infty\sup_{\underline{\mu} \in \Sigma_{n-n_1}}{\mathbb{P}_{n_1}(A_{p,\underline{\mu}} \in \caZ_n^{\mathrm{f}} | \caF_{n_1})}  <\infty, \quad   {\mathbb{P}-a.s.}
$$
\end{proposition}
Given all the machinery of this section, the proof of this proposition, starting from Proposition \ref{prop: main moment bound anti}, is completely analogous to the proof of Proposition \ref{prop: dissolution single patch}, and we therefore omit it. Thanks to Lemma \ref{lem: significance of set cag}, and the definition of $\caZ_{n_1,n}^{\mathrm{ff}}$, since $p \in \caZ_{n_1}^{\mathrm{f}}$, the above proposition means for any $p \in \caZ_{n_1}^{\mathrm{f}}$: 
\begin{equation} \label{eq: dissolve single patch}
    \sum_{n=n_1+1}^\infty\sup_{\underline{\mu} \in \Sigma_{n-n_1}}{\mathbb{P}_{n_1}(A_{p,\underline{\mu}} \in \caZ_{n_1,n}^{\mathrm{ff}} | \caF_{n_1})}  <\infty, \quad   {\mathbb{P}-a.s.}
\end{equation}

\subsection{Dissolution of most anti-resonant 2-tuples}

From Theorem \ref{thm: antires vanish}, recall the definition of $\widehat{\caZ}_n$. Similarly, we can define, $\widehat{\caZ}_n^{\mathrm{f}}:= \{\sigma \in \Sigma_n |  (\sigma,\sigma') \in \caZ_n^{\mathrm{f}} \: \text{for some } \: \sigma' \}$. We define $\widehat{\caZ}_{n_1,n}^{\mathrm{ff}}$, for proper $C<n_1<n$, similarly.  Notice that thanks to properties following \eqref{eq: def Z_n}, we have $|\widehat{\caZ}_n^{\mathrm{f}}| =  |\caZ_n^{\mathrm{f}}| $, and $|\widehat{\caZ}_{n_1,n}^{\mathrm{ff}}| =  |\caZ_{n_1,n}^{\mathrm{f}}|$.

Thanks to the basic  definition of anti-resonance set  (see properties following \eqref{eq: def Z_n}), we can observe that: 
\begin{equation} \label{eq: first trivial anti res bound}
    |\widehat{\caZ}_n| \leq |\widehat{\caZ}_n^{\mathrm{f}}|+ 2|\caR_n|, 
\end{equation}
where we recall the definition of the resonance set $\caR_n$ \eqref{eq: res def}.
Let $n>n_1>n_0>C(\alpha,\theta,B)$,  assuming $\caA_{n_0,n}\cap \caG_{n_0,n}$ holds,  similar to the above bound, appealing to Lemma \ref{lem: significance of set cag} and definition of $\widehat{\caZ}_n^{\mathrm{f}}, \widehat{\caZ}_{n_1,n}^{\mathrm{ff}}$, we have: 
\begin{equation} \label{eq: second trivial anti res bound}
    |\widehat{\caZ}_n^{\mathrm{f}}| \leq |\widehat{\caZ}_{n_1,n}^{\mathrm{ff}}| + 2 \cdot 2^{n-n_1} |\caR_{n_1}|. 
\end{equation}

On the other hand, assuming the event $\caG_{n_1,n}\cap\caA_{n_1,n}$ we can write $|\caZ_{n_1,n}^{\mathrm{ff}}|$ as :
\begin{equation} \label{eq: z_nff to Z_nf}
    |\widehat{\caZ}_{n_1,n}^{\mathrm{ff}}|= |\caZ_{n_1,n}
    ^{\mathrm{ff}}| = \sum_{p \in \caZ_{n_1}^{\mathrm{f}}}\sum_{\underline{\mu} \in \Sigma_{n-n_1}} \chi(A_{p,\underline{\mu}} \in \caZ_{n_1,n}^{\mathrm{ff}}).
\end{equation}
Combining the above expression with  Proposition \ref{prop: dissolution single pair}, bounding $|\caZ_{n_1}^{\mathrm{f}}| \cdot |\Sigma_{n-n_1}|$ by $2^n$, we deduce for any $n_1>C$: 

$$ \sum_{n>n_1} \mathbb{E}_{n_1}\left(\frac{|\widehat{\caZ}_{n_1,n}^{\mathrm{ff}}|}{2^n}\bigg| \caF_{n_1} \right) < \infty.$$
Then, by the Markov inequality and the Borel-Cantelli Lemma, we deduce for any $n_1>C$:
$|\widehat{\caZ}_{n_1,n}^{\mathrm{ff}}|/{2^n} \to 0,$  $\mathbb{P}_{n_1}- \text{almost-surely}.
$\\
Now, let $n_1>n_0>C$. Thanks to Lemma \ref{lem: prob event caa}, and Lemma \ref{lem: prob event cag} we have $\mathbb{P}(\caA_{n_0,\infty}\cap \caG_{n_0,\infty}) , \mathbb{P}(\caA_{n_1,\infty}\cap \caG_{n_1,\infty})>0$ for $\alpha$ small enough. Moreover, $\caA_{n_0,\infty}\cap \caG_{n_0,\infty} \subset \caA_{n_1,\infty}\cap \caG_{n_1,\infty}$ by definition. Therefore, we deduce that 
\begin{equation} \label{eq: Z_n ff limit}
|\widehat{\caZ}_{n_1,n}^{\mathrm{ff}}|/2^n \to 0, \qquad\mathbb{P}_{n_0}- \text{almost surely}, 
\end{equation}
for any $n_0>C$ . Combining \eqref{eq: first trivial anti res bound} and \eqref{eq: second trivial anti res bound}, then taking 
$\limsup_{n \to \infty}$ and using Theorem \ref{thm: dissolution of patches}, \eqref{eq: Z_n ff limit} and the fact that $\mathbb{P}(\caA_{n_0,\infty} \cap \caG_{n_0,\infty})>0$, we deduce that for any $n_1$ sufficiently large:
\begin{equation}
    \limsup_{n \to \infty} |\widehat{\caZ}_n|/2^n  \leq
    2 \times |\caR_{n_1}| /2^{n_1}, \qquad \mathbb{P}_{n_0}- \text{almost surely}.
\end{equation}
Since the LHS is independent of $n_1$, taking $\limsup_{n_1 \to \infty}$, and again using Theorem \ref{thm: dissolution of patches} and the fact that $\mathbb{P}(\caA_{n_0,\infty} \cap \caG_{n_0,\infty})>0$ we deduce that $|\widehat{\caZ}_n| /2^n\to 0$, $\mathbb{P}_{n_0}$- almost surely. Since this conclusion holds for arbitrary $n_0>C$, we deduce that the same result holds $\mathbb{P}$-almost surely. We omit this part since this last step uses the exact same reasoning as the last step in Section \ref{sec: dissolution of most}. The only difference is that one uses the combination of Lemma \ref{lem: prob event caa} and Lemma \ref{lem: prob event cag}  to deduce that $\mathbb{P}(\caA_{n_0,\infty}\cap \caG_{n_0,\infty}) \to 1 $ as $n_0 \to \infty$ in this case.  We therefore conclude the proof of Theorem \ref{thm: antires vanish}.

\section{Moments:  unperturbed process} \label{sec: factor1}

Given a $k$-tuple (of configurations) $\underline{\sigma}=(\sigma_1\dots,\sigma_k) \in \Sigma_n^{k}$, we denote: 
\begin{equation} \label{eq: Energy free config}
    E^o_{\sigma_i}:= \sum_{x=1}^n \sigma_i(x) h_x=\sigma_i.h. 
\end{equation}
The main question of this section is whether the random variables $\{E_{\sigma_i}^o\}_{i=1}^k$ become essentially independent for large $n$. In other words, is it true that, for appropriate intervals $\{\mathtt{I}_i\}_{i=1}^k$,  
\begin{equation} \label{eq: factor with ?}
\mathbb{P}(E_{\sigma_1}^o \in \mathtt{I}_1, \dots E_{\sigma_k}^o \in 
\mathtt{I}_k) \stackrel{?}{=} \mathbb{P}(E_{\sigma_1}^o \in \mathtt{I}_1)\dots \mathbb{P}(E_{\sigma_k}^o \in \mathtt{I}_k) \times (1+ o_n(1)).
\end{equation}
Since the pdf of $h_x$ is symmetric around zero, by local central limit theorem estimates, for any $\sigma$, $E_{\sigma}^o$ is well-approximated by a Gaussian random variable with mean zero and variance $\mathsf{s}^2n$.  Therefore, the above question is similar to asking whether $(E_{\sigma_1}^o,\dots, E_{\sigma_k}^o)$ is asymptotically a collection of independent Gaussian random variables. \\
Clearly, such independence needs assumptions on $(\sigma_1,\ldots,\sigma_k)$. For example, if $k=2$ and $\sigma_1(x)=\sigma_2(x)$ holds for all but a few $x$, then the independence fails.  However, we can prove the independence for  ``typical" $k$-tuples $\underline{\sigma}$, with the relevant notion of typicality to be defined in the next section.

\subsection{Setup}

\subsubsection{Abbreviations}
 We denote the pdf of the random variables $h_i$ by $\varphi(x)$. Recall that we've taken $\varphi(x)=\chi([-1/2,1/2])(x)$. 
 The following abbreviations of (un)normalized Gaussian densities will be used extensively in this section:
 \begin{equation} \label{eq: abbr gaussian}
     \mathfrak{g}_n(x):= \frac{\exp(-x^2/2\mathsf{s}^2 n)}{\mathsf{s} \sqrt{2 \pi n}}; \quad  \tilde{\mathfrak{g}}_n(x):= \exp(-x^2/2\mathsf{s}^2n).  
 \end{equation}
For any $n \in \mathbb{N}$, we denote by $\{{G}_n^{i}\}_{i=1}^{\infty}$ a collection of independent Gaussian random variables with mean zero and variance $n \mathsf{s}^2$. This collection is also independent of all other random variables that we consider. By abusing the notation, we denote for any interval $\mathtt{I}$:
$${G}_n^i(\mathtt{I}):=\mathbb{P}({G}_n^i \in \mathtt{I}). $$

The following events will also be abbreviated whenever intervals 
$\mathtt{I}_i$,  and configurations $\sigma_i$ are apparent from the context: 
\begin{equation} \label{eq: abbr event o}
    \mathcal{E}_i^o:= \{E_{\sigma_i}^o \in \mathtt{I}_i \}, \quad  \mathcal{E}^o_{[k]}:= \bigcap_{i=1}^k \mathcal{E}_i^o.
\end{equation}

\subsubsection{Typicality of $\underline{\sigma}$} \label{sec: typical def} 

Given a $k$-tuple $\underline{\sigma}=(\sigma_1, \dots,\sigma_k) \in \Sigma_n^k$ and a configuration $\tau \in \{\pm 1 \}^{k-1} $, we  define the following subsets of $[n]=:\{1,2,\dots,n\}$:
\begin{equation} \label{def: J_tau}
    \mathfrak{J}_{\tau}({\underline{\sigma}}) \equiv \mathfrak{J}_{\tau}=\{x \in [n] | \sigma_1(x).(\sigma_2(x),\dots,\sigma_{k}(x))=\tau \} \subset [n] ,
\end{equation}
where the above product is the usual product of the scalar $\sigma_1(x)$ with the vector $(\sigma_2(x),\dots,\sigma_k(x))$.
We note that $\{ \mathfrak{J}_{\tau}\}_{\tau \in \{\pm 1\}^{k-1}}$ is a partition of $[n]$. 
Then we define
\begin{itemize}
    \item $\underline{\sigma}= (\sigma_1,\dots, \sigma_k)$ is \textit{fully non-degenerate} if,  for  all  $\tau \in \{\pm 1\}^{k-1}$,  $\mathfrak{J}_{\tau} (\underline{\sigma}) \neq \varnothing$. 
({this is a much stronger requirement than the $k$-tuple $(\sigma_1,\dots, \sigma_k)$ being linearly independent}).
\item  A $k$-tuple $\underline{\sigma}=(\sigma_1 \dots,\sigma_k)$ is
\textit{typical}  iff.
\begin{equation}\label{def: typicality}
\left||\mathfrak{J}_{\tau}(\underline{\sigma})\right|- n/2^{k-1} |\leq  n^{1-1/4} \qquad \forall \tau \in \{\pm1 \}^{k-1}.
    \end{equation}
Depending on the context, we will use the adjectives \textit{$n$-typical} or \textit{$(k,n)$-typical} instead of \textit{typical}. 
\end{itemize} 
We note that a $(k,n)$-typical configuration is always  \textit{linearly independent}
 as a $k$-tuple of $\mathbb{R}^n$. 
 \begin{remark} \label{rmk: typical well defined}
   {Even though the definition of typicality seems to depend on the order in which the configurations are arranged in the $k$-tuple}, it is straightforward to check that this is not the case: if $(\sigma_1,\dots,\sigma_k)$ is typical then $(\sigma_{\pi(1)}.\dots,\sigma_{\pi(k)})$ is also typical for any permutation $\pi$ of $[k]$.
\end{remark}
It is useful to have a notation for the  set of all $(k,n)$-typical configurations: 
\begin{equation} \label{eq: def typical set}
    \mathcal{T}_{k,n}:= \{\underline{\sigma} \in \Sigma_n^k| \: \underline{\sigma} \:\: \text{is } (k,n)-\text{typical}  \}. 
\end{equation}
{We denote }$\mathcal{T}_{k,n}^c=\Sigma_n^k\setminus\mathcal{T}_{k,n}$. 
The term ``typical" is rationalized by  
\begin{lemma} \label{lemma: counting00}
    For any $0<a<1/2$ we have: 
    \begin{equation} \label{eq:countingII}
      \frac{ |\mathcal{T}^c_{k,n}|}{2^{nk}}\leq o_n(\exp(-n^a)). 
   \end{equation}
\end{lemma}
The proof of Lemma \ref{lemma: counting00} relies on standard tools, and it is in Appendix \ref{sec: app  LD estimate}.

\subsection{The factorization Lemma}
\begin{lemma}[Factorization lemma] \label{lem: factorization} 
  
For any collection of intervals $\{\mathtt{I}_i\}_{i=1}^k$, we have: 
       
\begin{equation} \label{eq: factor bound1}
   \sup_{\underline{\sigma}\in \mathcal{T}_{k,n}} \left|\mathbb{P}\left( \mathcal{E}^o_{[k]}\right) -\prod_{i=1}^k 
    {G}_n^i(\mathtt{I}_i)\right|  \leq o_n(1)
    \prod_{i=1}^k |\mathtt{I}_i|/\sqrt{n}.
\end{equation}
 where  $\mathcal{T}_{k,n}$ is the set of typical configurations, defined in \eqref{eq: def typical set}. {The factor $o_n(1)$} is independent of the intervals $\mathtt{I_i}$. In particular, $\mathtt{I}_i$ may depend on $n$. 
\end{lemma}
 The proof of the above lemma requires some preliminary tools, and it is postponed to the next sections.

The factorization anticipated in \eqref{eq: factor with ?} can be deduced from \eqref{eq: factor bound1} for any typical $k$-tuple, as long as $\mathtt{I}_i\subset [-n^{\rho},n^{\rho}]$  for some $\rho \in (0,1/2)$ and for all $i$.  This follows by a simple computation and the observation that for $k=1$, the set of typical $\sigma$ is equal to $\Sigma_n$. 

As a curiosity, we note that the error term in \eqref{eq: factor bound1} can be larger than the Gaussian product $\prod_{i=1}^k 
    {G}_n^i(\mathtt{I}_i)$ if $\mathtt{I}_i \subset [n_0^{1-\rho},\infty)$.  Nevertheless,  this bound remains useful even for such intervals.

Let us state a corollary, which is a variation on  Lemma \ref{lem: factorization}, to be used later.  Recall the size of the bath $n_B=O_n(1)$. For any $n$ we also let $$\breve{n}:= n+n_B.$$  Recall the collection of Gaussian r.v. $\{G_n^i \}_i$. Then we have:  

\begin{corollary} \label{cor: factorization with bath}
     Lemma \ref{lem: factorization} remains true if we replace $G_n$ with $G_{\breve{n}}$ i.e. if we replace \eqref{eq: factor bound1} with
 
 \begin{equation} \label{eq: factor bound 3 n_B}
       \sup_{\underline{\sigma} \in \mathcal{T}_{k,n}}  \left|\mathbb{P}\left( {\mathcal{E}}^o_{[k]} \right) -\prod_{i=1}^k 
    G_{\breve{n}}^i(\mathtt{I}_i)\right|  \leq o_n(1) 
    \prod_{i=1}^k |\mathtt{I}_i| /\sqrt{n}.
    \end{equation}
\end{corollary}
\subsection{Application of Factorization: Semi-perturbed process}
In this section, we recast the factorization Lemma \ref{lem: factorization} in  another useful form, namely Lemma \ref{lem: factorization conditional}. The motivation for doing so will become clear in the next section. 
\subsubsection{Semi-perturbed process: $n_0-n_1$ splitting} \label{sec: spliting}
Choose a constant $0<\varrho<1 $ such that 
\begin{equation} \label{eq: n_0 n_1 split 0}
    \alpha < \left(1/2\right)^{1/\varrho} <\alpha^{\theta}.
\end{equation}
For any $n$ such that $  \varrho n>2n_B$, we split $n$ by defining $$n_0:= \lfloor n \varrho \rfloor, \quad  n_1:=n-n_0.$$ 
This means that, for $n$ sufficiently large, 
\begin{equation} \label{eq: n_0 n scale}
    \alpha^{ n_0} \ll \sss_n \ll \alpha^{\theta n_0},
\end{equation}
where we recall $\sss_n$ from \eqref{sec: stat sn}. Given $\eta \in \Sigma_n$, we split $\eta=(\mu,\sigma)$ with $\mu \in \Sigma_{n_0}$ and $\sigma \in \Sigma_{n_1}$. Recalling the labeled eigenvalues of $H_{n_0}$ as $\{{E}_{\mu} \}_{\mu \in \Sigma_{n_0}}$,  we define the \textit{semi-perturbed} eigenvalues as: 
\begin{equation} \label{eq: tilde def}
    \widetilde{E}_{\eta} = \widetilde{E}_{(\mu,\sigma)}:= 
    {E}_{\mu}+ \sigma.h := {E}_{\mu}+ \sum_{x=n_0+1}^n \sigma(x) h_x.
\end{equation}
Thanks to Lemma \ref{lemma : labeling lemma}, we have: 
\begin{equation} \label{eq: tilde bar relation}
    |{E}_{\eta}-\widetilde{E}_{\eta}| \leq C \alpha^{n_0}. 
\end{equation}
The motivation for introducing these semi-perturbed eigenvalues is as follows: for large $n$, \eqref{eq: n_0 n scale} ensures that $\widetilde{E}_{\eta}$ approximates $E_{\eta}$ in a meaningful way, i.e., with error exponentially smaller than the mean level spacing. On the other hand, $\widetilde{E}_{\eta}$ is a ``regularized`` version of $E_{\eta}$ since its pdf is obtained by  convolution with a Gaussian (the pdf of $h.\sigma$ is well approximated by a Gaussian). This approximation is crucial in what will follow. 

\subsubsection{Intervals} \label{sec: intervals!}

In  \eqref{eq: factor bound1}, recall the expression $\prod_i G_n^i(\mathtt{I}_i)$. If we replace $\mathtt{I}_i$  with $\mathtt{I}'_i$ such that $|\mathtt{I}_i|=|\mathtt{I}_i'|$ as long as $\text{dist}(\mathtt{I}_i,\mathtt{I}'_i)<n^{\rho}$ with $\rho \in (0,1/2)$ the new expression depending on $\mathtt{I}'_i$ differs from the old expression at most by the same error as in \eqref{eq: factor bound1}. {Lemma \ref{lem: factorization conditional} substantiates this claim. To state it, we introduce three specific classes of intervals:}
\begin{enumerate}
\item{\textbf{Rescaled intervals}} We denote macroscopic \textit{$n$-independent} intervals of $\mathbb{R}$ by $\mathcal{I}$. Corresponding to each interval $\mathcal{I} \subset \mathbb{R}$, we define the rescaled interval $I$: 
\begin{equation} \label{eq: rescale intervals}
    I:= \sss_n \mathcal{I}.
\end{equation}
We reserve the notation $\mathcal{I}$ for macroscopic $n$-independent intervals, and $I$ for the rescaled counterpart of $\mathcal{I}$. 
\item{\textbf{Long intervals}}  Fix $0 < \rho <1/2$, and for any $\mathtt{t} \in \mathbb{Z} \times n_0^{\rho}=: \mathcal{J}$ define: 
\begin{equation} \label{eq: J interval def}
    J_{\mathtt{t}} := [\mathtt{t},\mathtt{t}+n_0^{\rho}).
\end{equation}
Choose a constant $C_{\mathcal J}$ such that $\text{spec}(H_{n_0}) \subset [-C_{\mathcal J} n_0,C_{\mathcal J} n_0]$. For example, $C_{\mathcal J}=2+\|H_B\|$ is a valid choice.   Let $n_0^-:= -C_{\mathcal J} n_0-2n_0^{\rho}$ then we denote:
$$ \mathcal{J}_{n_0}:=\mathcal{J}\cap [n_0^-,C_{\mathcal J} n_0].$$
Notice that $\{ J_\mathtt{t}\}_{\mathtt{t}\in \mathcal{J}_{n_0}}$ divide $\text{spec}(H_{n_0})$ into disjoint intervals $J_{\mathtt{t}}$ of size at most $n_0^{\rho}$.

\item{\textbf{Shifted intervals}} For any interval $I$, and for $t \in \mathbb{R}$ (possibly $n$ dependent) we denote 
$$I_i^{t}:= I_i-{t}.$$

\end{enumerate}

\subsubsection{Regularization by Free Spins}
Let us fix the $k$-tuples $\underline{\sigma}=(\sigma_1\dots,\sigma_k)\in \Sigma_{n_1}$ and $\underline{\mu}=(\mu_1,\dots,\mu_k) \in \Sigma_{n_0}$, and (macroscopic) intervals $\{\mathcal{I}_i\}_{i=1}^k$ and the corresponding rescaled intervals $\{I_i\}_{i=1}^k$, as defined above in  \eqref{eq: rescale intervals}.  We abbreviate: 
\begin{equation} \label{eq: mathcal abbr}
   \mathcal{E}_i:= \{{E}_{\mu_i} +\sigma_i.h \in I_i\}, \qquad \mathcal{E}_{[k]}:= \bigcap_{i=1}^k \mathcal{E}_{i}.
\end{equation}
Moreover, given long intervals $J_{\mathtt{t}_1},\dots J_{\mathtt{t}_k}$ as defined above, we denote: 
\begin{equation} \label{eq: abbr E J_t}
   \hat{\mathcal{E}}_i:= \{{E}_{\mu_i}  \in J_{\mathtt{t}_i}\}, \qquad \hat{\mathcal{E}}_{[k]}:= \bigcap_{i=1}^k \hat{\mathcal{E}}_{i}.
\end{equation} 

\begin{lemma} \label{lem: factorization conditional}
 For any collection of intervals $\{\mathcal{I}_i\}_{i=1}^k$,  we have: 
\begin{align} \label{eq:Factorizationestimate1}
 \sup_{\underline{\sigma}, \underline{\mu}, \underline{\mathtt{t}},\underline{J_{\mathtt{t}}} }\left| \mathbb{P}\left( \mathcal{E}_{[k]} \Big| \hat{\mathcal{E}}_{[k]}\right) -  \prod_{i=1}^k 
   G_{n_1}^i(I_i^{\mathtt{t}_i}) \right| \leq o_n(1)\prod_{i=1}^k |I_{i}|/\sqrt{n},
\end{align}
where the supremum is taken over all {$(k,n_1)$-typical} tuples $\underline{\sigma} \in \mathcal{T}_{k,n_1}$, all $\underline{\mu} \in \Sigma_{n_0}^k$,   all  $\underline{\mathtt{t}}=\{\mathtt{t}_i\}_{i=1}^k \subset \mathcal{J}_{n_0}$ and all the corresponding  long intervals   $\underline{J}_{\mathtt{t}}=\{J_{\mathtt{t}_i}\}_{i=1}^k$ such that $\hat{\caE}_{[k]}$ has nonzero probability. The factor $o_n(1)$ is independent of the  intervals $\mathcal{I}_i$.\\
The  bound also holds if we replace $\hat{\mathcal{E}}_{[k]}$ by $\hat{\mathcal{E}}_{[k]} \cap \mathcal{A}_{n_0}\neq \emptyset$, with $\mathcal{A}_{n_0}$ any $\mathcal{F}_{n_0}-$ measurable set such that $\hat{\mathcal{E}}_{[k]} \cap \mathcal{A}_{n_0}$ has nonzero probability.  
\end{lemma}
  
\subsection{Proof of the factorization Lemma I}
The first ingredient of the proof of Lemma \ref{lem: factorization} is an abstract inequality which is a direct consequence of the Brascamp-Lieb (BL) inequality.  
\subsubsection{Brascamp-Lieb inequality}
Let us first state the following version of the BL inequality, see\ \cite{Ball1989,barthe1998reverse,brascamp1976best}. 
\begin{lemma} \label{BL ineq lemma}
    Let $r,m \in \mathbb{N}$, with $r \leq m$. Let ${\mathfrak{u}_1},\dots, {\mathfrak{u}_m}$ be {unit vectors} in $\mathbb R^r$, and let $\lambda_1,\dots,\lambda_m>0$,
    satisfying $$\sum_{i=1}^m \lambda_i \ket{\mathfrak{u}_i}\bra{\mathfrak{u}_i}=\mathrm{Id}_r,$$ where $\mathrm{Id}_r$ is the identity operator in $\mathbb{R}^r$, and $\ket{\mathfrak{u}_i}\bra{\mathfrak{u}_i}$ denotes the one-dimensional projector associated to the vector $\mathfrak{u}_i$. Then for 
    $f_1,\dots, f_m \in L^1(\mathbb {R} ,[0,\infty))$ we have:
    \begin{equation} \label{BLineq}
         \int_{\mathbb{R}^r} \prod_{i=1}^m f_i^{\lambda_i}(\langle \mathfrak{u}_i,x \rangle) dx 
         \leq \prod_{i=1}^m\left(\int_{\mathbb{R}} f_i(y)dy \right)^{\lambda_i}.
    \end{equation}
    \end{lemma}
\begin{proof}
This is  Lemma 2 of \cite{Ball1989} or Theorem 5 of \cite{barthe1998reverse}. It follows from the BL  inequality in \cite{brascamp1976best} and a determinant inequality in Proposition 9 of \cite{barthe1998reverse}.
\end{proof}

\subsubsection{Application of the Brascamp-Lieb inequality}
We deduce a proposition, inspired by \cite{Ball1989}, that will be used in different forms later.

 We fix $k,m \in \mathbb{N}$ such that $k\leq m$. Let ${u_1}, \dots {u_k} $ be {linearly independent}  vectors in $\mathbb{R}^m$. We denote by $H^{\perp}$ the $k$-dimensional subspace spanned by ${u_1}, \dots, {u_k}$. $H$ is the orthogonal complement of $H^{\perp}$ i.e.\  $\mathbb{R}^{m}=H \oplus H^{\perp}$. Let $P$ be the orthogonal projection onto $H$.
\begin{proposition}  \label{prop: BL main app}
    Let  $\varphi_1,\dots ,\varphi_m\in L^1(\mathbb{R}) \cap L^{\infty} (\mathbb{R}) $, such that $\max_{i=1}^m\{ \|\varphi_i \|_{L^1}  \} \leq 1$. Let $\mathtt{I}_1,\dots,\mathtt{I}_k \subset \mathbb{R}$ be { arbitrary intervals}. We fix an 
 orthonormal basis $\{ {e_i}\}_{i=1}^m$ of $\mathbb{R}^m$ and we write 
     \begin{equation} \label{eq: def: lambda_i}
  \lambda_i := |P{e_i}|^2. 
\end{equation}
Then
    \begin{equation} \label{eq: BL main application}
     \left| \int_{\mathbb{R}^m} \chi\left(\bigcap_{i=1}^k \langle u_i,y \rangle\ \in \mathtt{I}_i \right) \prod_{i=1}^m \varphi_i(y_i) dy_i \right| \leq \frac{\prod_{i=1}^k |\mathtt{I}_i|} {(\det(SS^{tr}))^{\frac 12}} \times C(k)\prod_{i=1}^m  \|\varphi_i \|_{L^{\infty}}^{1-\lambda_i},  
    \end{equation}
    where $y=\sum_i y_i e_i$ for $y \in \mathbb{R}^m$,  $S$ is the $k \times m$ matrix, having ${u_i}$ in the $i$-th row, $|\mathtt{I}_i|$ denotes the length of the interval $\mathtt{I}_i$ and, finally,  $C(k)>0$  depends on $k$.
\end{proposition}

\begin{proof}
     For $1 \leq i \leq m$, recall $\lambda_i$ \eqref{eq: def: lambda_i} and let us first assume that all $\lambda_i$ are nonzero. We set 
    $${\mathfrak{v}_i} := P {e_i}/\sqrt{\lambda_i}. $$ 
Then
    $\{ {\mathfrak{v}_i}\}_{i=1}^{m}$ is a sequence of unit vectors in $H$, and $\dim(H)=m-k$. In addition, 
    it is straightforward to see  that: 
    \begin{equation} \label{eq:projectionBL2}
        P= \sum_{i=1}^{m} \lambda_i \ket{\mathfrak{v}_i}\bra{\mathfrak{v}_i},
    \end{equation}
    where $\ket{\mathfrak{v}_i}\bra{\mathfrak{v}_i}$ is the one-dimensional projector associated to $\mathfrak{v}_i$. 
    Any $y \in \mathbb{R}^{m}$ is written uniquely as $y=w+w^{\perp}$, with $w\in H$, 
    $w^{\perp} \in H^{\perp}$. {We write  $w_i,w^{\perp}_i$ for the coefficients of $w,w^{\perp}$ with respect to the basis $\{e_i\}_{i=1}^m$. By \eqref{eq:projectionBL2}, we then have 
    $w_i =\sqrt{\lambda_i} \langle \mathfrak{v}_i,y \rangle $ and  we obtain}
    \begin{align} 
         \left| \int_{\mathbb{R}^{m}} 
        \chi\left(\bigcap_{i=1}^k \langle u_i , y \rangle \in \mathtt{I}_i\right) \prod_{i=1}^{m} \varphi_{i}(y_i) dy_i  
        \right| 
        &  \leq \int_{H \oplus H^{\perp}} 
         \chi\left(\bigcap_{i=1}^k \langle u_i , w^{\perp} 
         \rangle \in \mathtt{I}_i \right) \prod_{i=1}^{m} 
         |\varphi_{i}(\sqrt{\lambda_i} \langle \mathfrak{v}_i,y\rangle+
         w^{\perp}_i)| dy_i 
          \nonumber \\
        & \leq \frac{\prod_{i=1}^k|\mathtt{I}_i|}{(\det{SS^{tr}})^{\frac12}}\sup_{w^{\perp}_*\in \widetilde{I}}
         \int_H \prod_{i=1}^{m} 
         |\varphi_{i}(\sqrt{\lambda_i} \langle \mathfrak{v}_i,y\rangle+
         (w^{\perp}_*)_i))|dw, 
        \label{eq:factorizationDeltai}
    \end{align}
    where $\widetilde{I}$ is the subset of $H^{\perp}$ where $\bigcap_{i=1}^k \langle u_i , w^{\perp} \rangle \in \mathtt{I}_i$.
  To get the first bound we used  the definition of $w,w^{\perp}$, 
    and $\mathfrak{v}_j$, in particular the fact that $\langle u_i,y\rangle=\langle u_i , w^{\perp} \rangle$. To get the second bound, we used Fubini's theorem and 
an $L^1-L^{\infty}$ bound. We also performed an integration in the subspace $H^{\perp}$ via a 
    change of variable; hence the factor of the determinant and the volume of the resulting $k$-cube. 

    For any $w_*^{\perp} \in \widetilde{I}$, we have:

      \begin{align}
  \int_H \prod_{i=1}^{m} 
     \Bigl| \varphi_i\!\bigl(\sqrt{\lambda_i}\,\langle \mathfrak{v}_i, y\rangle
        + (w^{\perp}_*)_i \bigr) \Bigr| \, dw
  &\;\leq\;
    \prod_{i=1}^{m}
      \| \varphi_i \|_{L^{\infty}}^{\,1-\lambda_i}
      \int_H \prod_{i=1}^{m}
      \Bigl| \varphi_i\!\bigl(\sqrt{\lambda_i}\,\langle \mathfrak{v}_i, w\rangle
        + (w^{\perp}_*)_i \bigr)\Bigr|^{\lambda_i} dw \nonumber \\
  &\;\leq\;
    \prod_{i=1}^{m} \| \varphi_i \|_{L^{\infty}}^{\,1-\lambda_i}
    \prod_{i=1}^{m} \left( \tfrac{1}{\sqrt{\lambda_i}} \right)^{\!\lambda_i},
  \label{eq: BL aux temp}
\end{align}

      where in the first bound we simply used the fact that $|\varphi_i(\cdot)| \leq \|\varphi\|_{L^{\infty}}^{1-\lambda_i} |\varphi(\cdot)|^{\lambda_i}$ since $\lambda_i \in [0,1]$. Then, we take advantage of the Brascamp-Lieb inequality \eqref{BLineq}: first we define $f_i(x):= |\varphi_i(\sqrt{\lambda_i}x + (w_*^{\perp})_i)|$; then, we use the definition of ${\mathfrak{v}_i}$, and the identity \eqref{eq:projectionBL2}. We also used the fact that $\|\varphi_i\|_{L^1} \leq 1$. \\

        In the last expression of \eqref{eq: BL aux temp},  $\prod_i \lambda_i^{-\lambda_i/2}$ can be bounded by $C(k)=e^{k/2}$. This is because $\prod_i\lambda_i^{\lambda_i}$ under the constraints  $\sum_{i=1}^m \lambda_i=\dim(H)=m-k$ and $0 < \lambda_i \leq 1$, is minimized by choosing all $\lambda_i=(m-k)/m$ (see \cite{Ball1989} Proposition 4). \\ 
       The above bound holds for all $(w_*)^{\perp} \in \widetilde{I}$, which means it holds for the $\sup$. Plugging this into \eqref{eq:factorizationDeltai}, we finish the proof of  \eqref{eq: BL main application}, save for the assumption that all $\lambda_i$ are nonzero.  The generalization to the case where some of the $\lambda_i$ are zero is straightforward: in this case $w_i=0$, which simplifies the second line of \eqref{eq:factorizationDeltai}. For each such $i$ we gain a factor of $\|\varphi_i\|_{L^{\infty}}$ which is already accounted for. The rest of the argument remains unchanged by modifying $m$ into $m-r$ with $r$ being the number of $\lambda_i$ that are zero, and redefining $\tilde{I}$ properly.
   \end{proof}
    \subsection{Proof of Factorization Lemma  II}

    \subsubsection{Set of independent random variables}
    Recall the abbreviation $\Sigma_{k-1}=\{\pm 1\}^{k-1}$.  We consider the vector space $\mathbb{R}^{\Sigma_{k-1}}$ with elements $a=(a(\tau))_{\tau \in \Sigma_{k-1}}$ and with canonical basis vectors $e_\tau$, defined by $e_\tau(\tau')=\delta_{\tau,\tau'}$, where $\delta$ denotes the Kronecker delta.  
   In what follows, we fix a $(k,n)-$typical configuration $(\sigma_1,\dots,\sigma_k) \in \Sigma_n^k$. 
    \begin{lemma} \label{lem: indep r.v. contruction}
        There is a random vector $y=(y(\tau)) \in \mathbb{R}^{\Sigma_{k-1}}$, and a set of vectors ${u_1},\dots {u_k} \in  \mathbb{R}^{\Sigma_{k-1}} $ such that 
        \begin{enumerate}
            \item For any $1 \leq j \leq k $,
            \begin{equation} \label{eq:changeofvariable}
    E_{\sigma_{j}}^o= \langle u_j , \underline{y} \rangle_{2^{k-1}}, 
\end{equation}
        where $\langle\cdot,\cdot\rangle_{2^{k-1}}$ denotes the inner product in $\mathbb{R}^{\Sigma_{k-1}}$.
        \item The random variables $(y(\tau))_{\tau\in \Sigma_{k-1}}$ are independent.
         \item Denoting by $S$ the $k \times 2^{k-1}$ matrix constructed by ${u_i}$, we have 
    \begin{equation} \label{eq: ON relation u}
    \langle u_i , u_{i'} \rangle_{2^{k-1}}= 2^{k-1} \delta_{ii'}, \quad \det(SS^{tr}) =2^{k(k-1)}.  
    \end{equation}
        \item Let $P$ be the orthogonal projection onto the subspace $\mathrm{span}({u_1},\dots,{u_k})^{\perp}$. Then we have for any $\tau$: 
        \begin{equation} \label{eq: lambda computation}
            |P{e_{\tau}}|^2= 1-k/2^{k-1}.
        \end{equation}

        \item Recall $\varphi$ as pdf of $h_i$ and let $\varphi^{*m} $ be the $m$-fold convolution of $\varphi$. Denote $\bar{m}:= n/2^{k-1}$ and let
        \begin{equation}
            \epsilon_n:=  \sup_{|m -\bar{m}|<n^{4/5}} \sqrt{n}\|\varphi^{*m} -\mathfrak{g}_{\bar{m}} \|_{L^{\infty}}.
        \end{equation}
        Then $\epsilon_n \to 0$ as $n \to \infty$. Moreover, denoting the pdf of $y(\tau)$ by $\varphi_{\tau}$ we have for all $\tau$, 
        \begin{equation} \label{eq: clt y var}
            \|\varphi_{\tau} -\mathfrak{g}_{\bar{m}} \|_{L^{\infty}} \leq \epsilon_n/\sqrt{n}.
        \end{equation}

        \end{enumerate}
    \end{lemma}
    \begin{proof}
        Given $(\sigma_1,\dots,\sigma_k) \in \Sigma_n^k$, for any $\tau \in \{\pm 1\}^{k-1}$ recall the set $\mathfrak{J}_{\tau}(\underline{\sigma})\equiv \mathfrak{J}_{\tau}$ \eqref{def: J_tau}.  Then for any $\tau \in \{\pm 1 \}^{k-1} $ define: 
        \begin{equation}
    y(\tau) := \sum_{x \in \mathfrak{J}_{\tau}} h_x \sigma_1(x).
    \end{equation}
       Since $\underline{\sigma}$ is typical, all $\mathfrak{J}_{\tau}$ are non-empty. Notice that thanks to the above expression, since  $\sigma_1(x)^2 =1$ i.e., $(\sigma_2(x),\dots,\sigma_k(x))=\sigma_1(x).(\tau_1,\dots,\tau_{k-1})$,  we have:
        \begin{equation} \label{eq: energy change of var}
   E_{\sigma_1}^o=\sum_{\tau} y(\tau), \qquad E_{\sigma_j}^o = \sum_{\tau} y(\tau) \tau(j-1), \qquad \text{if} \: \: j> 1.
\end{equation}
  For any $1 \leq j \leq k$, we may define 
 $(u_j(\tau))_{\tau \in \Sigma_{k-1}} \equiv u_j \in \mathbb{R}^{\Sigma_{k-1}}$ as  follows: 
   \begin{align} \label{eq:changeofvar2}
    \forall \tau \in \Sigma_{k-1}: \quad u_j(\tau)= \tau(j-1), \quad   \quad \text{if} \quad j>1, \quad 
     u_1(\tau)=1.  
\end{align}
    Then \eqref{eq:changeofvariable} (Item 1) follows from \eqref{eq: energy change of var},  \eqref{eq:changeofvar2}.
   For Item 2), the independence follows from the construction: Each $y(\tau)$ depends only on random variables in $\mathfrak{J}_{\tau}$, which is disjoint from all $\mathfrak{J}_{\tau'}$ with $\tau \neq \tau'$. Item 3), i.e.,  \eqref{eq: ON relation u}, follows from \eqref{eq:changeofvar2} by a simple computation. For Item 4, by definition of $\{e_{\tau} \}_{\tau}$ and $u_j$ one can observe that $\langle u_j ,e_{\tau} \rangle$ is either $\pm 1$ for all $j,\tau$. On the other hand, since $\{ u_j\}_j$ are orthogonal, we have  
   $$1-P= \sum_{j=1}^k \frac{1}{\|u_j \|^2}\ket{u_j}\bra{u_j},$$
   where $\ket{u_j}\bra{u_j}$ is the one-dimensional projector associated to $u_j$ and we denoted the identity operator by $1$. Combining this with the above fact and Item 3, conclude \eqref{eq: lambda computation}.



   To tackle the last item, we denote $m_{\tau}:=|\mathfrak{J}_{\tau}|$ for all $\tau$. Recall that the pdf of $h_i$ is $\varphi(x)$, and  that the pdf of $y(\tau)$ is given by $\varphi_{\tau}=\varphi^{*m_{\tau}}$, the $m_{\tau}$-fold convolution of $\varphi$ (since the  pdf of $h_i$ is symmetric around zero). Since $(\sigma_1,\dots,\sigma_k)$ is typical, $|m_{\tau}-\bar{m}|<n^{4/5}$ for all $\tau$. Therefore, we deduce \eqref{eq: clt y var} by definition of $\epsilon_n$. \\
    To prove $\epsilon_n \to 0$, we  first invoke the local CLT estimate \eqref{eq: LCLT1} to get 
    \begin{equation} \label{eq: m_i CLT}
    \sup_{|m-\bar{m}|<n^{4/5}}\|\varphi^{*m} -\mathfrak{g}_{m}\|_{L^{\infty}} =o_n(1/\sqrt{n}).
    \end{equation}
Then, for any $m$ such that $|m-\bar{m}|<n^{4/5}$ we observe uniformly in $y$:
   \begin{equation} \label{eq: m,n difference}
        |\mathfrak{g}_{\bar{m}}(y)-\mathfrak{g}_{m}(y)| \leq 
        C(k)\frac{|m-\bar{m}|}{n\sqrt{n}}.
   \end{equation}
To get the above bound, we add and subtract $\mathfrak{g}_{\bar{m}}(y)\sqrt{\bar{m}/m}$, and we take advantage of the fact that $|1-e^{-x}|\leq x$, and $x \exp(-x) \leq 1$. Combining \eqref{eq: m_i CLT} and \eqref{eq: m,n difference}, we finish the proof of \eqref{eq: clt y var}.

    \end{proof}
    \subsubsection{Proof of Lemma \ref{lem: factorization}}

    \begin{proof}[Proof of Lemma \ref{lem: factorization}]
        Given $(\sigma_1,\dots,\sigma_k) \in \mathcal{T}_{k,n}$, recall the random variables $(y(\tau))_{\tau \in \Sigma_{k-1}}$, and their corresponding pdf $\varphi_{\tau}$ from Lemma \ref{lem: indep r.v. contruction}. Also recall the desired event $\mathcal{E}_{[k]}^o$ from \eqref{eq: abbr event o}. Then, by these definitions, thanks to \eqref{eq:changeofvariable}, we rewrite $\mathbb{P}(\mathcal{E}_{[k]}^o)$ as follows:  
    \begin{equation} \label{eq: factordefp}
        p:= \mathbb{P}\left( \mathcal{E}_{[k]}^o \right) = 
        \int_{\mathbb{R}^{\Sigma_{k-1}}} \chi\left( \bigcap_{i=1}^k \langle u_i,y \rangle \in \mathtt{I}_i\right)
         \prod_{\tau\in \Sigma_{k-1}} \varphi_{\tau} (y(\tau)) dy(\tau),
    \end{equation}
   where ${u_i}$ is given in \eqref{eq:changeofvar2}, and $\langle\cdot,\cdot\rangle$ is the inner product in $\mathbb{R}^{\Sigma_{k-1}}$. Let us use the abbreviation $$\mathcal{U}_{[k]}:= \bigcap_{i=1}^k \langle u_i,y \rangle \in \mathtt{I}_i.$$ 
       Let us equip $\Sigma_{k-1}$ with a \textit{linear order (total order)} $\leq$.  For any $\tau,\tau' \in \Sigma_{k-1}$ we define: 
        \[f_{\tau,\tau'}(\cdot)= \begin{cases}
 \varphi_{\tau}(\cdot), \quad &\text{if } \: \tau'<\tau,\\
    (\varphi_{\tau}(\cdot)-\mathfrak{g}_{\bar{m}}(\cdot))/2 &\text{if} \: \tau=\tau', \\
    \mathfrak{g}_{\bar{m}} (\cdot) &\text{if} \: \tau<\tau'.
\end{cases} \]
  Then by a telescopic sum, we have: 
   
   \begin{align}
  \Delta 
  &:= \Bigl| \, \mathbb{P}(\mathcal{E}_{[k]}^o) 
         - \overbrace{\int_{\mathbb{R}^{\Sigma_{k-1}}} 
             \chi\!\bigl(\mathcal{U}_{[k]}\bigr)\,
             \prod_{\tau \in \Sigma_{k-1}} 
             \mathfrak{g}_{\bar{m}}(y(\tau)) \, dy(\tau)}^{=: \, p_2}
       \Bigr|\nonumber \\
  &= \Biggl| \, 2 \int_{\mathbb{R}^{\Sigma_{k-1}}} 
       \chi\!\bigl(\mathcal{U}_{[k]}\bigr)\,
       \sum_{\tau \in \Sigma_{k-1}} 
         \prod_{\tau' \in \Sigma_{k-1}} 
           f_{\tau,\tau'}(y(\tau')) \, dy(\tau') \Biggr| \nonumber  \\
  &=: \; 2 \Bigl| \sum_{\tau \in \Sigma_{k-1}} \Delta_{\tau} \Bigr| .
  \label{eq:sum2k-1}
\end{align}

   Thanks to Lemma \ref{lem: indep r.v. contruction}, we may apply Proposition \ref{prop: BL main app} to each $|\Delta_{\tau}|$ with the following choices and observations
    \begin{enumerate}
       \item In Proposition \ref{prop: BL main app}, we take $m= 2^{k-1}$ and we identify $\mathbb{R}^{2^{k-1}}$ with $\mathbb{R}^{\Sigma_{k-1}}$. 
       \item We note that for any $\tau, \tau'$, $\|f_{\tau,\tau'} \|_{L^1}\leq 1$. 
       \item By Item 4 of Lemma \ref{lem: indep r.v. contruction} we have $1- \lambda_{\tau'}=k/2^{k-1}$.
        \end{enumerate}  
        Therefore, we  get for any $\tau$:
      $$|\Delta_\tau| \leq C(k) \times \prod_{i=1}^k |\mathtt{I}_i| \times \prod_{\tau'\in \Sigma_{k-1}} \|f_{\tau,\tau'} \|_{L^{\infty}}^{k/2^{k-1}},  $$
      where we used Item 3 of Lemma \ref{lem: indep r.v. contruction} to absorb the determinant $\det(SS^{tr})$ inside $C(k)$. \\
    On the other hand, since $\|\mathfrak{g}_{\bar{m}} \|_{L^{\infty}}=C/\sqrt{n}$, by \eqref{eq: clt y var} we have $\prod_{\tau'} \|f_{\tau,\tau'} \|_{L^{\infty}} \leq C(k) \times \epsilon_n/n^{2^{k-2}}$. Therefore, 
           \begin{equation} \label{eq:delta'estimate}
        |\Delta|  \leq C(k) \times \frac{\tilde{\epsilon}_n}{\sqrt{n^{k}}}  \times
         \prod_{i=1}^k |\mathtt{I}_i|,
    \end{equation}
    where $\tilde{\epsilon}_n:= (\epsilon_n)^{k/2^{k-1}}$.\\
   To conclude the proof of \eqref{eq: factor bound1}, we compute $p_2$ appearing in \eqref{eq:sum2k-1} explicitly. To this end, recall $\{{u_i} \}_{i=1}^k$ \eqref{eq:changeofvar2} and let $H^{\perp}$ be the subspace spanned by these vectors.  For $1 \leq i \leq k$ define ${\tilde{u}_i}:= {u_i}/\|u_i\|_{L^2}$, where we recall $\|u_i \|_{L^2}=2^{(k-1)/2}$. Also, let 
   $$\{{\tilde{u}_i} \}_{i=k+1}^{2^{k-1}} \subset \mathbb{R}^{\Sigma_{k-1}}$$ be an orthonormal basis of $H$. Then, thanks to \eqref{eq: ON relation u}, $\{
   \tilde{u}_i \}_i$ is an orthonormal basis of $\mathbb{R}^{\Sigma_{k-1}}$. Define for any $i$
   $$z_i := \langle \tilde{u}_i , y\rangle,$$
   where $\langle\cdot,\cdot\rangle $ is the inner product in $\mathbb{R}^{\Sigma_{k-1}}$, and $y=(y(\tau))_{\tau\in \Sigma_{k-1}}$. By \eqref{eq: ON relation u}, this change of variables is unitary, i.e., $\langle z,z \rangle= \langle y,y \rangle$. Thanks to the Gaussian structure of $p_2$, with the covariance being a multiple of identity, $p_2$ can be factorized in $z$-representation. Therefore, integrating over $z_i$ for $i>k$ is straightforward and gives identity. Putting all these together, we get 
    \begin{align}
      p_2=\prod_{i=1}^k  \int_{\mathbb{R}} \chi\left(2^{(k-1)/2} z_i \in \mathtt{I}_i \right) \mathfrak{g}_{\bar{m}}(z_i)  dz_i = \prod_{i=1}^k 
      \int_{\mathbb{R}} \chi\left(\tilde{z}_i \in \mathtt{I}_i \right) \mathfrak{g}_{n}(\tilde{z}_i) d \tilde{z}_i =\prod_{i=1}^kG_n^i(\mathtt{I}_i),
         \label{eq:Gaussianfactor0}
   \end{align}
    where we did another change of variable in the second equality $\tilde{z}_i:= 2^{(k-1)/2} z_i$, and we used the fact that $\mathfrak{g}_{n}(\tilde{z}_i)d\tilde{z}_i=\mathfrak{g}_{\bar{m}}(z_i) dz_i$ since $\bar{m}=n/2^{k-1}$.  Recalling the definition of the collection of independent Gaussians $G_n^i$ finishes the proof of Lemma \ref{lem: factorization}, since $\epsilon_n \to 0$
    (and consequently $\tilde{\epsilon}_n \to 0$)
    as $n \to \infty$   thanks to Item 5 of Lemma \ref{lem: indep r.v. contruction}.
    
    \end{proof}

\subsubsection{Proof of Corollary  \ref{cor: factorization with bath} and Lemma \ref{lem: factorization conditional}}
\begin{proof}[Proof of Corollary \ref{cor: factorization with bath}]
    Comparing \eqref{eq: factor bound 3 n_B} from Corollary \ref{cor: factorization with bath} with the result of Lemma \ref{lem: factorization} i.e. \eqref{eq: factor bound1}, it is   clear that the following bound will imply the corollary
    \begin{equation}\label{eq: bound proving corollary twentyfour}
        \left| \prod_{i=1}^k G_n^{i}(\mathtt{I}_i) - \prod_{i=1}^k G_{\breve{n}}^{i}(\mathtt{I}_i)\right| \leq o_n(1) \prod_{i=1}^k |\mathtt{I}_i|/\sqrt{n},
    \end{equation}
    with $o_n(1)$ independent of the intervals. 
    To show the above bound, we note that for any $i$, $G_n^i(\mathtt{I}_i), G_{\breve{n}}^i(\mathtt{I}_i) \leq C/ \sqrt{n} |\mathtt{I}_i|$ ($C$: independent of the interval) since $n+n_B=\breve{n}$ and $n_B=O_n(1)$.  This follows from properties of Gaussians, in particular the fact that $\|\mathfrak{g}_n \|_{L^{\infty}} =1/\mathsf{s}\sqrt{2 \pi n}$. \\
    On the other hand, since $(\breve{n}-n)/n=O_n(1)$, a  similar consideration as in \eqref{eq: m,n difference} shows that $$\|\mathfrak{g}_{n}-\mathfrak{g}_{\breve{n}} \|_{L^{\infty}} \leq O_n(1/\sqrt{n^3})=o_n(1/\sqrt{n}).$$ This means that, for any $i$, $|G_{n}^i(\mathtt{I}_i)-G_{\breve{n}}^i(\mathtt{I}_i)| \leq o_{n}(1/\sqrt{n}) |\mathtt{I}_i|$. \\
    Combining these two observations, we deduce \eqref{eq: bound proving corollary twentyfour} by a simple telescopic sum ($\prod_i a_i -\prod_ib_i= \sum_j \prod_{i<j}a_i (a_j-b_j)\prod_{i>j} b_i$), which concludes the proof.  
\end{proof}

\begin{proof} [Proof of Lemma \ref{lem: factorization conditional}]
Given $\mathtt{t}_1,\dots,\mathtt{t}_k$, denote $\underline{\mathtt{t}}:= (\mathtt{t}_1,\dots,\mathtt{t}_k)$, and 
$\underline{J}:=J_{\mathtt{t}_1} \times\dots \times J_{\mathtt{t}_{k}}$. 
Notice that in a straightforward way, we have: 
\begin{equation} \label{eq: LU bounds}
L:= \inf_{\underline{t} \in \underline{J}} \mathbb{P}\left(\mathcal{E}_{[k]}^{o,\underline{t}}\right) \leq  \mathbb{P}\left( \mathcal{E}_{[k]} \big| \hat{\mathcal{E}}_{[k]} \cap \mathcal{A}_{n_0}\right)\leq \sup_{\underline{t} \in \underline{J}} \mathbb{P}\left(\mathcal{E}_{[k]}^{o,\underline{t}}\right)  =: U,
    \end{equation}
where we used the shorthand notation: 
$$\mathcal{E}^{o,\underline{t}}_{[k]}:= \bigcap_{i=1}^k \sigma.h \in I_i^{t_i} \quad \text{with} \quad I_i^{t_i}=I_i-t_i.$$
On the other hand,  we observe that: 
\begin{equation} \label{eq: Gntt' bound}
   \sup_{\underline{J}}\sup_{\underline{t}, \underline{t'} \in \underline{J}} \left|\prod_{i=1}^k G_{n_1}^i({I}_i^{t_i}) -  \prod_{i=1}^k G_{n_1}^i({I}_i^{t'_i})\right| \leq o_n(1/\sqrt{n^k}) \prod_{i=1}^k |I_i|.
\end{equation}
The above bound is direct. However, we give a very short proof of it in Appendix \ref{sec: app Gaussians}. In \eqref{eq: LU bounds}, we apply Lemma \ref{lem: factorization} to $\mathbb{P}\left(\mathcal{E}_{[k]}^{o,\underline{t}}\right)$ (note that here one should choose $\mathtt{I}_i$ $n$-dependent). Now apply \eqref{eq: Gntt' bound}, and taking $\sup$ and $\inf$ respectively we deduce 
$$|U- \prod_{i=1}^kG_{n_1}^i(I_i^{\mathtt{t}_i})| \leq o_n(1) \prod_{i=1}^k |I_i|/\sqrt{n}, \quad |L- \prod_{i=1}^kG_{n_1}^i(I_i^{\mathtt{t}_i})| \leq o_n(1) \prod_{i=1}^k |I_i|/\sqrt{n}.$$
Combining the above bound with \eqref{eq: LU bounds} finishes the proof.
\end{proof}

\subsection{Estimates for atypical $k$-tuples}

\begin{lemma} \label{lem:carrot}
Let $(\sigma_1,\dots, \sigma_k)\in \Sigma_{n_1}^k$ be {linearly independent} as vectors in $\mathbb{R}^{n_1}$. For any set of arbitrary intervals $\mathtt{I}_1,\dots,\mathtt{I}_k$ we have:
\begin{equation} \label{lemma:carrot0}
    \mathbb{P}\left( \bigcap_{i=1}^k \sigma_i.h \in \mathtt{I}_i \right)  \leq C(k) \prod_{i=1}^k |\mathtt{I}_i| \leq n_1^k C(k) \prod_i|\mathtt{I}_i|.
\end{equation}
In particular, for  a collection of bounded intervals $\{\mathcal{I}_i\}_{i=1}^k$, and any set of points $\{a_i\}_{i=1}^k \subset \mathbb{R}$, consider the rescaled shifted intervals $\{I_i^{a_i}=\sss_n\mathcal{I}_i-a_i\}_{i=1}^k$; then, we have:
\begin{equation} \label{lamme:carrot}
    p:=\mathbb{P}(\bigcap_{i=1}^k \sigma_i.h \in I_i^{a_i}) \leq C(\underline{\mathcal{I}},k) n^k \sss_n^k,
\end{equation}
where $C(\underline{\caI},k)$ depends on intervals $\caI_i$ and $k$, and does not depend on $a_i$.
\end{lemma}
Note that in the latter bound the RHS depends on $n$, whereas in the LHS $\sigma_i \in \Sigma_{n_1}$ is a spin configuration depending on $n_1$, rather than $n$. 
\begin{proof}
The first bound  is a direct consequence of Proposition \ref{prop: BL main app}:  Take ${u_i}=\sigma_i$, and $\varphi_i$ as the pdf of $h_i$. Then recalling the definition of $\lambda_i$ \eqref{eq: def: lambda_i} we see that $\sum_{i}(1-\lambda_i)= \dim(\text{span}(\sigma_1,\dots,\sigma_k))=k$; moreover, since $\|\varphi_i \|_{L^{\infty}} \leq C$,  we observe that $\prod_{i=1}^n \|\varphi_i \|^{1-\lambda_i}_{L^{\infty}} \leq C^k=  O_n(1)$. Moreover, $SS^{tr}$ is an invertible $k \times k$ matrix whose entries are  integers with absolute value at most $n$,  which implies that $1\leq |\det(SS^{tr})| <O_n(n^k)$. This concludes the proof of the first bound thanks to \eqref{eq: BL main application}.  
The second bound follows from the first bound by using $n\geq n_1$ and
$\prod_{i}|I_i^{a_i}| \leq C(\underline{\caI},k) \sss_n^k$, and setting $C(\underline{\caI},k)= C(k) \max_i |\caI_i|^k$, where $C(k)$ featured in the first bound.
\end{proof}
\begin{remark} \label{rmk: typically extension}
    In the above statements, we always assumed that $(\sigma_1,\dots,\sigma_k)$ is a typical $k-$tuple as in Definition \eqref{def: typicality}.   However, we have not used the particular value of the bound $n^{3/4}$ in \eqref{def: typicality}. In fact, from the proof, it is apparent that $n^{3/4}$ can be replaced by any other function of $o_n(n)$ as it appeared in \eqref{eq: m_i CLT}. We will use this subtlety later. 
\end{remark}

\section{Moments: typical cases for semi-perturbed process} \label{sec: factor 2}

\subsection{Resonance events}

Recall $\caR_n$ \eqref{eq: res def} as the set of resonant levels at scale $n$, $\caP_{n,k}$ as the set of patches of size $k$, and $\hat{\mathcal{Z}}_n$ as the set of anti-resonant levels. 
 Fix $\epsilon>0$ and $K \in \mathbb{N}$ and define the events
 \begin{equation}
      \mathcal{A}_n^{\epsilon}:= \{|\caR_n \cup \hat{\mathcal{Z}_n}|/2^n < \epsilon \}, \qquad \mathcal{A}_n^K:=  \{ \forall k>K: \caP_{n,k}=\emptyset  \}, \qquad  \mathcal{A}_{n}^{K,\epsilon} := \mathcal{A}_n^{\epsilon} \cap \mathcal{A}_n^K. 
 \end{equation}
The symbol $\mathcal{A}_{n,m}$ is used in previous sections for the ``no new resonance event". We recycled the symbol $\caA$ in the above definition for another event. This should not cause any confusion since in this section and subsequent sections, ``no new resonance" event will not appear except in Lemma \ref{lem: bound on caa}.
 \begin{lemma} \label{lem: bound on caa}
    \begin{equation}
        \mathbb{P}(\mathcal{A}_{n}^{K,\epsilon}) = 1-o_n(1)- \mathrm{o}_K(1). 
    \end{equation}
    
  \end{lemma}
\begin{proof}
   Recall $\mathcal{A}_{\lfloor\log_2 K \rfloor,\infty}$ as the event where there is no new resonance after scale $\lfloor\log_2K \rfloor$, see Section \ref{sec: non res event}. By definition, for any $n,K \in \mathbb{N}$ we have $ \mathcal{A}_{\lfloor\log_2 K \rfloor,\infty} \subset \mathcal{A}_{n}^K $ since maximal patch size at scale $\lfloor\log_2K \rfloor$ is at most $K$ and  maximal patch size will not increase after scale $\lfloor\log_2 K \rfloor$ in the event $\mathcal{A}_{\lfloor\log_2 K \rfloor,\infty}$.  The above inclusion, combined with Lemma \ref{lem: prob event caa} and the total expectation formula, means 
   \begin{equation} \label{eq: PAnK}
     \mathbb{P}(\mathcal{A}_{n}^K) \geq 1- \mathrm{o}_K(1),
   \end{equation}
for any $n$. From Theorem \ref{thm: dissolution of patches} and Theorem \ref{thm: antires vanish} we deduce that $|\caR_n \cup \hat{\mathcal{Z}}_n|/2^n \to 0 $ almost surely, meaning that for all $\epsilon>0$, $\mathbb{P}(\mathcal{A}_n^{\epsilon})\geq  1-o_n(1)$. Combining this with \eqref{eq: PAnK} and the inequality $\mathbb{P}(A\cap B) \geq \mathbb{P}(A) + \mathbb{P}(B)-1$ concludes the lemma.
\end{proof}

\subsection{Main result}

Recall the notion of an $n$-typical $k$-tuple  from Section \ref{sec: typical def}, and recall the set $\mathcal{T}_{k,n}$ from \eqref{eq: def typical set}. For any $k$-tuple $(\eta_1,\dots,\eta_k)\in \Sigma_n^k$, we split each $\eta_i=(\mu_i,\sigma_i) \in \Sigma_{n_0} \times \Sigma_{n_1}$ as in Section \ref{sec: spliting}. Then we define: 
\begin{equation} \label{eq: def: D_k}
D_k:=\{ (\eta_1,\dots, \eta_k) \in \Sigma_n^k | \: (\sigma_1,\dots, \sigma_k ) \in \mathcal{T}_{k,n_1}, \: (\mu_1,\dots,\mu_k) \in \mathcal{T}_{k,n_0}\}.
\end{equation}
From now on, we will use the abbreviated notation 
\begin{equation} \label{eq: notation expectation res}
    \mathbb{P}_{\caA}(\cdot):= \mathbb{P}(\cdot \, \cap \mathcal{A}_{n_0}^{K,\epsilon}),\qquad    \mathbb{E}_{\caA}(\cdot):=   \mathbb{E}(\cdot \, \chi(\mathcal{A}_{n_0}^{K,\epsilon})) .
\end{equation}
where $\chi(E)$ is the indicator function of an event $E$.

Let $\{\mathcal{I}_i\}_{i=1}^k$ be a collection of bounded macroscopic intervals as in \ref{sec: intervals!}, and let $\{I_i\}_{i=1}^k$ be the collection of corresponding rescaled intervals. Also, recall the semi-perturbed eigenvalues $\widetilde{E}_{\eta}$ from \eqref{eq: tilde def}. Then the main  result of the present section reads: 
\begin{lemma} \label{lem: moments typical case} 
\begin{equation} \label{eq: kmomment1}
   \sum_{\underline{\eta} \in D_k} 
    \mathbb{P}_{\mathcal{A}}\left(\bigcap_{i=1}^k 
   \widetilde{E}_{\eta_i} \in I_i \right)  = \prod_{i=1}^k
    |\mathcal{I}_i| \left(1+o_n(1)+o_K(1) \right).
\end{equation}

\end{lemma}
The {main idea} of the proof is as follows: $E_{\mu_i}$ and $\sigma_i.h$ are independent. Moreover, by the results in Section \ref{sec: factor1}, for a typical $\underline{\sigma}:= (\sigma_1,\dots,\sigma_k)$,
 $\underline{\sigma}$ can be approximated by a collection of $k$ independent Gaussians with variance of order $n_1$. Therefore, in view of $\underline{\sigma}$, changing the value of $E_{\mu_i}$ by  $n_0^\rho$ with $0<\rho<\frac12$ will have a \textit{sub-leading effect}. This means that $E_{\mu}+\sigma.h$ is well approximated by a Gaussian with variance of order $n$. We devote the next section to this proof. 

\subsection{Proof of Lemma \ref{lem: moments typical case}}
\subsubsection{Preliminaries}
In this section, we state a basic lemma involving simple computations concerning Gaussian random variables. Recall ``long intervals" from \ref{sec: intervals!} $\{J_{\mathtt{t}}\}_{\mathtt{t}\in \mathcal{J}_{n_0}}$.  We also abbreviate $\sum_{\mathtt{t}\in \mathcal{J}_{n_0}}=: \sum_{\mathtt{t}}$
\begin{lemma} \label{lem: dis convolution}
For any macroscopic interval $\mathcal{I}$, let   $I$ be its rescaled counterpart, then
\begin{align} 
   |G_n(I)-|I|/\sqrt{2\pi n \mathsf{s}^2 }| \leq |I|o_n(1/\sqrt{n}), \label{eq: max entropy Gaussian} \\
    \left|\sum_{\mathtt{t}} G_{n_1}(I^{\mathtt{t}}) G_{n_0}(J_{\mathtt{t}}) - G_{n_0+n_1}(I)\right| \leq |I| o_n(1/\sqrt{n}), \label{eq: dis convolution}\\
       n_0^{\rho}/\sqrt{n}  \sum_{\mathtt{t}} G_{n_1}(I^{\mathtt{t}}) \leq C  |I|/\sqrt{n} \label{eq: simple gaussina bound}.
\end{align}
where the error term $o_n(1/\sqrt{n})$ is independent of  $I$.

\end{lemma}
In the above lemma, \eqref{eq: max entropy Gaussian} is rather direct, \eqref{eq: dis convolution} is a simple discretization of the convolution of two Gaussians, and \eqref{eq: simple gaussina bound}  is a Riemann sum error computation. Although this lemma is direct and only involves basic Gaussian computations, we give a short proof in Appendix \ref{sec: app Gaussians}, for  reader's convenience. 

\subsubsection{Main part of the proof}
\begin{proof} [Proof of Lemma \ref{lem: moments typical case}]
\textbf{Step 1.} Given $(\underline{\mu},\underline{\sigma})\in D_k$, recall $\{J_{\underline{\mathtt{t}}} \}_{\underline{\mathtt{t}}\in \mathcal{J}_{n_0}^k}$ and the events $\hat{\mathcal{E}}_{[k]}$ \eqref{eq: abbr E J_t}, and $\mathcal{E}_{[k]}$. We make the dependence of the former event on $\underline{\mathtt{t}}$ more explicit by writing $\hat{\mathcal{E}}_{[k]}(\underline{\mathtt{t}})$.  By the total probability formula, we have:
\begin{equation} \label{eq: total prob conditional}
    p(\underline{\mu},\underline{\sigma}):=\mathbb{P}_{\mathcal{A}}\left(\bigcap_{i=1}^k \widetilde{E}_{\eta_i}  \in I_i \right) = \sum_{\underline{\mathtt{t}}} \mathbb{P}\left(\mathcal{E}_{[k]}\big|\hat{\mathcal{E}}_{[k]}(\underline{\mathtt{t}})\cap \mathcal{A}\right) \mathbb{P}_{\mathcal{A}}\left( \hat{\mathcal{E}}_{[k]}(\underline{\mathtt{t}})\right),
\end{equation}
where only terms with nonzero $\mathbb{P}_{\caA}(\cdot)$ should be considered.
Since $\sum_{\mathtt{t}} \mathbb{P}_{\mathcal{A}}(\cdots)=\mathbb{P}(\mathcal{A})\leq 1$, by \eqref{eq:Factorizationestimate1}
 from Lemma \ref{lem: factorization conditional}, we perform an $L^1-L^{\infty}$ estimate and get
 \begin{equation} \label{eq: k moment L1 Linfty1}
\left|p(\underline{\mu},\underline{\sigma})-\sum_{\underline{\mathtt{t}}} \prod_{i=1}^kG_{n_1}^i(I_i^{\mathtt{t}_i}) \mathbb{P}_{\mathcal{A}}\left(\hat{\mathcal{E}}_{[k]}(\underline{\mathtt{t}})\right) \right| \leq o_n(1) \prod_{i=1}^k |I_i|/\sqrt{n},
 \end{equation}
 where the error is independent of $I_i$,  $\underline{\sigma},\underline{\mu}$.
    For any $\underline{\mu}$, let us denote 
 \begin{equation} \label{def: temp p1}
p_1(\underline{\mu}):=\sum_{\underline{\mathtt{t}}} \prod_{i=1}^kG_{n_1}^i(I_i^{\mathtt{t}_i}) \mathbb{P}\left(\hat{\mathcal{E}}_{[k]}(\underline{\mathtt{t}})\right).
 \end{equation}
 Note that there is no conditioning on $\caA$ anymore.  Then
\begin{equation} \label{eq: p p1 error}
    \sum_{(\underline{\mu},\underline{\sigma})\in D_k} \left| p_1(\underline{\mu})-p(\underline{\mu},\underline{\sigma})\right| \leq C(\mathbb{P}(\mathcal{A}^c)+o_n(1)) \prod_i|\mathcal{I}_i|.
\end{equation} 
 To  get the above estimate, we compare $p_1(\underline{\mu})$ with the sum appeared in the LHS of \eqref{eq: k moment L1 Linfty1} as  follows: First, we note that $\mathbb{P}(\cdot)-\mathbb{P}_{\mathcal{A}}(\cdot)\geq 0$, and we perform an $L^1-L^{\infty}$ bound for $\sum_{\mathtt{t}} \cdots$. Then we bound 
  $|\prod_i G_{n_1}^i(I_i^{\mathtt{t}_i})| \leq  C \prod_i |I_i|/\sqrt{n}$ by properties of Gaussians and the definition of $n,n_1$. Then, by the total probability formula,  we observe that:
  $$\sum_{\underline{\mathtt{t}}} \mathbb{P}(\hat{\mathcal{E}}_{[k]})-\mathbb{P}_{\mathcal{A}}(\hat{\mathcal{E}}_{[k]})=\mathbb{P}(\mathcal{A}^c).$$
 Finally, we take advantage of \eqref{eq: k moment L1 Linfty1} and we use the definition of $|I_i|=\sss_n|\mathcal{I}_i|$ with $\sss_n\leq C \sqrt{n}/2^n$, and  using that $|D_k| \leq 2^{nk}$. \\
 Note that we did not take advantage of the conditioning on $\caA $, and this is the only place where such a conditioning appears. In the absence of this conditioning, the error term $\mathbb{P}(\caA^c)$ would be absent. \\
 
\textbf{Step 2.} In this step,  we show  
\begin{equation} \label{eq: factorization temp}
   \sup_{\underline{\mu} \in \mathcal{T}_{k,n_0}} \sup_{\underline{\mathtt{t}}} \left|\mathbb{P}\left(\hat{\mathcal{E}}_{[k]}(\underline{\mathtt{t}}) \right) -   \prod_{i=1}^k G_{n_0}^i(J_{\mathtt{t}_i})\right| \leq o_n\left(n^{k\rho}/\sqrt{n^k}
   \right).
\end{equation}
For any $\mu_i \in \Sigma_{n_0}$, we slightly abuse the notation and denote (only here): 
$$E^o_{\mu_i}:= \sum_{x>n_B}^{n_0} \mu_i(x) h_x.$$
Let us mention a straightforward observation: there exists an absolute constant ${C}_{\mathcal{\mathcal{B}}}>0$ (independent of $n_0$) such that for any $\mu_i \in \Sigma_{n_0}$ we have: $|E_{\mu_i}^o-E_{\mu_i}|< C_{\mathcal{B}}$.
This is thanks to Lemma \ref{lemma : labeling lemma} and the fact that $\|H_B \|<C$ with $C>0$ being an absolute constant. 

Then, we denote for any $\mathtt{t}_i \in \mathcal{J}_{n_0}$
$$J_{\mathtt{t}_i}^{\pm}:= [\mathtt{t}_i \mp 2C_{\mathcal{B}}, \mathtt{t}_i+n_0^{\rho}\pm 2C_{\mathcal{B}}), \quad \mathcal{E}^{\pm}_{[k]}(\underline{\mathtt{t}}):= \bigcap_{i=1}^k E_{\mu_i}^o \in J_{\mathtt{t}_i}^{\pm}. $$
Given the fact that $|E_{\mu_i}^o- E_{\mu_i}|< C_{\mathcal{B}}$, we have:

\begin{equation} \label{eq:tirivial bound 22}
    B_{l}:=\mathbb{P}\left(\mathcal{E}^-_{[k]}(\underline{\mathtt{t}})  \right) \leq \mathbb{P}\left(\hat{\mathcal{E}}_{[k]}(\underline{\mathtt{t}}) \right)  \leq  \mathbb{P}\left(\mathcal{E}^+_{[k]}(\underline{\mathtt{t}})  \right) =:B_{u}.
\end{equation}
For any $i$ (given the relation between $n_0,n$, and basic Gaussian properties) we know $|G_{n_0}^i(J_{\mathtt{t}_i}^{\pm})| \leq (C/\sqrt{n}) \times n^{\rho}$. Similarly,  $$|G_{n_0}^i(J_{\mathtt{t}_i}^{\pm})-G_{n_0}^i(J_{\mathtt{t}_i})| \leq C/\sqrt{n}=n^{\rho} o_n(1/\sqrt{n}).$$ Combining this relation with Corollary \ref{cor: factorization with bath}, i.e.\ the bound \eqref{eq: factor bound 3 n_B}, we get
$$\left| B_{\sharp} -\prod_{i=1}^k G_{n_0}^i(J_{\mathtt{t}_i})\right| \leq o_n(1)n^{k\rho}/\sqrt{n^k}, \quad \text{for} \quad \sharp=l,u,$$
where the error term is uniform in $\underline{\mathtt{t}}, \underline{\mu} \in \mathcal{T}_{k,n_0}$. 
To obtain the above bound, we used Corollary \ref{cor: factorization with bath} and  Remark \ref{rmk: typically extension}. Combining the above bound with \eqref{eq:tirivial bound 22}, thanks to the fact that error bounds are uniform in $\underline{\mu}, \underline{\mathtt{t}}$,  we deduce the desired estimate \eqref{eq: factorization temp}.\\

 \textbf{Step 3.} Plugging the estimate \eqref{eq: factorization temp} inside $p_1(\underline{\mu})$ from \eqref{def: temp p1} we get: 
\begin{align}
\sum_{(\underline{\mu},\underline{\sigma}) \in D_k}\left| p_1(\underline{\mu}) -
\sum_{\underline{\mathtt{t}}} \prod_{i=1}^k G_{n_1}^i(I_i^{\mathtt{t}_i}) G_{n_0}^i(J_{\mathtt{t}_i})\right| &= o_n(1) |D_k| \sum_{\underline{\mathtt{t}}} \prod_{i=1}^k G_{n_1}^i(I_i^{\mathtt{t}_i}) n^{\rho}/\sqrt{n} \nonumber  \\ &\leq  o_n(1)|D_k|\prod_{i=1}^k |I_i|/\sqrt{n} \nonumber \\ 
& \leq o_n(1) \prod_{i=1}^k |\mathcal{I}_i| \label{eq: p1 estimate},
\end{align}
where in the second line we used the simple Gaussian bound \eqref{eq: simple gaussina bound} from Lemma \ref{lem: dis convolution}, and in the last line we used the bound $\prod_i|I_i||D_k|\leq C\prod_i|\mathcal{I}_i|\sqrt{n}$ by the definition of $I_i$ and  $D_k$. \\
Combining \eqref{eq: p1 estimate}, with  the discrete convolution estimate \eqref{eq: dis convolution},  realizing that $G_{n_0+n_1}(I_i^{\mathtt{t}_i})\leq C |I_i|/\sqrt{n}$, and using  the definition of the rescaled interval and $D_k$ as before, we get: 

\begin{equation} \label{eq: p(mu sigma) bound}    \sum_{(\underline{\mu},\underline{\sigma})\in D_k}\left|p_1(\underline{\mu})- \prod_{i=1}^k G_{n}^i(I_i)\right| \leq o_n(1)\prod_{i=1}^k|\mathcal{I}_i|.
\end{equation}

 Thanks to \eqref{eq: max entropy Gaussian}, large deviation estimate \eqref{eq:countingII}, and definition of $I_i$ 
($|I_i| =|\mathcal{I}_i| \sss_n$, $\mathcal{I}_i$ being bounded), by a simple computation, we get: 
$$\sum_{(\underline{\mu},\underline{\sigma})\in D_k} \prod_{i=1}^kG_n^i(I_i)=\prod_{i=1}^k |\mathcal{I}_i|(1 + o_n(1)).$$
Combining the above bound with  \eqref{eq: p(mu sigma) bound}, \eqref{eq: p p1 error}, using the estimate $\mathbb{P}(\mathcal{A}^c)=o_n(1)+o_K(1)$ from Lemma  \ref{lem: bound on caa},  finishes the proof of \eqref{eq: kmomment1}. 
\end{proof}

\subsection{Density of States}

For any macroscopic interval $\mathcal{I} $, recall the rescaled interval $I:=\sss_n \mathcal{I}$.  Recall $\{ E_{\mu} \}_{\mu \in \Sigma_n}$ as eigenvalues of $H_n$. Then, we define:  
\begin{equation} \label{eq: def N_n}
    N_n(\mathcal{I}):= |\{\eta| E_{\eta} \in \sss_n\mathcal{I} \}|.
\end{equation}
\begin{lemma} \label{lem: a priori bound}

 For any bounded interval $\mathcal{I}$, we have:
\begin{equation} \label{eq:Firstmoment}
    \lim_{n\to \infty} \mathbb{E}(N_n(\mathcal{I})) = |\mathcal{I}|.
\end{equation}
\end{lemma}
\begin{proof}
We first highlight that the proof of Lemma \ref{lem: moments typical case} works identically if we replace $\mathbb{P}_{\caA}$ with $\mathbb{P}$ in \eqref{eq: kmomment1}, as we mentioned in the bulk of the proof. In fact, the latter case is even easier, and the error term will only depend on $n$: $o_n(1)$. Moreover, if $k=1$, then $D_1=\Sigma_n$. All this means: 
    \begin{equation} \label{eq:first moment semi}
        \sum_{\eta \in \Sigma_n} \mathbb{P}(\widetilde{E}_{\eta} \in (\sss_n \mathcal{I})) = |\mathcal{I}|(1+o_n(1)).
    \end{equation}
     Fix $\epsilon>0$, and let $\mathcal{I}_- \subset \mathcal{I} \subset \mathcal{I}_+$ be such that $|\mathcal{I}_{\pm}|=|\mathcal{I}|\pm \epsilon$ and $\inf \mathcal{I}_{\pm}= \inf\mathcal{I}\mp \epsilon/2$. We let $$I_{\pm}= \sss_n\mathcal{I}_{\pm}, \quad I=\sss_n\mathcal{I}.$$ Recall that for $\eta=(\mu,\sigma)$,  $\widetilde{E}_{\eta}={E}_{\mu} + \sigma.h $. We also recall that  $|\widetilde{E}_{\eta} -E_{\eta}| < C \alpha^{n_0}$ from \eqref{eq: tilde bar relation}. Given  \eqref{eq: n_0 n scale}, \eqref{eq: n_0 n_1 split 0} (i.e. $\alpha^{n_0} \ll \sss_n$), we deduce that for  $n$ large enough: 
    $$\mathbb{P}(\widetilde{E}_{\eta} \in I_-) \leq \mathbb{P}(E_{\eta} \in I) \leq \mathbb{P}(\widetilde{E}_{\eta} \in I_+)$$
   The  above inequality as well as \eqref{eq:first moment semi} means $\sum_{\eta} \mathbb{P}({E}_{\eta} \in I)$ is  lower and upper bounded by $(|\mathcal{I}|-\epsilon)(1+o_n(1))$ and $(|\mathcal{I}|+\epsilon)(1+o_n(1))$. This concludes the proof, since this is true for any $\epsilon>0$, and $\mathcal{I}$ is bounded.
\end{proof}
Let us restate the content of the above lemma in another form; such an explicit form will be useful later.  Let us denote: 
\begin{equation} \label{eq: DOS tilde approx}
  \widetilde{N}_n(\mathcal{I})= \sum_{\eta \in \Sigma_n}  \chi(\widetilde{E}_{\eta} \in \sss_n\mathcal{I}). 
\end{equation}
Notice the difference between $\widetilde{N}_n$ and $N_n$ \eqref{eq: def N_n} where we replaced the actual levels $E_{\eta}$ with the approximate ones $\widetilde{E}_{\eta}$.
Then we have: 
\begin{corollary} \label{cor: uniform bound first moment}
    For any bounded interval $\mathcal{I}$ we have: 
    \begin{equation} \label{eq: DOS approx limit}
Moments: atypical cases for semiperturbed process    \end{equation}
    Moreover, for any bounded interval $\mathcal{I}$ we have: 
    \begin{equation} \label{eq: first moment bound}
        \mathbb{E}(\widetilde{N}_n(\mathcal{I})), \mathbb{E}({N}_n(\mathcal{I}))
        \leq C(\mathcal{I}),
    \end{equation}
    uniformly in $n$.
\end{corollary}
\begin{proof}
\eqref{eq: DOS approx limit} is a consequence of \eqref{eq:first moment semi} stated in the proof of the above lemma. \eqref{eq: first moment bound} is deduced from \eqref{eq: DOS approx limit}, \eqref{eq:Firstmoment}. 
\end{proof}
\section{Moments: atypical cases for semiperturbed process}
\label{sec:kth moment main}
Recall the abbreviation $[k]=\{1,\ldots,k\}$.

\subsection{Classes of multiconfigurations}\label{sec: classes of multiconfigs}

We will define a covering of $\Sigma_{n}^{k} $ by classes $\caC_{n}$.
To do that, we first construct abstract classes $\caC$:
An abstract class $\caC$ is
specified by
\begin{enumerate}
    \item A partition of  $[k]$ into two sets $D, F$, i.e.\ with 
    $D\cap F=\emptyset$, and $D\cup F=[k]$, with $|D|<k$.
    \item   A  $|D|\times |F|$ real matrix $v=(v_{i,j})_{i \in D, j\in F}$. In the case $|D|=0$, we set $v=0$.
\item A binary variable $\varsigma \in \{0,1\}$. 
\end{enumerate}
Whenever necessary, we write $\caC=\caC(D,F,v,\varsigma)$. 
Given an abstract class $\caC$, we define the class $\caC_n=\caC_n(D,F,v,\varsigma)$ as a subset of $\Sigma_n^k$, in the following way:
Given $\underline{\sigma} \in \Sigma^k_n$, we write $\sigma_F$ for the collection $(\sigma_i)_{i\in F}$ and $\sigma_D$ for the collection $(\sigma_i)_{i\in D}$. Then, $\caC_n(D,F,v,\varsigma)$ is defined as the set of $\underline{\sigma} \in \Sigma_n^k$ satisfying 
\begin{enumerate}
    \item $\sigma_F$ is linearly independent (when viewed as a $|F|$-tuple of elements in $\mathbb{R}^n$). If $\varsigma=1$, $\sigma_F$ is $(|F|,n)$-typical, as defined in \eqref{def: typicality}.   If $\varsigma=0$, it is non-typical.
    \item For all $i\in D$:  $\sigma_{i}=\sum_{j \in F}v_{i,j}\sigma_j$. In case $|D|=0$, this condition is absent.
\end{enumerate}


We note that the set of classes $\caC_n$ is a covering of $\Sigma_{n}^k$, but the classes are not disjoint. 
At first sight, one might think that the number of classes is uncountable, because any choice of a real matrix $v$ corresponds to a different class. However, it turns out that all but a finite number of matrices $v$ yield empty classes.

\begin{lemma}\label{lem: coeff in module group}
There is a constant $C(k)$ such that the following holds: 
If a class $\caC_n=\caC_n(D,F,v,\varsigma)$ is  non-empty, then all entries of $v$ are of the form  $q_1/q_2$ with $q_1,q_2\in \mathbb Z$ and $|q_1|,|q_2| \leq C(k)$.
\end{lemma}

\begin{proof}[Proof of Lemma \ref{lem: coeff in module group}]
Choose an $F$-tuple $\sigma_j, j \in F$ in the non-empty class. It is linearly independent by definition. Since the column rank of a matrix equals its row rank, one can find an $|F|$-tuple of sites $x_i, i=1,\ldots, |F|$ such that the square $|F| \times |F|$ matrix
 $$
 M_{i,j}=\sigma_j(x_i)
 $$
 is non-singular. 
 This means that, for any $z \in D$, the vector $(v_{z,j})_{j \in F}$ is the unique solution of 
 $$
 \sum_{j \in F} M_{i,j} v_{z,j} = \sigma_{z}(x_i) \quad \forall i.
 $$
 By basic algebra, we can construct this solution. All the entries of the matrix $M_{i,j}$ are either $\pm1 $ by construction. Therefore, the determinant of this matrix and all of its minors are integers whose absolute values are between $1$ and $|F|! \leq k!$. Hence the solution is of the claimed form. 
\end{proof}

{Naively, the number of multiconfigurations $\underline{\sigma}$ in a class $\caC$, is at most $|\Sigma_n^{F}|=2^{n|F|}$. However, for most classes, the number of multiconfigurations is exponentially smaller, as we show now:}

\begin{lemma}\label{lem: counting} Consider an abstract class $\caC(D,F,v,\varsigma)$ with $|D|>0$, such that there is  at least one $i \in D$ such that  $v_{i,j}$ is nonzero for at least two values of $j\in F$, then 
\begin{equation} \label{eq: ld bound}
   |\caC_n(D,F,v,\varsigma) | \leq 
   (2^{|F|}-1)^n =  
  (1-2^{-|F|})^n |\Sigma_{n}^{F}|.  
\end{equation}
\end{lemma}

\begin{proof}
Let us abbreviate $w_j=v_{i,j}$ for (one of) the $i$ as in the statement, {and all $j \in F$}.

For any $x\in [n]$, introduce $\tau_x \in \{\pm 1\}^F$  by 
$$\tau_x(j)=\sigma_j(x), \qquad j\in F.$$

We argue that there is at least one element in $\{\pm 1\}^F$  that does not occur, i.e.\ there is no $x$ such that $\tau_x$ equals that element.  
First, note that for any site $x$, 
\begin{equation}\label{eq: basic restriction config}
   (\sum_{j \in F}  w_j \tau_x(j))^2=(\sum_{j \in A} w_j-\sum_{j\in A^c}  w_j)^2= 
  {(\sigma_i(x))^2}=1, 
\end{equation}
where $A=A_x=\{j\in F:  \tau_x(j)=1\}$ and $A^c=F\setminus A$. 
 Assume now that for any subset $A \subset F$, there is a $x_A \in[n]$ such that 
$\tau_{x_A}(j)=\chi(j\in A)-\chi(j\in A^c)$.  We will show that this leads to a contradiction. 
Applying \eqref{eq: basic restriction config} once at $x_A$ and once at $x_F$, we get  
\begin{align}\label{eq: basic restriction config2}
 0&= (\sum_{j \in A} w_j-\sum_{j\in A^c}  w_j)^2 -    (\sum_{j \in A} w_j+\sum_{j\in A^c}  w_j)^2  \nonumber \\
 &=  -4(\sum_{j \in A} w_j) (\sum_{j\in A^c}  w_j).
\end{align}
Since this has to hold for any choice of $A$, we conclude that there is at most one $j\in F$ such that $w_j\neq 0$. We have thus arrived at a contradiction given the initial assumption on $\mathcal{C}$. 
We conclude hence that there is at least one element in $\{\pm 1\}^F$ that is not realized at any $x$. This means that the number of multiconfigurations $\tau$ (which is equal to the number of multiconfigurations $\sigma$) in the class is bounded by  $(2^{|F|}-1)^n$.

\end{proof}

\subsection{Leading contribution to moments}\label{sec: kth moment}

Let us recall from Section \ref{sec: spliting}:  we write  $n=n_0+n_1$, and $\eta=(\mu,\sigma)$ for elements in $\Sigma_{n}$, with $\mu \in \Sigma_{n_0},\sigma\in \Sigma_{n_1}$.  We will often deal with $k$-tuples of $\eta$ and we denote them by $\underline{\eta}$ and we write $\underline{\eta}=(\eta_i)_{i\in [k]}$. \\

Let $\{\mathcal{I}_{i}\}_{i=1}^k$ be a collection of bounded intervals in $\mathbb{R}$. Let $\{I_i\}_{i=1}^k$ be the corresponding rescaled intervals (see Section \ref{sec: intervals!}: $I_i=\sss_n\mathcal{I}_i$). For a given $\underline{\eta} \in \Sigma_n^k$, recall $\widetilde{E}_{\eta}$ \eqref{eq: tilde def}, and   the events 
$$
\mathcal E_i = \{\widetilde{E}_{\eta_i} \in I_i\},
$$
and also 
$$
\caE_{[k]}= \bigcap_{i=1}^k 
    \caE_i.
$$
from \eqref{eq: mathcal abbr}. \\

Let an abstract class $\caC=\caC(D,F,v,\varsigma)$ be  \emph{simple} iff.
\begin{enumerate}
    \item $\varsigma=1$
    \item For any $i \in D$, there is a unique $j(i) \in F$, such that $v_{i,j}=\delta_{j,j(i)}$ {$\forall j \in F$}. Here, $\delta_{j,j'}$ is the Kronecker delta: $\delta_{j,j'}=1$ iff. $j=j'$. This is in particular satisfied if $|D|=0$.  
\end{enumerate}

For any $\epsilon>0$, and $K \in \mathbb{N}$, recall the event $\mathcal{A}=\mathcal{A}_{n_0}^{K,\epsilon}$, and the abbreviation $\mathbb{P}_{\caA}(\cdot)= \mathbb{P}(\cdot \cap \mathcal{A})$. Then we have 
\begin{proposition}\label{lem: reduction to pairings}
Given $(F,D,v,\varsigma)$, if the abstract class $\caC=\caC(D,F,v,\varsigma)$ is not \emph{simple}, then 
  \begin{equation} \label{eq:pk}
   \left| \sum_{(\underline{\mu},\underline{\sigma}) \in \Sigma_{n_0}^k \times \mathcal C_{n_1}} 
    \mathbb{P}_{\caA}\left( 
    \caE_{[k]} \right) \right| \leq  \epsilon C(K,k) + o_n(1) .
\end{equation}  
For a simple abstract class $\caC$, we have 
  \begin{equation} \label{eq:pk2}
   \left| \sum_{(\underline{\mu},\underline{\sigma}) \in \Sigma_{n_0}^k \times \mathcal C_{n_1}} 
    \mathbb{P}_{\caA}\left( 
    \caE_{[k]}    \right)  -   \sum_{(\underline{\mu},\underline{\sigma}) \in \caC_{n_0} \times \mathcal C_{n_1}} 
    \mathbb{P}_{\caA}\left( 
    \caE_{[k]}    \right)
    \right| \leq  \epsilon C(K,k) + o_n(1).
\end{equation} 

\end{proposition}

The above proposition is phrased in a way that is most suitable for its proof, which is postponed to the next sections.  We now recast the proposition in a way that is easy to use later. 
 Let  $\caP$ be  a partition of the index set $[k]$, i.e.\ $\caP=\{A_1,\ldots, A_p\}$ where $A_{1,\ldots,p}$ are disjoint non-empty subsets of $[k]$, such that $\cup_{i=1}^p A_i=[k]$. Since in this section there is no trace of resonant patches, this notation should not cause any confusion. We denote by $|\caP|=p$ the cardinality of the partition.
To any partition $\caP$, we associate a subset $\widetilde\caP \subset \Sigma_n^k$ as follows: $(\eta_1,\ldots,\eta_k)$ is in $\widetilde\caP$ iff.\ the following two conditions hold
\begin{enumerate}
    \item If $j,j' \in A_i$, then $\eta_j=\eta_{j'}$. 
    \item For any $i_1 \in A_1, i_2 \in A_2\ldots, i_p \in A_{p}$, the $p$-tuple $(\eta_{i_1},\ldots, \eta_{i_p}) \in D_{|\mathcal{P}|}$, where for any $p$, $D_p$ is defined in \eqref{eq: def: D_k}.
\end{enumerate}
Notice that $\widetilde{\mathcal{P}}$ is well-defined thanks to Remark \ref{rmk: typical well defined}.

The connection with the formalism used above is that, for a simple abstract class $\caC$, there is a unique partition $\caP$ such that
\begin{equation}\label{eq: association partition and class}
   \caC_{n_0} \times \caC_{n_1} =\widetilde\caP, 
\end{equation}

where the partition $\caP$ is defined as follows from the data $(F,D,v,\varsigma)$: $\caP$ consists of $p=|F|$ sets;   every element $ i \in F$ sits in a different partitioning set $A_{i'}$, and the other elements of $A_{i'}$ are the elements $j$ of $D$ such that $v_{i,j}=1$. 
For a given simple class $\caC_{n_1}$, there is  indeed a unique partition satisfying this. 
Conversely, for a given partition $\caP$, the number of classes $\caC_{n_1}$ such that \eqref{eq: association partition and class} holds, equals $|A_1|\times |A_2| \times \ldots \times |A_p|$: This is  the number of ways to choose an element from each of the partitioning sets; this choice will construct the set $F$. Fixing $F$, $\caC_{n_1}$ is determined uniquely by \eqref{eq: association partition and class}.   Indeed, for all these classes, the sets $\caC_{n_0}\times \caC_{n_1}$ are simply the same. 
Additionally, for any simple class $\caC'_{n_1}$ that does not satisfy \eqref{eq: association partition and class} for a given partition $\caP$, we have 
$$
\caC_{n_0}\times \caC_{n_1} \cap \widetilde
\caP =\emptyset,
$$
since for any $(p,n_1)$- typical configuration $(\sigma_{i_1},\dots,\sigma_{i_p})$, by definition, $\sigma_{i_j}\neq \sigma_{i_{j'}}$ whenever $i_j\neq i_{j'}$.  We mentioned that classes form a cover of $\Sigma_n^k$, i.e., two classes are not necessarily disjoint. However,  notice that if $\underline{\mu}$ belongs to a simple class $\caC_{n_0}$ then it cannot belong to another non-simple class. Same holds for any $\underline{\sigma}$ belonging to a simple class $\caC_{n_1}$. The above remarks can be summarized into the following simple observation:  
\begin{equation} \label{eq: partition classes}
    \Sigma_{n_1}^k=  \left(\bigcup_{\caC :\text{non-simple}} \caC_{n_1} \right) \bigsqcup \left(\bigcup_{\caP} \widetilde{\caP}_1
    \right),
\end{equation} \\
where for any partition $\caP$, $\widetilde{\caP}_1 \subset \Sigma_{n_1}^k$ is constructed from $\widetilde{\caP}$ as follows: $\underline{\sigma} \in \widetilde{\caP}_1$ iff. for some $\underline{\mu} \in \Sigma_{n_0}^k$, $(\underline{\mu},\underline{\sigma}) \in \widetilde{\caP}$;  moreover, $\sqcup$ denotes a disjoint union, i.e., the corresponding sets are disjoint.


Combining these remarks, we see that Proposition \ref{lem: reduction to pairings}  establishes 
\begin{corollary}\label{cor: pairings}
   \begin{equation} \label{eq: cor:pairings}
       \left| \sum_{\underline{\eta} \in \Sigma_n^k} 
    \mathbb{P}_{\caA}\left( 
    \caE_{[k]} \right)  -\sum_{\caP}   \sum_{\underline{\eta} \in \widetilde \caP} 
    \mathbb{P}_{\caA}\left( 
    \caE_{[k]} \right)     \right|  \leq o_n(1)+ \epsilon C(K,k).
\end{equation} 
\end{corollary}
This corollary will be crucial in Section \ref{sec: computation of moments}. 
\begin{proof}
    For a non-simple abstract class $\caC$, denote the LHS of \eqref{eq:pk} by $\Delta_{n_1}(\caC)$; to any simple abstract class $\caC$ we can associate a unique partition of $[k]$ via \eqref{eq: association partition and class}. Notice that if two abstract class $\caC\neq\caC'$ correspond to the same partition $\caP$, then the LHS of  
  \eqref{eq:pk2} 
   will be identical for $\caC$ and $\caC'$. Therefore, for any abstract class $\caC$, we denote the LHS of \eqref{eq:pk2} by  $\Delta_n(\caP)$, where $\caP$ is the partition associated to $\caC$ and this notation is well defined.  Then the remarks above this corollary, specially \eqref{eq: partition classes},  mean: 
    
    $$\text{RHS of \eqref{eq: cor:pairings}} \leq \sum_{\caC \text{ : not simple}} \Delta_{n_1}(\caC) + \sum_{\caP} \Delta_n(\caP),$$
    
    where the sum is over all classes such that the corresponding set $\caC_{n_1}$ is non-empty.  Combining the above expressions with \eqref{eq:pk}, \eqref{eq:pk2},   Lemma    \ref{lem: coeff in module group}, and the fact that number of partitions of $[k]$ only depends on $k$ and not $n$, finishes the proof. 
\end{proof}

\subsection{Preparation of the proof of Proposition \ref{lem: reduction to pairings}}

We fix a class $\mathcal C_{n_1}=\caC_{n_1}(D,F,v,\varsigma)$.  We introduce the random sets, indexed by $i \in D$ and $\mu_F \in \Sigma_{n_0}^F$:  
$$
\mathcal R_i(\mu_F)=\left \{\mu \in \Sigma_{n_0} \Big{|} 
| E_{\mu} -\sum_{j\in F} v_{i,j} E_{\mu_j}  | \leq C(|\underline{\caI}|,k) \sss_n
 \right\}
$$
where $C(\underline{\caI},k)$ depends on intervals $\{\caI_i\}_i$ and on $k$. In particular, we take $C(\underline{\caI},k)$ 
equal to $$|\mathrm{Conv}(\cup_{i} \mathcal{I}_i \cup_{i}(-\mathcal{I}_i))| kC_0(k),$$ where $C_0(k)$ is an upper bound for $C(k)$ featuring in the statement of  Lemma \ref{lem: coeff in module group}, and $\mathrm{Conv}$ denotes the convex hull. 
We note that this random set is $\caF_{n_0}$-measurable, i.e.\ it depends only on $h_i,g_i, \:i\leq n_0$.  We set $\mathcal{R}_i=\emptyset$ if $D= \emptyset$.

We then have 
\begin{lemma}\label{lem: properties of car}
Recall the above fixed class $\caC$, and corresponding set $\caC_{n_1}$. For any  multi-configuration $\underline{\eta}=(\underline{\mu},\underline{\sigma})$ recall $\mu_F=(\mu_i)_{i\in F}$. Let the (\textit{random}) multi-configuration $\underline{\eta}=(\underline{\mu},
\underline{\sigma})$ be such that the event $\caE_{[k]}$ holds true, and such that $\underline{\sigma} \in \caC_{n_1}$, then:
$$\mu_i \in \mathcal R_i(\mu_F)\quad \text{for any} \quad  i\in D.$$ 
If, additionally, the event $\mathcal{A}_{n_0}^{K,\epsilon} \equiv \caA $  holds true, then for $n$ sufficiently large we have: 
   \begin{enumerate} 
    \item  $|\mathcal R_i(\mu_F)| \leq 2K$. 
 
    \item If $v_{i,j}=-\delta_{j,j'}$ for some $i\in D$, some $j'\in F$ and all $j\in F$, then,  
$$
\frac{\left|\{\mu_F \in \Sigma_{n_0}^F | \:|\mathcal R_i(\mu_F)| > 0       \} \right| }{|\Sigma_{n_0}^F|} \leq C \epsilon.
$$ 
    \item If $v_{i,j}=\delta_{j',j}$ for some $i\in D$, some $j'\in F$ and all $j\in F$, then, 
$$
\frac{\left|\{\mu_F \in \Sigma_{n_0}^F | \:|\mathcal R_i(\mu_F) \setminus \{\mu_{j'}\}| > 0   \} \right| }{ |\Sigma_{n_0}^F|} \leq C \epsilon.
$$

\end{enumerate}

\end{lemma}
\begin{proof}
Let us emphasize that $\mu_F, \caR_i(\mu_F),$ and above sets are \textit{random}; the above expressions are true given the event $\caE_{[k]}\cap \caA$. \\ 
The fact that $\mu_i \in \caR_i(\mu_F)$ then follows by an explicit computation, given our choice of constants. Item 1) follows because any two elements of $\mathcal R_i(\mu_F)$ are necessarily in resonant with each other, since $\alpha^{\theta n_0} \gg  C \sss_n$ for sufficiently large $n$; therefore all elements of $\caR_i(\mu_F)$ belong to a resonant patch. The claim follows by definition of $\mathcal{A}^K_n$. 
For item 2), we observe that  $v_{i,j}=-\delta_{j',j}$ implies that 
$$
\mathcal R_i(\mu_F)= \{\mu \in \Sigma_{n_0} | 
| E_{\mu} + E_{\mu_{j'}} | \leq C \sss_n
 \},
$$
which means that any element of $\mathcal R_i(\mu_F)$ is anti-resonant with $\mu_{j'}$ for large enough $n$.
Hence, if $|\mathcal R_i(\mu_F)| >0$, then $\mu_{j'} \in \widehat{Z}_{n_0}$ .  The claim follows now by using the definition of $\caA$. Similarly, for item 3), the relation  $v_{i,j}=\delta_{j',j}$ implies that any element of $\mathcal R_i(\mu_F)$ is resonant with $\mu_{j'}$ (unless it is equal to $\mu_{j'}$), from which the claim follows. 
\end{proof}
We now establish a general bound that will be the starting point for the next section. 
\begin{lemma}\label{lem: basic splitting}
Let $\caC_{n_1}= \caC_{n_1}(D,F,v,\varsigma)$. Then for any $\underline{\mu}\in \Sigma_{n_0}^k$
    \begin{align}
\sum_{\underline{\sigma} \in \caC_{n_1} }  \mathbb{P}_{\caA}(  \caE_{[k]} ) 
    & \leq     \mathbb{E}_{\caA} \left(\prod_{i\in D}  \chi(\mu_i \in \mathcal R_i(\mu_F)) \right)    \times \sum_{\underline{\sigma} \in \caC_{n_1}}  
    \sup_{\underline{a} \in \mathbb{R}^F}  
    \mathbb{P}\left(\bigcap_{j \in F} a_j + h\cdot\sigma_j \in I_j\right),  \label{eq: basic strategy}
\end{align}
where by convention the first term on the RHS is equal to one if $D=\emptyset$.

\end{lemma}

Note that the second factor on the right-hand side makes no reference to the $\mu$-configurations. The first factor refers to the $\sigma$-configuration only via the partition  $[k]=D\cup F$. 
\begin{proof}
We estimate
\begin{align}
   \mathbb{P}_{\caA}( \caE_{[k]})  & =  \mathbb{P}_{\caA}(\caE_{[k]}  \bigcap (\bigcap_{i\in D}  \mu_i \in \mathcal R_i(\mu_F)   ) ) \\ 
  & \leq \mathbb{P}_{\caA}(\bigcap_{j \in F} \mathcal E_j \bigcap (\bigcap_{i\in D}  \mu_i \in \mathcal R_i(\mu_F)  ) ) \\  
    & \leq   \mathbb{E}_{\caA}(\prod_{i\in D}  \chi(\mu_i \in \mathcal R_i(\mu_F)) )  \sup_{\underline{a}} \mathbb{P}(\bigcap_{j \in F} a_j + h\cdot\sigma_j \in I_j).
\end{align}
The first equality follows from the first statement of Lemma \ref{lem: properties of car}. The first inequality is by $\caE_{[k]} \subset \bigcap_{j\in F} \caE_j$.
The last inequality follows by a $L^1-L^{\infty}$-bound with respect to the variables $g_{1},\dots,g_{n_0},h_{1},\ldots,h_{n_0}$.  The claim of the lemma now follows by taking the sum over $\underline{\sigma}\in\caC_{n_1}$.

\end{proof}

\subsection{Final steps of the proof of Proposition \ref{lem: reduction to pairings}} 

Every class that is not simple and non-empty satisfies at least one of the following criteria
\begin{enumerate}
    \item $\varsigma=0$
    \item The matrix $v$  has nontrivial relations. By this we mean that there is at least one $i \in D$ such that  $v_{i,j}$ is nonzero for at least two values of $j\in F$.
    \item $\varsigma=1$ and the matrix $v$ has at least one anti-pairing.  That is, there is at least  one $i \in D$  such that 
 $v_{i,j} = b\delta_{j,j'}$ for some $j'\in F$ and some $b\neq 1$.
Of course, $b$ then needs to be $-1$, else the class is empty.
\end{enumerate}

To prove the first claim of Proposition \ref{lem: reduction to pairings}, it hence suffices to consider the above three cases one-by-one. We treat them in Sections \ref{sec: first case first statement}, \ref{sec: perhaps a new case} and \ref{sec: second case first statement}, respectively. 
%
Finally, Section \ref{sec: second case} proves the second claim of Proposition  \ref{lem: reduction to pairings}.

\subsubsection{Classes $\caC_{n_1}$ with $\varsigma =0$} \label{sec: first case first statement}
We recall that our aim is to bound 
$$
\sum_{\underline{\mu}\in \Sigma^k_{n_0}, \underline{\sigma} \in \caC_{n_1}} \mathbb{P}_{\caA}(\mathcal E_{[k]}) 
$$
with the class $\caC_{n_1}$ having $\varsigma =0$. 
We will use the bound \eqref{eq: basic strategy}. We first estimate the second factor on the right-hand side of that bound as 
\begin{equation} \label{eq: bound second factor atypical}
 \sum_{\underline{\sigma} \in \caC_{n_1}}  
    \sup_{\underline{a}}  
    \mathbb{P}(\bigcap_{j \in F} a_j + h\cdot\sigma_j \in I_j)  \leq |\mathcal C_{n_1}| \times C(\underline{\caI},k)n^{|F|} \times  \sss_n^{|F|},
\end{equation}
which follows by Lemma \ref{lem:carrot}, where we notice that $\{\sigma_i\}_{i \in F}$, are linearly independent by definition of $F$. 

Next, we focus on the first factor on the right-hand side of \eqref{eq: basic strategy}. We split the sum
$$\sum_{\underline{\mu}\in \Sigma_{n_0}^k} =
\sum_{\mu_F\in \Sigma_{n_0}^F}  
\sum_{\mu_D\in \Sigma_{n_0}^D}.  $$ and we move the sum over 
 $\mu_D$ inside the expectation. Given the event $\caA$, this sum is then estimated as
\begin{align}
       \sum_{\mu_D}  \prod_{i\in D} \chi( \mu_i \in \mathcal R_i(\mu_F))) 
     & \leq      \prod_{i\in D}  |\mathcal R_i(\mu_F)|  \leq  (2K)^{|D|},  \label{eq: acro res sets}
\end{align}
where we used item 1 of Lemma \ref{lem: properties of car}. 
Summing over $\mu_F$ and taking the expectation $\mathbb{E}_{\caA}$, we get then
$$
 \sum_{\underline{\mu}\in\Sigma_{n_0}^k } \mathbb{E}_{\caA} (\prod_{i\in D}  \chi(\mu_i \in \mathcal R_i(\mu_F)) ) \leq (2K)^{|D|} |\Sigma_{n_0}^F|.
$$
Notice that the RHS of the above expression is equal to $|\Sigma_{n_0}^k|$ if $D=\emptyset$.
We now combine this with the bound \eqref{eq: basic strategy} and the bound  \eqref{eq: bound second factor atypical} to get
\begin{equation}\label{eq: final bound first case}
\sum_{\underline{\mu}\in \Sigma^k_{n_0}, \underline{\sigma} \in \caC_{n_1}} \mathbb{P}_{\caA}(\mathcal E_{[k]})  \leq   (2K)^{|D|} |\Sigma_{n_0}^{F}| 
\times  |\mathcal C_{n_1}|  \times C(\underline{\caI},k)n^{|F|} \times \sss_n^{|F|}.
\end{equation}
By the definition of 'non-typical' classes, the large deviation estimate of Lemma \ref{lemma: counting00}, and the lower bound $n_1\geq cn$, we see that 
$$|\Sigma_{n_0}^{F}|  \times |\mathcal C_{n_1}| \times  \sss_n^{|F|} \leq C \exp(- cn^{1/4}). $$ 
Therefore, the right-hand side of \eqref{eq: final bound first case}
is $o_n(1)$.
\subsubsection{Classes $\mathcal C_{n_1}$  where $v$ has a non-trivial relation}\label{sec: perhaps a new case}

We proceed as in the case treated in the preceding section, but in the very last step the application of  Lemma \ref{lemma: counting00} is replaced by Lemma \ref{lem: counting}, providing a bound on $|\mathcal C_{n_1}|$.

\subsubsection{Classes $\mathcal C_{n_1}$ with $\varsigma=1$ and $v$ containing an anti-pairing}\label{sec: second case first statement}

As in the previous sections, we start by bounding the second factor of \eqref{eq: basic strategy}, this time using the sharper bound from Lemma \ref{lem: factorization}, which is applicable because $\sigma_F$ is typical: 
\begin{equation}\label{eq: typical bound second factor}
  \sum_{\underline{\sigma} \in \caC_{n_1}}  
    \sup_{\underline{a}}  
    \mathbb{P}(\bigcap_{j \in F} a_j + h\cdot\sigma_j \in I_j)  \leq    |\Sigma_{n_1}^{F}| \times C n_1^{-|F|/2} \sss_n^{|F|},    
\end{equation}
where we also used the trivial bound \eqref{eq: max entropy Gaussian} regarding Gaussians,  we bounded $\prod_i|\caI_i|$ by a constant (we dropped the dependence of the constants on $\underline{\caI}$, since it does not matter here), and we trivially bounded $|\caC_{n_1}|$ by $|\Sigma_{n_1}^F|$.  
Next, to bound the first factor on the right-hand side of \eqref{eq: basic strategy}, we estimate for $n$ large enough:
\begin{align}
   \sum_{\mu_F} \sum_{\mu_D}  \prod_{i\in D} \chi( \mu_i \in \mathcal R_i(\mu_F)) 
     & \leq      \sum_{\mu_F}  \prod_{i\in D}  |\mathcal R_i(\mu_F)|  \\
     &\leq
      (2K)^{|D|}     \sum_{\mu_F}    \chi(|\mathcal R_{i_o}(\mu_F)| >0)   \\
     &\leq   C (2K)^{|D|}   \epsilon |\Sigma_{n_0}^{F} |;   \label{eq: acro res sets bis}
\end{align}
since $\caC_{n_1}$ contains an anti-pairing, there exists $i_o$ such that $v_{i_o,j}=-\delta_{j',j}$. This $i_o$ features in the second line above. Given the event $\caA$,  we used item 1) of Lemma \ref{lem: properties of car} and the definition of $i_o$ to get the penultimate inequality; item 2) of  Lemma \ref{lem: properties of car} is used to get the last inequality. 
We now combine the above bounds to obtain 
\begin{align}
\sum_{\underline{\mu}\in \Sigma^k_{n_0}, \underline{\sigma} \in \caC_{n_1}}\mathbb{P}_{\caA}( \mathcal E_{[k]}) 
     & \leq   \sum_{\underline{\mu} \in \Sigma^k_{n_0}} 
  \mathbb{E}_{\caA} (\prod_{i\in D}  \chi(\mu_i \in \mathcal R_i(\mu_F)) )    \times \sum_{\underline{\sigma}\in \caC_{n_1}}\sup_{\underline{a}} \mathbb{P}(\bigcap_{j \in F} a_j + h\cdot\sigma_j \in I_j)    \\
        & \leq   C \epsilon |\Sigma_{n_0}^{F} |  (2K)^{k}  
        \times   |\Sigma_{n_1}^{F}| \times  n_1^{-|F|/2} \sss_n^{|F|}   \\
        &  \leq     \epsilon C(k,K). 
\end{align}
The second inequality follows from the bounds \eqref{eq: typical bound second factor} and \eqref{eq: acro res sets bis}  and the bound $|D|\leq k$. 
The third inequality follows from the expression for $\sss_n$ and the fact that $n_1 \geq cn$.

Now, the first statement of Proposition  \ref{lem: reduction to pairings}, i.e.,\ the bound \eqref{eq:pk}, is proven. 

\subsubsection{Second claim of Proposition  \ref{lem: reduction to pairings}} \label{sec: second case}
Let $\underline{\sigma} \in \caC_{n_1}$ with $\caC$ a simple class and let $\caC_{n_0}$ be the corresponding class, as in the second claim of  Proposition  \ref{lem: reduction to pairings}. 
Take a multiconfiguration $\underline{\mu}$ such that $\underline{\mu}\notin \caC_{n_0} $. This implies that either  $\mu_F$ is not typical, or there exists at least one $i\in D$ such that $\mu_i\neq \mu_{j(i)}$, see the definition of simple classes at the beginning of Section \ref{sec: kth moment}.

Therefore, if the event $\caE_{[k]} \cap \caA$ holds and $\underline{\mu}\notin \caC_{n_0} $, then either $\mu_F$ is not typical, or there is $i\in D$ such that  $|\caR_i(\mu_F) \setminus \{\mu_{j(i)}\}| >0$, since $\mu_i \in \caR_i(\mu_F)$ by Lemma \ref{lem: properties of car}.



Therefore, the second claim of Proposition \ref{lem: reduction to pairings} follows from the following two estimates
\begin{equation}\label{eq: new basic expression first}
  \sum_{\substack{\underline{\mu} \in \Sigma_{n_0}^k \\  \mu_F \, \text{non-typical}}}  \sum_{\underline{\sigma} \in \caC_{n_1} }  \mathbb{P}_{\caA}(  \caE_{[k]}    )  =o_n(1), 
\end{equation}
\begin{equation}\label{eq: new basic expression}
  \sum_{\underline{\mu} \in \Sigma_{n_0}^k}  \sum_{\underline{\sigma} \in \caC_{n_1} }  \mathbb{P}_{\caA}(  \caE_{[k]}   \cap \{ \exists i \in D| \:|\caR_i(\mu_F) \setminus \{\mu_{j(i)}\}| >0\} )  \leq \epsilon C(K,k).
\end{equation}
\noindent \textbf{Proof of the bound \eqref{eq: new basic expression first}}. 
This is very similar to the reasoning in the previous Section \ref{sec: second case first statement}. 
We first invoke Lemma \ref{lem: basic splitting} and we use the bound \eqref{eq: typical bound second factor} for the second factor of \eqref{eq: basic strategy}. Notice that we are allowed to use this estimate, since $\caC$ is simple, meaning $\sigma_F$ is typical. This yields 
\begin{equation}
  \sum_{\substack{\underline{\mu} \in \Sigma_{n_0}^k \\  \mu_F \, \text{non-typical}}}  \sum_{\underline{\sigma} \in \caC_{n_1} }  \mathbb{P}_{\caA}(  \caE_{[k]}    )  \leq   
  \mathbb{E}_{\caA}  \Big( \sum_{\substack{\underline{\mu} \in \Sigma_{n_0}^k \\  \mu_F \, \text{non-typical}}} 
 (\prod_{i\in D}  \chi(\mu_i \in \mathcal R_i(\mu_F)) )   \Big) \times  
  |\Sigma_{n_1}^{F}| \times C n_1^{-|F|/2} \sss_n^{|F|} . 
\end{equation}
 We again split $\sum_{\underline{\mu}}=\sum_{\mu_F} \sum_{\mu_D}$ and we bound the sum over $\mu_D$  by $(2K)^{|D|}$, as in  \eqref{eq: acro res sets}.   This yields that the above line is bounded by
 $$
\big(   \sum_{ \mu_F \in \Sigma_{n_0}^F} \chi( \mu_F \, \text{non-typical})\big)
\times 
    (2K)^{|D|} \times  |\Sigma_{n_1}^{F}| \times C n_1^{-|F|/2} \sss_n^{|F|}. 
 $$
Now the desired result follows by the large deviation estimate of Lemma \ref{lemma: counting00} applied to the sum over $\mu_F$.  




\vspace{2mm}

\noindent \textbf{Proof of the bound \eqref{eq: new basic expression}}
We proceed as in the proof of Lemma \ref{lem: basic splitting}, obtaining 

\begin{align}
\text{LHS  of \eqref{eq: new basic expression}}    &\leq     \sum_{\underline{\mu} \in \Sigma_{n_0}^k} \mathbb{E}_{\caA}    \prod_{i'\in D} \chi( \mu_{i'} \in \mathcal R_{i'}(\mu_F)) \chi(\exists i \in D| \: |\caR_i(\mu_F) \setminus \{\mu_{j(i)}\}| >0) \nonumber  \\
&\qquad \times \sum_{\underline{\sigma} \in \caC_{n_1}}  
    \sup_{\underline{a}} 
    \mathbb{P}(\bigcap_{j \in F} a_j + h\cdot\sigma_j \in I_j).  \label{eq: basic strategy again} 
\end{align}

We now focus on the first factor of the right-hand side of \eqref{eq: basic strategy again}. We first split  $\sum_{\underline{\mu}}=\sum_{\mu_F}\sum_{\mu_D}$ and we bound 

\begin{align}
&\sum_{\mu_D} \chi(\exists i \in D| \: |\caR_i(\mu_F) \setminus \{\mu_{j(i)}\}| >0)\prod_{i'\in D} \chi( \mu_{i'} \in \mathcal R_{i'}(\mu_F))  \\ & \leq   (2K)^{|D|}  \sum_{i\in D}  \chi(|\caR_i(\mu_F) \setminus \{\mu_{j(i)}\}| >0), \label{eq: bound last case}
\end{align}
where we used the fact that in the first line, the first expression in $\chi(\cdots)$ does not depend on $\mu_D$; then we bound similarly to  the bound \eqref{eq: acro res sets}.

The sum of the expectation $   \sum_{\mu_F} \mathbb{E}_{\caA}(\cdot)$ of the last  expression of \eqref{eq: bound last case} is bounded by $ \epsilon C  k (2K)^k |\Sigma^{F}_{n_0}|$, because of  
 item 3) of Lemma \ref{lem: properties of car}. Notice that we may use this lemma for all $i \in D$, since $\caC$ is simple.
 The second factor in \eqref{eq: basic strategy again} is bounded exactly as in Section \ref{sec: second case first statement}, or as in the proof of the bound \eqref{eq: new basic expression first}. 
We then obtain the bound \eqref{eq: new basic expression}.
 This ends the proof of the second statement of Proposition \ref{lem: reduction to pairings}.




 



\subsection{Computation of moments}\label{sec: computation of moments}
 In this section, we put the machinery of the previous sections to work to actually compute some moments.  For an interval $\mathcal{I}$, recall its rescaled version as $I=\sss_n\mathcal{I}$. Then we recall $\widetilde N_n(\mathcal{I})$ \eqref{eq: DOS tilde approx} is the number of ``approximate levels" inside that interval, i.e.\
$$ 
  \widetilde{N}_n(\mathcal{I})= \sum_{\eta \in \Sigma_n}  \chi(\widetilde{E}_{\eta} \in \sss_n\mathcal{I}). 
$$
  
  Previously, we had a collection of intervals $\{\mathcal{I}_{i}\}_{i=1}^k$. Let us first take all these intervals to be equal, i.e.\ $\mathcal{I}_i=\mathcal{I}$, and we simply write $I$ for $\sss_n\mathcal{I}$.  We will compute the $k$-th moment
  $\mathbb{E}_{\caA}(  \widetilde{N}^k_n(\mathcal{I}))$, using Corollary \ref{cor: pairings} and the terminology introduced before that corollary. This computation will not be needed in what follows, but we find it useful to present it first, because the later computation (ending with \eqref{eq: conclusion moment computation}) will rely on the same ideas, but with much more notational overhead. 
\begin{align}
    \mathbb{E}_{\caA}(  \widetilde{N}^k_n(\mathcal{I})) &= \sum_{\underline{\eta} \in \Sigma_n^k}  \mathbb{P}_{\caA}\left( 
    \caE_{[k]} \right)  \\
    &=  \sum_{\caP_k}   \sum_{\underline{\eta} \in \widetilde \caP_k}  \mathbb{P}_{\caA}\left( 
    \caE_{[k]} \right) + o_n(1)+\epsilon C (K,k)  \\
    &= 
      \sum_{\caP_k}  |\mathcal{I}|^{|\caP_k|} + o_n(1)+o_K(1)+\epsilon C(K,k), \label{eq: last line of exercise computation}
\end{align}
where we made the dependence on $k$ in the partitions explicit by writing  $\caP_k$, and where $|\caP_k|$ is the number of partitioning sets in the partition $\caP_k$. The second equality is Corollary \ref{cor: pairings}, and the last equality follows from
Lemma \ref{lem: moments typical case}. 
We recognize the expression $\sum_{\caP_k}  |\mathcal{I}|^{|\caP_k|} $ as the $k$-th moment of a Poisson distribution with rate $|\mathcal{I}|$. Indeed, all cumulants of a Poisson distribution are equal to its rate.

Next, we choose $p$ intervals $\mathcal{J}_j, j=1,\ldots, p$ such that the intervals are \textit{pairwise disjoint}.  These  intervals are just like the intervals $\mathcal{I}_i$ considered earlier, but we need to denote them by a different symbol to avoid a clash of notation in what follows. 
Our goal is to compute
\begin{equation} \label{eq: product of moments}
\mathbb{E}_{\caA}((  \widetilde{N}_n(\mathcal{J}_1))^{k_1}\ldots (  \widetilde{N}_n(\mathcal{J}_p))^{k_p} ),
\end{equation}
for naturals $k_1,\ldots, k_p \in \mathbb{N}$. 

First, we define the index sets
$$
K_1= [k_1], \qquad K_j= [k_1+\ldots +k_j]\setminus [k_1+\dots+k_{j-1}] \quad \text{for $j=2,\ldots,p$}.
$$
We can now cast the expectation \eqref{eq: product of moments} in a sum of  $\mathbb{P}(\caE_{[k]})$ with $k=k_1+\ldots+k_p$ and with the rescaled intervals $I_i, i \in [k]$ determined by 
$$
I_i = \sss_n \mathcal{J}_j \quad \text{iff.\  $i \in K_j$}.
$$
Furthermore, we note that any partition of $[k_j]$ can be interpreted as a partition of $K_j$ by shifting by $\min K_j-1$, i.e.\ the partition $\{A_1,\ldots, A_z\}$ of $[k_j]$ gives rise to the partition  $\{A_1+\min K_j-1,\ldots, A_z+\min K_j-1\}$ of $K_j$. Given partitions $\caP_{k_j}$ of $[k_j]$, for all $j=1,\ldots, p$, we thus construct a partition of $[k]$ by taking the union of all the shifted partitions, and we write the resulting partition of $[k]$ as
$$
\caP_k(\caP_{k_1},\ldots, \caP_{k_p} ).
$$
We are now ready to compute \eqref{eq: product of moments}, i.e.\ :
\begin{align}
\mathbb{E}_{\caA}((  \widetilde{N}_n(\mathcal{J}_1))^{k_1}\ldots (  \widetilde{N}_n(\mathcal{J}_p))^{k_p} ) & =
\sum_{\underline{\eta} \in \Sigma_n^k} 
    \mathbb{P}_{\caA}\left( 
    \caE_{[k]} \right) \\[1mm]
    &=
\sum_{\caP_k}   \sum_{\underline{\eta} \in \widetilde \caP_k}  \mathbb{P}_{\caA}\left( 
    \caE_{[k]} \right)  +o_n(1)+ \epsilon C(K,k)
    \\[1mm]  & =   \sum_{\caP_{k_1},\ldots, \caP_{k_p} }   \sum_{
    \substack{\underline{\eta} \in \widetilde \caP_k \\
 \caP_k=\caP_k(\caP_{k_1},\ldots, \caP_{k_p} ) }}  \mathbb{P}_{\caA}\left( 
    \caE_{[k]} \right)  +o_n(1)+ \epsilon C(K,k)
    \\[1mm]
     & = \prod_{j=1}^p \sum_{\caP_{k_j}} |\mathcal{J}_j|^{|\caP_{k_j}|}     + o_n(1)+o_K(1)+\epsilon C(K,k). \label{eq: conclusion moment computation}
\end{align}
The first equality was explained above, and the second equality is 
a consequence of Corollary \ref{cor: pairings}. To get the third equality, we note 
that any partition $\caP_k$ that is not of the form $\caP_k=\caP_k(\caP_{k_1},\ldots, \caP_{k_p} )$ leads to an event $\caE_{[k]} =\emptyset$. This follows since the intervals $\mathcal{J}_j$ are pairwise disjoint.   Finally, the last bound follows from Lemma \ref{lem: moments typical case}, in a straightforward generalization of the reasoning used to obtain \eqref{eq: last line of exercise computation}. 

This result will be the starting point of the next section.
\section{ Proof of Theorem \ref{thm: main theorem}} \label{sec: proof of them 2}
Theorem \ref{thm: main theorem} is concerned with the comparison between a Poisson point process $\bar\xi$ with unit rate and the empirical processes $\xi_n$, introduced in Section \ref{sec: level statistics}, i.e., \eqref{sec: stat def: point process}.   It is convenient to define first an approximation to $\xi_n$, namely 
\begin{equation}
    \tilde{\xi}_n:= \sum_{\eta \in \Sigma_n}\delta_{\sss_n^{-1}\widetilde{E}_{\eta}}(dx).
\end{equation}
This is not equal to $\xi_n$ because we have the approximate levels $
\widetilde E_\eta$ \eqref{eq: tilde def} instead of $E_\eta$. \\
Finally, we recall the pairing of a point process $\xi$ with a test function $\varphi$: $\xi(\varphi)=\int \xi(dx)\varphi(x)$.



\subsection{The variable $\tilde \xi(\psi)$ for simple functions $\psi$. }

Consider a simple positive test function 
\begin{equation} \label{eq: def test function}
    \psi= \sum_{a=1}^M b_a 
    \chi_{\mathcal  I_a}, 
\end{equation}
where $b_a \geq 0$, $\max_a{b_a} <C$,  $\psi$ is supported in $[-C_{\psi},C_{\psi}]$ (all constants are $O(1)$; and $\psi$ is compactly supported), and $\mathcal I_a$ are \textit{disjoint intervals}. Let us recall that $\chi_{\caI_a }$ denotes the indicator function of the interval $\caI_a$.

\begin{lemma}\label{lem: comparison moments kth moment}
Let $\psi$ be as in \eqref{eq: def test function}, then  for any $k \geq 1$:
    \begin{equation} \label{eq: moment to Laplace step 1} \left|\mathbb{E}_{\caA}\left(\tilde{\xi}_n(\psi)^k \right) - \bar{\mathbb{E}}\left(\bar{\xi}(\psi)^k \right) \right| \leq C(k,\psi,K) \epsilon +o_n(1)+o_K(1), 
    \end{equation}
where we omit the dependence of  error terms ($o_n(1)$,$o_K(1)$) on $\psi$.
\end{lemma}
\begin{proof}
Recall $  \widetilde{N}_n(\cdot)$ from \eqref{eq: DOS tilde approx}; by definition  we can write
$$
\tilde{\xi}_n(\chi_{\mathcal I_a}) =   \widetilde{N}_n( \mathcal{I}_a).
$$
For any $k-$tuple $\underline{a}\equiv (a_1,\dots, a_k)\in [M]^k$ denotes $b_{\underline{a}}:= \prod_{i=1}^k b_{a_i}$. Then we have:
 \begin{align}      \mathbb{E}_{\caA}\left(\tilde{\xi}_n(\psi)^k \right) &= \sum_{\underline{a}\in [M]^k} b_{\underline{a}}
 \mathbb{E}_{\caA} \left(\prod_{i=1}^k   \widetilde{N}_n( \mathcal{I}_{a_i})\right)  \\
& =\sum_{\underline{a}\in [M]^k} 
b_{\underline{a}}\bar{\mathbb{E}}\left(\prod_{i=1}^k \bar{\xi}(\mathcal{I}_{a_i}) \right) + \epsilon C(K,k,\psi) +  o_n(1)+o_K(1) \label{eq: k moment temp}.
    \end{align}
The first equality follows by definition and linearity; the second equality follows from \eqref{eq: conclusion moment computation} and the following observation: for a fixed $\underline{a}$, let us partition $[k]$ into $B_1,\dots,B_p$ such that $a_i=a_{i'}$ whenever $i,i'$ belong to the same set $B_{p'}$ and $a_i\neq a_{i'}$ otherwise. Let us take $a_{i_1},\dots,a_{i_p}$ such that $a_{i_j}\in B_j$ for all $1\leq j\leq p$. Then, we denote $\mathcal{J}_j:=\mathcal{I}_{a_j}$. Notice that all $\{\mathcal{J}_j\}_{j=1}^p$ are disjoint. Now,  applying \eqref{eq: conclusion moment computation}, we get the expression 
$$
\prod_{j=1}^p \sum_{\caP_{k_j}} |\mathcal{J}_j|^{|\caP_{k_j}|},
$$ as the leading order, where $k_j=|B_j|$.  On the other hand, this expression is equal to the expectation of the product of moments of Poisson variables
$$
\bar{\mathbb{E}}\left(\prod_{j=1}^p (\bar{\xi}(\mathcal{J}_j))^{k_j}\right),
$$
where the intervals $\mathcal{J}_j$ are defined above.  This observation follows by the same logic as the corresponding observation below \eqref{eq: last line of exercise computation} as well as basic properties of Poisson point process with rate one: i.e., $\bar{\xi}(\mathcal{J})$, $\bar{\xi}(\mathcal{J}')$ are independent Poisson r.v. with rate $|\mathcal{J}|$,$|\mathcal{J}'|$ for disjoint intervals $\mathcal{J},\mathcal{J}'$.
 Now \eqref{eq: k moment temp} follows, noticing that the above expression coincides with $\bar{\mathbb{E}}(\prod_{i=1}^k \bar{\xi}(\mathcal{I}_{a_i}))$ thanks to the definition of $\mathcal{J}_j$. Notice that we sued the fact that $M$ and $\sup_a b_a$ are of order one. Finally, we conclude the proof by noticing that the leading expression in \eqref{eq: k moment temp} is exactly equal to $\bar{\mathbb{E}}(\bar{\xi}(\psi)^k)$ by definition. 
\end{proof}

We now use the moment bounds to prove 
\begin{lemma} \label{lem: comparison moments}
\begin{equation} \label{eq: moment Laplace step 2}
\left|\mathbb{E}_{\caA}\left[\exp(-\tilde{\xi}_n(\psi))\right]- 
\bar{\mathbb{E}}\left[\exp(-\bar{\xi}(\psi))\right] \right|= C(k_*,\psi,K) \epsilon +o_n(1)+o_K(1)+o_{k_*}(1).
\end{equation}
\end{lemma}
\begin{proof}
    
By the bounds on remainders in a Taylor series, we have
    $$
    \left|e^{-x}-\sum_{k=0}^{k_*} (-x)^k/k!\right| \leq \frac{x^{k_*+1}}{(k_*+1)!}, \qquad x\geq 0.
    $$
    Taking advantage of the above bound once for $x=\tilde{\xi}_n(\psi)$, and once for $x= \tilde{\xi}(\psi)$, then
    using Lemma \ref{lem: comparison moments kth moment} both for $1 \leq k\leq k_*$ and for $k=k_*+1$, and finally using Lemma \ref{lem: bound on caa} for $k=0$, we deduce:
     \begin{align} 
         \left|\mathbb{E}_{\caA}\left[\exp(-\tilde{\xi}_n(\psi))\right]- 
\bar{\mathbb{E}}\left[\exp(-\bar{\xi}(\psi))\right] \right| 
 & \leq      C(k_*,\psi,K) \epsilon +o_n(1)+o_K(1) +2 \bar{\mathbb{E}}(\bar{\xi}(\psi)^{k_*+1})/(k_*+1)! .  \nonumber
     \end{align}
    Note that we used in a crucial way the fact that $\psi$ is non-negative.  We now bound $\bar{\xi}(\psi) \leq C \bar{\xi}(\chi_{[-C_\psi,C_\psi]})$; notice that this bound holds surely between the above random variables. Moreover, we note that $\bar{\xi}(\chi_{[-C_\psi,C_\psi]})$ is a Poisson random variable with rate $2C_\psi$.  As an application of Dobinski's formula, the $k$-th moment of the Poisson r.v. with rate $C$ is bounded by $k^k/(\ln(k/C+1)^k = o_k(1)k!$ (see \cite{Poissonbound}), 
   which yields our claim
    
\end{proof}

In the next lemma, we get rid of the dependence on the non-resonance event $\caA=\caA_n^{\epsilon,K}$.

 \begin{lemma}\label{lem: comparison of exp without caa}
     \begin{equation} \label{eq: main limit simple functions}
         \lim_{n \to \infty} {\mathbb{E}\left(\exp(-\tilde{\xi}_n(\psi)) \right)}={\bar{\mathbb{E}}\left(\exp(-\bar{\xi}(\psi)) \right)}.
     \end{equation}
 \end{lemma}
\begin{proof}
For any random variable $X$ such that $|X|\leq 1$, $\mathbb{P}$-almost surely, we have 
$$
|\mathbb{E}(X)-\mathbb{E}_{\caA}(X)| \leq 1- \mathbb{P}(\caA_n^{\epsilon,K}).
$$
Applying this bound with $X=\exp(-\tilde{\xi}_n(\psi)) $, and using Lemma \ref{lem: comparison moments}, we get 
  \begin{align}
      \left| \mathbb{E}\left(\exp(-\tilde{\xi}_n(\psi)) \right) -\bar{\mathbb{E}}\left(\exp(-\bar{\xi}(\psi)) \right)\right| \leq 
      C(\psi,k_*,K) \epsilon +o_n(1)+o_{k_*}(1)+o_K(1),
   \end{align}
 where we used  Lemma \ref{lem: bound on caa} to bound $1- \mathbb{P}(\caA_n^{\epsilon,K})=o_n(1)+o_K(1)$.\\ 
Given $\varepsilon>0$,  we first take $k_*$ sufficiently large so $|o_{k_*}(1)|< \varepsilon/4$  (constants in this term only involve the test function, and not $K,\epsilon$). Then we take $K$ sufficiently large such that $|o_K(1)|<\varepsilon/4$. Fixing the above $k,K$, we take $\epsilon$ sufficiently small such that the third term involving $\epsilon$ becomes bounded by $\varepsilon/4$. Fixing $k,K,\epsilon$ as above and taking $n$ large enough finishes the proof (note that the term with $o_n(1)$ depends on $K,k_*$; the term  $o_K(1)$ on $k_*$, and $o_{k_*}(1)$ is independent of $K,n,\epsilon$, and that is the reason for the above order). 
\end{proof}

\subsection{General test functions $\varphi$}
Everything up to now was stated for nonnegative simple functions $\psi$. The next lemma extends this. 
\begin{lemma}\label{lem: main theorem with tilde} The conclusion of Lemma \ref{lem: comparison of exp without caa} holds also for a general $\varphi \in C^+_K(\mathbb{R})$, i.e.\ 
  \begin{equation} \label{eq: main limit general functions}
         \lim_{n \to \infty} {\mathbb{E}\left(\exp(-\tilde{\xi}_n(\varphi)) \right)}={\bar{\mathbb{E}}\left(\exp(-\bar{\xi}(\varphi)) \right)}.
     \end{equation}
\end{lemma}
\begin{proof}
Since simple functions are dense in $C_K^+(\mathbb{R})$, we can find for any $\delta>0$ a nonnegative simple function $\psi_\delta$ such that 
$$
||\varphi-\psi_\delta||_{\infty} \leq \delta. 
$$
Then, we note that by definition
$$
| \tilde \xi_n(\varphi)-\tilde \xi_n(\psi_\delta) | \leq ||\varphi-\psi_\delta||_{\infty} \tilde \xi_n(\chi_{\mathcal I}) \leq \delta \tilde \xi_n(\chi_{\mathcal I}),  
$$
where $\mathcal I$ is a bounded interval such that $\mathrm{supp}(\varphi)\subset \mathrm{int}(\mathcal I)$, and $\mathrm{supp}(\psi_{\delta})\subset \mathrm{int}(\mathcal{I})$ for any $\delta>0$, where $\mathrm{int}(\cdot)$ denotes the interior of an interval. 
Then, we abbreviate $A=\tilde \xi_n(\varphi), B=\tilde \xi_n(\psi_\delta) $; thanks to the fact that $|1-e^{-x}|<x$ for $x>0$, and since $A,B\geq 0$ by a simple computation we have: 
\begin{align}
|e^{-A}-e^{-B}| \leq  |A-B|  \leq  \delta \tilde \xi_n(\chi_{\mathcal I}) =\delta \widetilde{N}_n(\mathcal{I}), 
\end{align}
where we used the definition of $\widetilde{N}_n(\mathcal{I})$ \eqref{eq: DOS tilde approx}. On the other hand, 
 $\mathbb{E}(\widetilde{N}_n(\mathcal{I})  ) \leq C$ uniformly in $n$ and $\delta$, thanks to \eqref{eq: first moment bound}. Then,  we get:  
$$
\left|\mathbb{E} \left( \exp(-\tilde \xi_n(\varphi))\right) - \mathbb{E}\left(\exp(-\tilde \xi_n(\psi_\delta)) \right) \right| \leq C \delta, \qquad \text{uniformly in $n$}.
$$
A similar argument can be made for the Poisson process $\bar\xi$, except that, in this case, the random variable $\bar\xi(\chi_{\mathcal I}) $ is simply a Poisson random variable with rate $|\mathcal I|$. Since $\delta$ was arbitrary, our claim is now proven by a so-called ${\varepsilon}/3$ argument. 
\end{proof}

\subsection{End of proof of Theorem \ref{thm: main theorem}}

Comparing the statement of Theorem \ref{thm: main theorem} with Lemma \ref{lem: main theorem with tilde}, we see that all that remains is to show that 
\begin{equation}\label{eq: what remains}
    \lim_{n \to \infty} \mathbb{E} \left[\exp(-\xi_n(\varphi)) - \exp(-\tilde \xi_n(\varphi)) \right]=0.   
\end{equation}

To achieve this, we need the following two remarks. 
\begin{enumerate}
    \item   Let $\mathcal I$ be a bounded interval such that $\mathrm{supp}(\varphi)\subset \mathrm{int}(\mathcal I)$, where $\mathrm{int}$ denotes the interior of the set. Recall the random variables $N_n(\mathcal I)$ \eqref{eq: def N_n}, introduced before Lemma \ref{lem: a priori bound}, then 
 $$ \xi_n(\chi_{\mathcal I})  = N_n(\mathcal{I}), \qquad \mathbb{E}(N_n(\mathcal I)) \leq C_{\mathcal{I}},$$
 where the second bound is by \eqref{eq: first moment bound}.
 
\item    From \eqref{eq: tilde bar relation}, and the fact that  $\alpha^{n_0} \ll \sss_n$ we know that  
   $|E_{\eta}\sss_n^{-1}-\widetilde{E}_{\eta}\sss_n^{-1}|< \alpha^{cn}$, with $c>0$. Therefore
$$
|\tilde \xi_n(\varphi) -\xi_n(\varphi)| \leq  (\widetilde N_n(\mathcal I)+ N_n(\mathcal I )) \sup_{x,y: |x-y|\leq \alpha^{cn} }  |\varphi(x)-\varphi(y)|. 
$$
The function $\varphi$ is uniformly continuous (continuous on a compact interval); therefore, the above $\sup_{x,y}(\ldots)$ is $o_n(1)$ (because $\alpha^c<1$). 
Furthermore, thanks to Corollary \ref{cor: uniform bound first moment}, i.e., \eqref{eq: first moment bound}, $N_n(\mathcal I)$ has a bounded expectation, and the same holds for $\widetilde N_n(\mathcal I)$.  This yields
\begin{equation} \label{eq: comparison processes}
    \lim_n \mathbb{E}\left(|\tilde \xi_n(\varphi) -\xi_n(\varphi)|\right) =0.
\end{equation}
\end{enumerate}

Finally, we abbreviate  $A=\xi_n(\varphi), B= \tilde \xi_n(\varphi) $ and as before we observe
\begin{align}
|e^{-A}-e^{-B}| \leq  |A-B|, 
\end{align}
where we used that $A,B\geq 0$. Now,  by the above bound and using \eqref{eq: comparison processes}, we get \eqref{eq: what remains} and hence Theorem \ref{thm: main theorem} is proven.

\appendix

\section{Non-degeneracy}\label{sec: app Non-degeneracy}

Let Assumption \ref{ass: nondegeneracy} hold. Fix $n\geq n_B$, and recall the Hamiltonian  $H_n$ \eqref{eq: main model1}, and its corresponding eigenvalues $\{ E_{\mu}\}_{\mu \in \Sigma_n}$.  We define: 
\begin{equation} \label{eq: def gap antigap}
\Delta_+(n):= \min_{\mu,\mu'} \left |E _{\mu} + E_{\mu'}  \right| , \qquad  \Delta_-(n):= \min_{\mu \neq \mu'} | E _{\mu} - E_{\mu'}|  .
\end{equation}
 Notice that  $\Delta_{-}(n)$ is the minimal gap of the Hamiltonian. We also denote $\Delta_{\upsilon}^B:= \Delta_{\upsilon}(n_B)$ for $\upsilon =\pm1$.  

\begin{lemma} \label{lemma: minimal gap antigap}
    For any $ n> n_B$, and any $\upsilon \in \{\pm1 \}$ we have: 
    \begin{equation}
        \mathbb{P}(\Delta_{\upsilon}(n)>0)=1.
    \end{equation}
\end{lemma}
The rest of this section is devoted to the proof of the above lemma.  One may provide a shorter proof. However, we decided to include the following proof because it gives a weak quantitative bound. 
         
\subsection{Preliminaries: Matrix-valued Cartan Lemma}
Before proceeding, we recall the following multi-dimensional version of the Cartan Lemma (see Lemma 1 of \cite{bourgain2009Cartan}): 
    \begin{lemma} \label{lem:Cartan multi}
        Let $F$ be a real analytic function on $[-1/2,1/2]^n$, which extends to an analytic function on $D^n$, with $D$ being the open unit disk in the complex plane, with the bound 
        $\sup_{\underline{z} \in D^n} |F(\underline{z})| <1$. If there exists $a \in [-1/2,1/2]^n$ such that $|F(a)|>\epsilon$ for $0 < \epsilon<1/2$ then we have for any $\delta>0$: 
        \begin{equation} \label{eq: Cartan n dim}
            |\{ x \in [-1/2,1/2]^n | |F(x)|<\delta \} | < Cn \delta^{c/\ln(1/\epsilon)},
        \end{equation}
        $|\cdot|$ denotes the $n$-dimensional real Lebesgue measure.
    \end{lemma}

    \subsection{Proof of Lemma \ref{lemma: minimal gap antigap}}

    \begin{proof}[Proof of Lemma \ref{lemma: minimal gap antigap}]
       
        \textbf{Step 1.} Definition of  $F_{\upsilon},A_{\upsilon}$:\\ The Hamiltonian $H_n$ is defined over the Hilbert space $\mathcal{H}$. Let 
        $$\mathcal{H}_{+}:= \mathcal{H} \otimes \mathcal H,  \quad \mathcal{H}_{-}:=\mathcal{H} \wedge \mathcal{H}; \quad  \quad N_+:=2^{2n} ,\quad N_-:= 2^{n}.(2^{n}-1)/2,$$ where $\cdot \wedge \cdot$ denotes the usual antisymmetric tensor product. Denote the $N_+ \times N_+$ identity matrix by $I$. We define $A_{\upsilon}$ over $\mathcal{H}_{\upsilon}$: 
        \begin{equation}
            A_{\upsilon} := \left(I \otimes H_n + \upsilon H_n \otimes I \right)^2. 
        \end{equation}
       By $A_{-}$ we mean the restriction of the above operator to $\mathcal{H}_{-}$.
        $A_{\upsilon}$ can be viewed as a function of disorder (recall $\tilde{n}=n-n_B$):
        
        $$A_{\upsilon}(\underline{h},\underline{g}):[-1/2,1/2]^{2\tilde{n}} \to \mathbb{C}^{N_{\upsilon}}.$$ 
         We extend this function to $D^{2 \tilde{n}}$, where $D$ is the unit complex disk. We also have for $\upsilon \in \{\pm 1 \}$: 
       \begin{equation} \label{eq: minimal lemma bound 1}
        \sup_{z \in {D}^{2\tilde{n}}} \|A_{\upsilon}(z) \| \leq C n^2.
       \end{equation}
          Define $F_{\upsilon}:D^{2\tilde{n}} \to \mathbb{C}$:
        \begin{equation} \label{eq: def: Ftheta}
            F_{\upsilon}(z) := \frac{\det(A_{\upsilon}(z))}{(Cn^2)^{N_{\upsilon}}}.
        \end{equation}
        Observe that $A_{\upsilon}$ is entry-wise analytic, hence $F_{\upsilon}$ is analytic on $D^{2 \tilde{n}}$. Also, $|F_{\upsilon}(z)|\leq 1$, for any $z \in D^{2\tilde{n}}$ since $|\det(A_{\upsilon}(z))| \leq \|A_{\upsilon}(z)\|^{N_{\upsilon}}$. \\
       
        \textbf{Step 2.} We have for any $\delta>0$: 
        \begin{equation}\label{eq: Cartan upper bound}
        \mathbb{P}(\Delta_{\upsilon}(n) < \delta )   \leq \mathbb{P}(|F_{\upsilon}|< {\delta^2 N_{\upsilon}}/Cn^2)=:p_{n,\upsilon}(\delta).
        \end{equation}
        To prove \eqref{eq: Cartan upper bound}, first, we recall that for any $N_{\upsilon} \times N_{\upsilon}$ matrix, $A$ we have 
        $$\|A^{{-1}} \| \leq \frac{N_{\upsilon}\|A \|^{N_{\upsilon}-1}}{|\det(A)|}. $$
        The above bound is a direct consequence of Cramer's rule (see Lemma 10.4 of \cite{kruger2012localization}). Therefore, by a simple computation, we have 
        $$\mathbb{P}(\|A_{\upsilon}^{-1}\| >\delta^{-2} ) \leq p_{n,\upsilon}(\delta).$$ 
        On the other hand, if the set of eigenvalues and 
    corresponding normalized eigenvectors of $H_n$ is given by $\{E_{\mu},\psi_{\mu} \}_{\mu \in \Sigma_n}$, then    
        $$\{(E_{\mu}+ E_{\mu'})^2,  {\psi_{\mu}}\otimes {\psi_{\mu'}} \}_{\mu,\mu'} \quad  \text{and} \quad \{(E_{\mu}-E_{\mu'})^2, \psi_{\mu} \wedge\psi_{\mu'} \}_{\mu < \mu'} $$ 
        are the sets of eigenvalues and eigenvectors of $A_{+}$ and $A_{-}$ respectively, where we equipped $\Sigma_n$ by a total order $<$, and $\wedge$ denotes the usual antisymmetric tensor product.  This means $\|A_{\upsilon}^{-1}(z) \|= \Delta^{-2}_{\upsilon}(n)$ for real $z$. 
        This finishes the proof of this step, i.e., \eqref{eq: Cartan upper bound}. \\

        \textbf{Step 3.} For $\upsilon \in \{\pm1 \} $, there exists $(\underline{h}^{\upsilon},\underline{g}^{\upsilon}) \in [-1/2,1/2]^{2 \tilde{n}}$ such that 
        \begin{equation} \label{eq: Cartan multi lower bound}
            F_{\upsilon}(\underline{h}^{\upsilon},\underline{g}^{\upsilon}) > (c^{-n} \Delta_{\upsilon}^B)^{2N_{\upsilon}}=: \varepsilon_{\upsilon}, 
        \end{equation}
        for some $c>1$. Here we separate the two cases:
        \begin{enumerate}
            \item In  case $\upsilon=+$, taking $(\underline{h}^+,\underline{g}^{+})=0$, one can observe that evaluated at this point $\Delta_{+}(n)=\Delta_{+}^B$. 
            
            \item In case $\upsilon=-$, we set for all $n_B<i\leq n$:
       
        \begin{equation} \label{eq: cute lower bound}
            \underline{g}^-=0;  \quad   h_{i}^-=b\Delta_{-}^B/2^{i+2}  \implies \Delta_-(n) \geq b\Delta_-^B/2^{{n}+2},
        \end{equation} 
        where $b$ is equal to one if $\Delta_-^B<2^{n_B-1}$ and is equal to $2^{n_B}/2\Delta^B_-$ otherwise.
        We postpone the proof of this simple observation to the end of this section.
        \end{enumerate}
        
        Recalling the fact that $\|A_{\upsilon}^{-1} \|= \Delta_{\upsilon}^{-2}(n)$, and using the well-known inequality $|\det(A_{\upsilon})| \geq 1/\|A_{\upsilon}^{-1}\|^{N_{\upsilon}}$ finishes the proof of \eqref{eq: Cartan multi lower bound}.\\

        \textbf{Step 4.} Now, we may apply Cartan Lemma \ref{lem:Cartan multi} with $a=(\underline{h},\underline{g})$, $\epsilon=\varepsilon_{\upsilon}$ and $F= F_{\upsilon}$. Recalling the definition of $p_{n,\upsilon}(\delta)$ \eqref{eq: Cartan upper bound} we get 
        $$p_{n,\upsilon}(\delta) \leq C \tilde{n} (\delta^2 N_{\upsilon}/Cn^2)^{(c/N_{\upsilon}\ln(c^n(\Delta_B^{\upsilon})^{-1})},$$ which converges to $0$ as $\delta \to 0$ whenever $\Delta_{\upsilon}^B>0$ (the expression in the $\ln$ is greater than one by construction). By taking $\delta \to 0$, 
  this finishes the proof of Lemma \ref{lemma: minimal gap antigap}.

        \end{proof}
      \begin{proof}[Proof of \eqref{eq: cute lower bound}]
        For $\underline{g}=0$, eigenvalues of $H_n$ are  given by $\{E_i+ \sigma.h\}$ for $i=1,\dots, 2^{n_B}$, and $\sigma \in \Sigma_{\tilde{n}}$, where $\{E_i\}_{i=1}^{2^{n_B}}$ are energy levels of the bath Hamiltonian $H_B$. Then 
        $$\Delta_-(n)=\min_{(i,\sigma) \neq (j,\sigma ')} \{|E_i-E_j+(\sigma-\sigma').h| \}.$$  Thanks to the choice of $\underline{h}$, $|(\sigma-\sigma').h|\leq b\Delta_-^B/2 \leq \Delta_-^{B}/2$. This means
        $|E_i-E_j+(\sigma-\sigma').h| \geq \Delta^B_-/2$ if $i \neq j$ unifromly in $\sigma,\sigma'$. 
        
        Now take $i=j$. Notice that the set $G_n:= \{q \in \mathbb{Q}| 2^n \times q \in \mathbb{Z} \}$ is a group under addition. Notice that $(\sigma-\sigma').h^{-}/2b\Delta_-^B \in G_{n+2}$. This means $|(\sigma-\sigma').h^{-}/2b\Delta_-^B|\geq 1/2^{n+2}$ unless it is zero. Let $x \leq n $ be the largest index such that $\sigma(x) \neq \sigma'(x)$. If the 
        above expression becomes zero it means 
        $$(\sigma'(x)-\sigma(x))h_x^-/2b\Delta_-^B= \sum_{y < x} (\sigma(y)-\sigma'(y))h_y^-/2b\Delta_-^B.$$ 
        However, this is not possible because the right-hand side belongs to $G_{x+1}$, whereas, the left-hand side does not belong to this set. Hence $\sigma =\sigma'$. This proves the claim.           
        \end{proof}

\section{Local central limit theorem estimate} \label{sec: LCLT}

\begin{lemma} \label{lemma : LCLT}
Let $X_1, X_2,\dots$ be a sequence of i.i.d. random variables, with an absolutely continuous (w.r.t Lebesgue measure) density $\varphi(x) dx$. Furthermore, assume that $\varphi$ is bounded and compactly supported,  $\mathbb{E}(X_i)=0$, and $\mathbb{E}(X_i^2)=\mathsf{s}^2$.
Denote the pdf  of $ X_1+\dots+X_n$  by $\varphi^{*n}$, meaning $n$-fold convolution of $\varphi$. Then we have  
\begin{equation} \label{eq: LCLT1}
    \|\varphi^{*n} -\mathfrak{g}_n\|_{L^{\infty}} = o_n(1/\sqrt{n}), 
\end{equation}
where we recall $\mathfrak{g}_n$ as the pdf of the centered Gaussian with variance $n\mathsf{s}^2$. 
\end{lemma}
\begin{proof}
   Abbreviate $\varphi_n:= \varphi^{*n}$, by definition we have: 
   $$ \sup_x \sqrt{n}|\varphi_n(x) -\mathfrak{g}_n(x)|= 
   \sup_x \left|\sqrt{n} \varphi_n(\mathsf{s} \sqrt{n}x) - 
   \sqrt{n}\mathfrak{g}_n(\mathsf{s}\sqrt{n} x)\right|=: \Delta_n. $$
    Thanks to the Gnedenko version of the Local central limit theorem for densities, $\mathsf{s} \Delta_n \to 0$ as $n \to \infty$  (see Theorem 1,  Section 46, Chapter 8, Part III of \cite{gnedenko1968limit}). We also refer to this link of \href{https://terrytao.wordpress.com/2015/11/19/275a-notes-5-variants-of-the-central-limit-theorem/}{Terry Tao's blog} for a more pedagogical version of this theorem. Notice that the assumptions of the mentioned theorem (existence of the second moment of $X_i$, and the fact that $\varphi_n\in L^r$, $1<r \leq2$ for some $n$) are obviously satisfied by our more restrictive assumptions on $\varphi$. This finishes the proof.
\end{proof}

\section{LD estimate} \label{sec: app  LD estimate}

\begin{lemma}[Counting Lemma] \label{lem: counting 00}
For $\underline{\sigma}=(\sigma_1,\dots, \sigma_k) \in \Sigma_n^k$, recall the definition of $\mathfrak{J}_{\tau}(\underline{\sigma})$ for 
$\tau \in  \{ \pm1 \}^{k-1} \equiv \Sigma_{k-1}$ \eqref{def: J_tau}.  
Then, given any $\underline{\ell}=(\ell_{\tau})_{\tau\in \Sigma_{k-1}} \in \mathbb{N}_0^{\Sigma_{k-1}}$ such that 
 $\sum_{\tau} \ell_{\tau}=n$, we have: 
\begin{equation} \label{eq:countingI}
   \left| \left\{\underline{\sigma} \in \Sigma_n^k \big | \: |\mathfrak{J}_{\tau}(\underline{\sigma})| 
    =\ell_{\tau}, \: \forall \tau \right\} \right|= 2^n
    {n\choose \underline{\ell}}= 
    2^n \frac{n!}{\prod_{\tau }\ell_{\tau} !},
\end{equation} 
where the above product is for $\tau \in \{\pm1\}^{k-1}$.
\end{lemma}
\begin{proof}
    Thanks to the construction of $\{ \mathfrak{J}_{\tau} \}_{\tau}$, given  $\sigma_1$, and $\{ \mathfrak{J}_{\tau}\}_{\tau}$, the configuration $(\sigma_1, \dots, \sigma_k)$ is determined uniquely;  moreover, given the configuration $(\sigma_1,\dots,\sigma_k)$, $\sigma_1$ and $\{ \mathfrak{J}_{\tau} \}_{\tau}$
    can be constructed uniquely (a bijection exists here). Therefore, the above counting problem is equivalent to choosing $\sigma_1$ with $2^n$ choices and then counting all sets $\{ \mathfrak{J}_{\tau}\}_{\tau}$, with $\mathfrak{J}_{\tau} \subset \{1,\dots,n\} $, and $|\mathfrak{J}_{\tau}|=\ell_{\tau}$. This is a rather well-known counting problem, and the solution is given by the so-called multinomial coefficients (see \cite{stanley2011enumerative} Section 1.2) as written in \eqref{eq:countingI}. 
\end{proof}
\begin{proof}[Proof of Lemma \ref{lemma: counting00}]

  For $\underline{\sigma} \in \Sigma_n^k$, recall the definition of $\mathfrak{J}_{\tau}(\underline{\sigma})$ \eqref{def: J_tau}. Denote the set of \textit{typical} configurations by $A$: 
  \begin{equation} \label{eq:typical2}
      A:= \left\{  (\sigma_1,\dots, \sigma_k) \in \Sigma_n^k  \big| \: ||\mathfrak{J}_\tau| - n/2^{k-1}|<n^{1-1/4}\right\}.
  \end{equation}
  Then for any $0<a<1/2$ we should prove: 

   \begin{equation} \label{eq:countingIII}
       \frac{\left|A^c \right|}{(2^{n})^{k}}\leq o_n(\exp(-n^a)).
   \end{equation}
Let $0<b<3/4$,  and define 
$$\Gamma_b:= \{\underline{\ell} \in\{0,1,\dots,n\}^{\Sigma_{k-1}} | \exists \tau \quad \text{such that,} \quad |\ell_{\tau} -n/2^{k-1}| > n^b \},$$

Notice that $|\Gamma_b| \leq (n+1)^{2^{k-1}}$. By the above definitions, we have: 

\begin{equation}  \label{eq:obviousinclusion}
A^c \subset \bigcup_{\underline{\ell} \in \Gamma_b} \{  \underline{\sigma} \in \Sigma_n^k  \big| \: |\mathfrak{J}_\tau(\underline{\sigma})|= \ell_{\tau} \}= : \bigcup_{\underline{\ell} \in \Gamma_b} A_{\underline{\ell}}. 
\end{equation}
Notice that $|A_{\underline{\ell}}|$ is computed in \eqref{eq:countingI}.
Denoting $\lambda_{\tau}:= \ell_{\tau}/n$ (by convention set $0 \ln(0)=0$), and applying the 
Stirling's approximation we get for $\underline{\ell} \in 
\Gamma_b $: 
\begin{equation} \label{mainineqaulity counting}
\frac{|A_{\underline{\ell}}|}{2^{nk}} \leq C\sqrt{n} 
\exp(-n\left[\sum_{\tau}\lambda_{\tau} \ln(\lambda_{\tau}) -
\ln(1/2^{k-1}) \right]) \leq C\sqrt{n} \exp(-2n^{2b-1}),
\end{equation}
where in the first inequality we used Stirling's approximation. The second estimate is obtained in the following manner:
notice that $\lambda_{\tau} \in [0,1]$, and $\sum_{\tau} \lambda_{\tau} =1$. Therefore, the expression $\sum_{\tau }\lambda_{\tau} \ln (\lambda_{\tau}) - \ln(1/2^{k-1})$ is the relative entropy (KL divergence) of two discrete probability distributions: $P(\tau)= \lambda_{\tau}$, and 
$Q(\tau)= 1/2^{k-1}$ for $\tau \in \{\pm1\}^{k-1}$. Therefore, applying Pinsker's inequality, the above expression is lower bounded by $2 (\delta(P,Q))^2$, where $\delta(P,Q)$ is the total variation norm, which itself is lower bounded by $n^b/n$ thanks to the choice of $\underline{\ell} \in \Gamma_b$.   
Taking $0<a<2b-1<1/2$, performing a union bound on \eqref{eq:obviousinclusion}, knowing that $|\Gamma_b| \leq (n+1)^{2^{k-1}}$, and taking advantage of \eqref{mainineqaulity counting} we conclude \eqref{eq:countingII}.

\end{proof}

\section{Perturbation Theory} \label{sec:PTtheory}

Here, we essentially recall Chapter Two, Sections 1 and 4 of Kato's book \cite{kato1995perturbation}  in our notation, for the reader's convenience:
Recall  for $n>n_B$
$$H_n = H_{n-1/2}+ g_n\alpha^{n-n_B}X_1 X_n, \quad \text{with} \quad H_{n-1/2}= H_{n-1}+h_nZ_n. $$
Let us denote: 
$$H_0:= H_{n-1/2}, \quad g\equiv g_n, \quad V:= X_1X_n,\quad  H_g:=H_0 +g_n\alpha^{n-n_B}V.$$
Assume that the event $\mathcal{A}_{n-1,n}$ holds,  fix a patch $p \in \mathcal{P}_{n-1,k}$, and  $\mu \in \{ \pm 1\}$. As we defined $\caA_{n-1,n}$ in Section \ref{sec: term}, this means $p+\mu h_n$ is a spectral patch of $H_0$ such that 
\begin{equation} \label{eq: gap PT}
\mathrm{dist}(\mathrm{Conv}(p+\mu h_n), \mathrm{spec}(H_0)\setminus \mathrm{Conv}(p+\mu h_n)) > \alpha^{\theta(n-1) }.
\end{equation}
Thanks to the above expression,  we may choose $\Gamma$ to 
be a counter-clockwise oriented stadium-shaped curve in the complex plane encircling  $p+\mu h_n$,  such that
\begin{equation} \label{eq: PT basic bound 00}
\sup_{z \in \Gamma} \| (z-H_0)^{-1}\| \leq 2/\alpha^{\theta (n-1)},\quad  |\Gamma|\leq (\pi+2(k-1)) \alpha^{\theta (n-1)} , 
\end{equation}
where we used the fact that the distance among consecutive levels inside $p+\mu h_n$, is at most $\alpha^{\theta(n-1)}$ since $p \in \mathcal{P}_{n-1,k}$.

 For all $g$ in a unit disk and for all $z \in \Gamma$, $\|(H_g-H_0)(z-H_0)^{-1} \| < 1$, if we take $n$ large enough. This is because $\alpha^{n-n_B}=C\alpha^n <\alpha^{\theta(n-1)}/4$ (recall that  $\alpha^{1-\theta}<1/2$), where we also  recall that $g_n\alpha^{n-n_B}$ is the norm of the perturbation.  Then thanks to a Neumann series analysis we deduce that $z-H_n$ is invertible for all $z \in \Gamma$ (it is sufficient to consider the Neumann series of $(1+(z-H_0)^{-1}(H_g-H_0))^{-1}$). Therefore, for all $g$ in  a unit disk, if we define: 
\begin{equation}
    P_g:=\frac{1}{2 \pi i} \int_{\Gamma} \frac{1}{z-H_n} dz, 
\end{equation}
$P_g$ is analytic and has constant rank. Moreover, it represents the projection onto the eigenspace of the patch after the perturbation. \\
Expanding the resolvent ($1/(z-H_n)$) thanks to the Neumann series, we get for all $g_n\equiv g$ in a unit disk (see \cite{kato1995perturbation} Chapter 2, (1.14), (1.18)\footnote{We have a different convention for the sign of the resolvent compared to \cite{kato1995perturbation}.}): 
\begin{equation} \label{eq:seriesexpansion0}
    P_g= P_0+\sum_{m=1}^{\infty} g^m B^o_m, \quad \text{with} \quad B_m^o:= \frac{(\alpha^{n-n_B})^m}{2\pi i} \int_{\Gamma} 
    \frac{1}{z-H_0}V\frac{1}{z-H_0}\dots V\frac{1}{z-H_0} dz,
\end{equation}
 where in the above expression $V=X_1X_n$, and it appears $m$ times. Then, we have: 
\begin{equation} \label{eq:expansion bound0}
\|B_m^o\| \leq \frac{(\alpha^{n-n_B})^m}{2\pi} |\Gamma| \sup_{z \in \Gamma} \|(z-H_0)^{-1} \|^{m+1} \leq  
\left(C(2 \alpha^{1-\theta})^n  \right)^m,
\end{equation}
where we used \eqref{eq: PT basic bound 00},  the fact that $k\leq 2^n$, and we did a simple computation. Notice that $C$ depends on the bath and $\alpha$.\\
From \cite{kato1995perturbation} Chapter Two, Sections 4 and 5 we recall:
since $P_g$ is analytic on the unit disk, there exists  $U_g$, which is analytic in the same domain, $U_g^{-1}$ exists and is analytic in the same domain. Moreover, we have
 $U_gP_0U_g^{-1}=P_g$. Furthermore, denote
 $$\mathcal{R}_g:= P_gU_g =U_g P_0, \quad \mathcal{L}_g:= U_g^{-1}P_g=P_0U_g^{-1}.$$ 
 Then $\mathcal{L}_g,\mathcal{R}_g$ are given by the following power series: (see \cite{kato1995perturbation}, Chapter 2, 4.20, 4.21, 4.22): 
\begin{align}
     \mathcal{R}_g&= P_0+\sum_{m=1}^{\infty} g^m R^{(m)}, \quad \quad  \mathcal{L}_g = P_0+\sum_{m=1}^{\infty} g^m L^{(m)}, \text{with}, \nonumber  \\
     & mR^{(m)}=mB_m^o P_0 + \sum_{i=1}^{m-1} (m-i) B_{m-i}^oR^{(i)}, \nonumber \\
     & m L^{(m)}=mP_0B_m^o+ \sum_{i=1}^{m-1} (m-i)L^{(i)}B_{m-i}^o, \label{eq:seriesexpansion1}
 \end{align}
where $B_m^o$ is given in \eqref{eq:seriesexpansion0}. Combining the above expressions and rearranging the series via a direct computation, we get (denote $L^{(0)}=R^{(0)}=P_0$): 
\begin{align}
   & H_P(g):=U_g^{-1}P_gH_gP_gU_g= \mathcal{L}_g(H_0+g\alpha^{n-n_B}V) \mathcal{R}_g = P_0H_0P_0 + \sum_{k=1}^{\infty} \tilde{B}_k g^k, \quad \text{with}  \nonumber \\
   &\tilde{B}_k:= \sum_{i=0}^k L^{(i)} H_0 R^{(k-i)} + \alpha^{n-n_B} \sum_{i=0}^{k-1} L^{(i)}VR^{(k-1-i)}. \label{eq:seriesexpansion3}
\end{align}
Given the bound \eqref{eq:expansion bound0}, and the recursion relations in \eqref{eq:seriesexpansion1} thanks to a direct induction (strong induction,  using the fact that $\sum_{i=1}^n2^i <2^{n+1}$),  taking advantage of the sub-multiplicativity of the norm, and adjusting $C$ by multiplying it by $2$ we get:
\begin{equation} \label{eq:expansionbound1}
    \|R^{(m)} \|, \|L^{(m)}\| \leq (C(2\alpha^{(1-\theta)})^n)^m.
\end{equation}
Inserting the above bound into the expression of $\tilde{B}_k$ \eqref{eq:seriesexpansion3}, and using the fact that $\| H_0\| \leq Cn$, for $n$ sufficiently large  we get: 
\begin{equation} \label{eq:boundPT}
    \|\tilde{B}_k \| \leq   ((4\alpha^{(1-\theta)})^n)^k.
\end{equation}
Since $H_0$, and $V$ are self-adjoint, for real $g$, $H_{P}(g)$ is Hermitian, and one can choose $\tilde{B}_k$ to be self-adjoint(\cite{kato1995perturbation} Chapter 2, Section 6). Finally, notice that since $P_0H_P(g)P_0=H_P(g)$, using properties of $X_1X_n$, one may deduce that $\tilde{B}_k=0$ for odd $k$; however, we do not use this fact. Let us emphasize that in \eqref{eq:expansion} one should set $B_k=: \tilde{B}_k(4\alpha^{(1-\theta)})^{-nk}$.

\section{Basic Gaussian properties} \label{sec: app Gaussians}
\subsection{Basic Lemma}
\begin{lemma}
Recall the Gaussian densities $\mathfrak{g}_{n}$, and $0<\rho<1/2$ introduced in Section \ref{sec: intervals!}. Then for any  $0< \delta_1<1/2-\rho$, we have:  
  \begin{align} \label{eq: properties of m}
   \sup_{|x-y|<n^{\rho}} \left|\mathfrak{g}_{n_1}(x) -\mathfrak{g}_{n_1}(y)\right| = O_n(n^{\rho+\delta_1}/n) =o_n(1/\sqrt{n}).
\end{align}
\end{lemma}
\begin{proof}
Without loss of generality, assume that $|x|\leq|y|$. If $|y|>n^{1/2+\delta_1}$, then both terms are of order $O_n(\exp(-n^{2\delta_1}))$ and we are done. Otherwise, combining the following observations:
$$ \forall \: z>0, \: \: |1-\exp(-z)| \leq z, \qquad|y^2-x^2| =O_n(n^{1/2+\delta_1+\rho}), \qquad  \mathfrak{g}_{n_1}(x) \leq C/\sqrt{n}, $$  finishes the proof.
\end{proof}
\begin{proof}[Proof of \eqref{eq: Gntt' bound}]
From \eqref{eq: properties of m}, by an integration, change of variable, and an $L^1-L^{\infty}$ bound, we deduce \eqref{eq: Gntt' bound} for $k=1$. For $k>1$, by obtaining a standard telescopic sum and using  the above observation, as well as the fact that for any $i$ trivially we have $\sup_{t_i} G_{n_1}^i(I_i^{t_i})\leq C/\sqrt{n}|I_i|$, we deduce \eqref{eq: Gntt' bound}. 
\end{proof}

 \subsection{Proof of Lemma \ref{lem: dis convolution}}

 \begin{proof} [Proof of Lemma \ref{lem: dis convolution}]
\textbf{Proof of \eqref{eq: dis convolution}}:
Let us assume $\mathcal{I}$ to be bounded. The unbounded case is trivial since the right-hand side of the bound becomes infinite. 
 In the expression \eqref{eq: dis convolution}, $\sum_{\mathtt{t}}$ is understood as sum over $\mathtt{t}\in \mathcal{J}_{n_0}$. Let us  first extend this sum. Only in this section of the proof, from now on $\sum_{\mathtt{t}}$ should be understood as a sum over $\mathtt{t}\in\mathcal{J}$ (defined in \ref{sec: intervals!}). First, observe that: 
 \begin{equation} \label{eq: extended t bound}
    \Delta_o:=\left| \sum_{|\mathtt{t}|> cn_0} G_{n_1}(I^{\mathtt{t}})G_{n_0}(J_{\mathtt{t}})\right|  \leq O_n(\exp(-cn_0)) |I| \leq o_n(1/\sqrt{n})|I|,
 \end{equation}
 where we  bound $G_n(I^{\mathtt{t}})$ by $O_n(\exp(-cn_0))|I|$ for $|\mathtt{t}|>O_n(n_0)$; then, we used the fact that $\sum_{|\mathtt{t}|>cn_0}G_{n_0}(J_{\mathtt{t}}) \leq 1.$ Notice that the term $o_n(\cdots)$ is independent of $I$.\\
 
On the other hand,  density of $G_n$ is given by $(\mathfrak{g}_{n_0}*\mathfrak{g}_{n_1})(z)$, where $*$ stands for the convolution. Using this fact, by definition, we have: 
 \begin{align}
 \Delta_1 := &\left| \sum_{\mathtt{t}\in \mathcal{J}}
 G_{n_1}(I^{\mathtt{t}}) G_{n_0}(J_{\mathtt{t}})-G_{n_0+n_1}(I)\right|  \nonumber\\ 
\leq & \:   \sum_{\mathtt{t}\in \mathcal{J}} \int\int \chi(z\in I) \chi(x \in J_{\mathtt{t}}) 
 |\mathfrak{g}_{n_1}(z-\mathtt{t}) -\mathfrak{g}_{n_1}(z-x)|\mathfrak{g}_{n_0}(x) dz dx  \nonumber \\ 
 \leq & \:   o_n(1/\sqrt{n})|I|\sum_{\mathtt{t}} \int \chi(x \in J_{\mathtt{t}}) \mathfrak{g}_{n_0}(x)dx \leq o_n(1/\sqrt{n})|I|,
 \end{align}
 where, in the second line, for the first term we simply rewrote it in terms of the pdf of $\mathfrak{g}_{n_0}$ and $\mathfrak{g}_{n_1}$; then we made a change of variable $z'=z+\mathtt{t}$. For the other term, we used the explicit pdf $\mathfrak{g}_{n_0}*\mathfrak{g}_{n_1}$, then we used the partition of unity $1=1(x)= \sum_{\mathtt{t}}\chi(x \in J_{\mathtt{t}})$. \\
 In the third line, we bounded the term $|\mathfrak{g}_{n_1}(z-x)-\mathfrak{g}_{n_0}(z-\mathtt{t})|$  by $o_{n}(1/\sqrt{n})$ thanks to \eqref{eq: properties of m} since $|x-\mathtt{t}|<n^{\rho}$. Then we performed the integral in $z$ and got $|I|$. The last bound is trivial by partition of unity. Notice again that the error term is independent of $I$. This, in combination with \eqref{eq: extended t bound} finishes the proof of 
 \eqref{eq: dis convolution}. \\

 \textbf{Proof of \eqref{eq: max entropy Gaussian}}: We omit the straightforward proof.\\

 \textbf{Proof of \eqref{eq: simple gaussina bound}}:
Fix $\delta>0$ such that $\rho+2\delta <1/2$. Thanks to \eqref{eq: properties of m}, and an $L^1-L^{\infty}$ bound, we have: 
     
     $$\left|\int_{\mathtt{t}}^{\mathtt{t+{n_0}}^{\rho}} \mathfrak{g}_{n_1}(x/2)dx -n_0^{\rho} \mathfrak{g}_{n_1}(\mathtt{t}/2)\right| \leq O_n(n^{2\rho +\delta}/n) .$$
 Using the above bound for $|x|,|\mathtt{t}|\leq n^{1/2+\delta}$, and using the exponential decay of $\mathfrak{g}_{n_1}$ we get: 
  \begin{equation} \label{eq: temop reimann saum}
     \Delta:=\left|n_0^{\rho}\sum_{\mathtt{t}} \mathfrak{g}_{n_1}(\mathtt{t}/2) - \int \mathfrak{g}_{n_1}(x/2)dx \right| \leq o_n(1), \implies   n_0^{\rho} \sum_{\mathtt{t}} \mathfrak{g}_{n_1}(\mathtt{t}/2) \leq C.
 \end{equation}
    where we used the fact that $\|\mathfrak{g}_{n_1} \|_{L^1}=1$, for the above implication. On the other hand, by definition of $I$,  $G_{n_1}$, $\mathfrak{g}_{n_1}$ for large $n$, we have $(x-\mathtt{t})^2\geq\mathtt{t}^2/4$ for any $x\in I$ (thanks to the fact that $\mathcal{I}$ is bounded). This means  $G_{n_1}(I^{\mathtt{t}}) \leq |I|\mathfrak{g}_{n_1}(\mathtt{t}/2)$. This, in combination with \eqref{eq: temop reimann saum} finishes the proof. 
 \end{proof}

\section*{Declarations}
AI declaration: The technical core of this work was developed before the widespread adoption of large language models (LLMs). In particular, the technical content of the present manuscript is substantially unchanged from that of the first arXiv version, posted in June 2025. LLMs, primarily ChatGPT (GPT-5.6), were used in the preparation of the present version for proofreading, identifying mathematical inconsistencies, improving clarity and readability, and assisting in the creation of illustrations. The authors reviewed and verified all AI-assisted material.

\section*{Acknowledgments}
We are grateful to François Huveneers for enlightening discussions on the topic of this paper as well as his encouragement.  
\subsection*{Funding}
W.D.R. and A.H. (while the latter was a postdoctoral fellow at KU Leuven)  were supported by the FWO and F.R.S.-FNRS under the Excellence of Science (EOS) program through the research project G0H1122N EOS 40007526 CHEQS, the KULeuven Runner-up Grant No. iBOF DOA/20/011, and the internal KULeuven Grant No. C14/21/086.\\
A.H. was also supported by the ERC Consolidator Grant ProbQuant (jointly with the Swiss State Secretariat for Education, Research and Innovation).\\

\subsection*{Competing Interests}
The authors have no relevant financial or non-financial interests to disclose.
\subsection*{Data availability.} Data sharing is not applicable.
We do not analyze or generate any datasets, because our work proceeds within a theoretical and mathematical approach.

\bibliographystyle{plain}
\bibliography{bib_QSM}

\end{document}